# A Theory of Composing Protocols

Laura Bocchi[a], Dominic Orchard[a,b], and A. Laura Voinea[c]

a University of Kent, UK

b University of Cambridge, UK

c University of Glasgow, UK

**Abstract** In programming, protocols are everywhere. Protocols describe the pattern of interaction (or communication) between software systems, for example, between a user-space program and the kernel or between a local application and an online service. Ensuring conformance to protocols avoids a significant class of software errors. Subsequently, there has been a lot of work on verifying code against formal protocol specifications. The pervading approaches focus on distributed settings involving parallel composition of processes within a single monolithic protocol description. However we observe that, at the level of a single thread/process, modern software must often implement a number of clearly delineated protocols at the same time which become dependent on each other, e.g., a banking API and one or more authentication protocols. Rather than plugging together modular protocol-following components, the code must re-integrate multiple protocols into a single component.

We address this concern of combining protocols via a novel notion of 'interleaving' composition for protocols described via a process algebra. User-specified, domain-specific constraints can be inserted into the individual protocols to serve as 'contact points' to guide this composition procedure, which outputs a single combined protocol that can be programmed against. Our approach allows an engineer to then program against a number of protocols that have been composed (re-integrated), reflecting the true nature of applications that must handle multiple protocols at once.

We prove various desirable properties of the composition, including behaviour preservation: that the composed protocol implements the behaviour of both component protocols. We demonstrate our approach in the practical setting of Erlang, with a tool implementing protocol composition that both generates Erlang code from a protocol and generates a protocol from Erlang code. This tool shows that, for a range of sample protocols (including real-world examples), a modest set of constraints can be inserted to produce a small number of candidate compositions to choose from.

As we increasingly build software interacting with many programs and subsystems, this new perspective gives a foundation for improving software quality via protocol conformance in a multi-protocol setting.

# 1 Introduction

Protocols are everywhere. Whenever two entities need to communicate (perhaps via function calls, or messages sent over a channel), a protocol can be used to ensure that both parties effectively exchange information. Protocols can be seen as a *specification* of communication, and as such have been leveraged for the purposes of verification in programming languages, e.g., session types [9, 25, 26, 27], choreographies [13, 14, 39], typestate [42], behavioural types in general [21, 30], and more.

There may be many protocols that a program has to conform to, capturing different interactions between different parts of a system. Here we use the term *protocol* to denote a specification of the interaction patterns between different system components. For example, when considering distributed systems, a protocol may describe the causalities and dependencies of the communication between processes. To give a more concrete intuition, an informal specification of a protocol for an e-banking system may be as follows: *The banking server repeatedly offers a menu with three options: (1) request a banking statement, which is sent back by the server, (2) request a payment, after which the client will send payment data, or (3) terminate the session.* We elaborate on this example later, using it as a motivating example.

Much of the work on systematising the process of programming against a specification assumes a monolithic view of protocols: a protocol is often given for the entire system, explaining the communication between all parties involved. This up-front, single point of definition runs contrary to the human aspects of real-world programming, in which a programmer gradually pieces together their code, perhaps heavily leveraging libraries, to reach their intended goal; programs are gradual *compositions*.

A view that is globally defined once does not reflect the real process of software composition. In contrast, a view that defines lots of local protocols or sub-protocols places the burden of configuring their interaction on the programmer: programmers must themselves work in a situation where they have to consider many smaller protocols and work out how they want dependencies between them to be resolved. Instead, we propose that a flexible, non-monolithic notion of *protocol composition* (and possibly recomposition, when a piece of code is refactored and rewritten, or reused) is needed to support the engineering of protocol-dependent code. Ideally, such a notion should support well-founded semi-automated protocol composition and support implementation with formal guarantees.

This work lays a foundation for compositional protocol engineering based on a notion of *interleaving composition* of protocols. An interleaving composition of two protocols 'weaves' them together into a single unified protocol. This differs from sequential composition, in which one protocol follows the other or one's inputs are coupled to the other's outputs. It differs from parallel composition, which traditionally (e.g., in CCS or CSP) describes a semantic interleaving of programs; our approach calculates a single syntactic protocol specification.

We address, in general terms, the question of what a correct protocol composition is, and introduce a syntactic definition of composition that characterises finite sets of *correct* interleaving compositions, each representing a 'good way' to interleave the component protocols with respect to domain-specific user-specified constraints.

The resulting approach gives a theoretical basis for protocol (re-)engineering based on a process calculus with constraint annotations. Interleaving composition has the purpose of enhancing the awareness of what a protocol means, and facilitating reasoning about its properties. We give an algorithmic implementation of interleaving composition supporting the process of defining protocols and inspecting the generated compositions, and code generation of skeletons of processes following a given protocol (composite or not). Code generation is based on Erlang/OTP gen_statem behaviour [1] allowing code to be migrated in subsequent compositions and reused. Correspondence of our protocol language with Finite State Machines (FSM) via directed graphs yields straightforward links between protocols and FSM-structured code.

A related line of work defines composition as *run-time weaving*, for example applying principles of aspect-oriented programming to protocol composition [43]. Unlike [43], we *statically* derive protocol compositions that enable (human/automated) reasoning and verification of their properties. Another related line of work is *automata composition* [7, 20, 23]. Team Automata [7, 20] provide several means of composing machines via synchronization on their common actions, and give a formal framework for composition. Unlike Team Automata, we express composition constraints orthogonally to communication: instead of synchronization on common actions, we use 'asserts'/'requires' as contact points for composition, and reason about the properties of a composite protocol from the perspective of the application logic. The resulting composition relation given in this work is not characterizable as one of the synchronizations of Team Automata (discussed further in Section 6).

Unlike in aforementioned works, our protocols are mono-threaded. This is not unusual in literature, e.g., session types are essentially mono-threaded [9, 25, 26]. Also real-world protocols, such as POP2, POP3, and SMTP, are described in their RFCs as single state machines and have been modelled, without parallel composition, as session types [11, 22, 29]. Still, one could use parallel composition as a basis for defining protocol compositions (as in Team Automata), and this would yield general and syntactically concise concurrent specifications. These concurrent specifications, with all their interleavings, would be harder for a human to understand than a well-specified interleaving composition. We explore an unusual approach to composition, with the purpose of supporting a process of human understanding of what protocol composition should be. Our novel approach is also reflected in the tool. The code for the composition of two protocols is not the composition of the existing implementations (plus some adaptor code) – as one would expect. The tool generates new code via: (1) automated generation of a stubs of the new composite protocol, and (2) migration of relevant parts of the old code – besides the stub infrastructures – into the new code. This yields simple mono-thread code that are still close to the protocol's structure.

### 1.1 Motivating Example

The banking protocol discussed earlier in this section can be formally specified as $S_B$ in Figure 1 using a process calculus notation. $S_B$ repeatedly (via a fixed point $\mu\mathtt{t}$) offers (denoted &) three options: option statement is followed by a send action (denoted !) of a message with the bank statement, option payment is followed by a receive action

$$S_B := \mu\mathtt{t}.\&\begin{Bmatrix} \mathsf{statement} : \mathtt{!statement.t} \\ \mathsf{payment} : \mathtt{?details.t} \\ \mathsf{logout} : \mathtt{end} \end{Bmatrix}$$

$$S_A := \mathtt{?pin}.\oplus\begin{Bmatrix} \mathsf{ok} : & \mu\mathtt{r}.\mathtt{!id.?tan}.\oplus\begin{Bmatrix} \mathsf{ok} : & \mathtt{r} \\ \mathsf{fail} : & \mathtt{r} \end{Bmatrix} \\ \mathsf{fail} : & \mathtt{end} \end{Bmatrix}$$

**Figure 1** Banking ($S_B$) and PIN/TAN authentication ($S_A$) protocols. The arrows show the desired dependencies: entering the loop in $S_B$ requires correct PIN authentication (i.e., at ok, first occurrence in $S_A$) and each payment iteration in $S_B$ requires TAN authentication. (i.e., at ok, second occurrence in $S_A$).

(denoted ?) with details of the payment, and option logout is followed by termination of the protocol (denoted `end`). After each of the first two options, the control flow goes back to the initial state (via `t`).

Assume now that we want to extend $S_B$ with two-level authentication: one level for accessing the service and one additional level for each payment transaction. Concretely, we wish to compose $S_B$ with the PIN/TAN (Personal Identification Number/Transaction Authentication Number) protocol modelled in Figure 1 as $S_A$ which offers two-stage authentication. The first stage is pin authentication: the server receives a `pin` and decides (⊕) whether to continue (i.e., ok) or terminate (i.e., fail). If ok is chosen, the protocol enters a loop (i.e., $\mu$`r`) that manages multiple TAN authentications, supporting multiple transactions requiring an additional level of security. In the loop, the server sends an identifier `id` for which the client must send back a `tan`. The server notifies the client about the correctness of the `tan` with either ok or fail.

We want to compose the banking and authentication protocols into a single protocol where their actions follow a specific interleaving: access to the banking service requires a PIN authentication, and each payment instance/iteration requires an extra TAN authentication (see dotted arrows in Figure 1). This specific interleaving entails an authorization property, which we later express and ensure by using assertion annotations. Moreover, we want tools that facilitate engineering of programs implementing interleaving compositions. For example, we want to obtain a skeleton implementation for the banking and PIN/TAN protocol, and in a second stage we want to reuse the code when composing banking with a different multi-factor authentication protocol, e.g., offering other options besides TAN, such as keycard authentication.

### 1.2 Contributions

In Section 2, we define a process-calculus-based notation for protocols with 'assertions'. Assertions specify contact points and constraints between component protocols, to be checked statically. In Section 3, we give a definition of interleaving composition that is relational, as there may be many valid interleaved protocols (or even none). In Sec-tion 3.1.1 we provide two less restrictive definitions of interleaving composition via two additional rules, *weak branching* and *correlating branching* that capture more scenarios but enjoy a weaker fairness properties. In Section 4, we prove that our composition relation returns *correct* interleaving compositions, namely: (*behaviour preservation*) interleaving compositions only perform sequences of actions that may be performed by either of the component protocols; (*fairness*) interleaving compositions eventually execute the next available action of each protocol; (*well-assertedness*) interleaving compositions always satisfy requirements prescribed by the assertions in the protocols being composed. Thus, we establish that the composition relation produces sets of protocol compositions that are correct-by-construction. Our definition is sound but not complete, as discussed in Section 4.4. In Section 5, we introduce a tool for protocol engineering in Erlang, which implements interleaving composition, generation and protocol extraction to/from Erlang gen_statem code. Section 6 discusses related work.

## 2 Asserted Protocols

We introduce a language of protocol specifications to abstractly capture essential features of sequential computation: sequencing, choice, and looping. Our protocol language somewhat resembles Milner's CCS [38] or the $\pi$-calculus [41], but without parallel composition or name restriction, and has some relation to Kleene algebras [34] but we provide more general patterns of recursion via recursive binders rather than a single closure operator. Generally, two protocols can be composed in several ways, each reflecting a possible interleaving of the actions of the two protocols. Not all such interleavings are meaningful depending on the scenario or domain. The protocol language therefore includes a notion of 'assertions' which can be used to capture the behavioural constraints of a protocol to guide interleaving composition in a meaningful way; they act as a specification of minimal 'contact points' between protocols akin to pre- and post-conditions. Following an explanation of the syntax and various examples, we give an operational model to the protocol language which serves to explain both the program semantics which it abstracts, and the meaning of the assertion actions.

**Definition 1 (Asserted protocols)** *Asserted protocols, or just* protocols *for short, are ranged over by $S$ and are defined as the following syntax rules:*

| | | | | |
|---|---|---|---|---|
| $S$ | $::=$ | $p.S$ | *action prefix* | |
| | $\mid$ | $+\{\mathrm{l_i} : S_i\}_{i \in I}$ | *branching* | |
| | $\mid$ | $\mu\mathtt{t}.S$ | *fixed-point* | |
| | $\mid$ | $\mathtt{t}$ | *recursive variable* | |
| | $\mid$ | $\mathtt{end}$ | *end* | |
| | $\mid$ | $\mathsf{assert}(n).S$ | *assert (produce)* | assertion fragment |
| | $\mid$ | $\mathsf{require}(n).S$ | *require* | assertion fragment |
| | $\mid$ | $\mathsf{consume}(n).S$ | *consume* | assertion fragment |

*where $p \in \mathscr{P}$ ranges over prefixing actions, $\mathrm{l} \in \mathscr{L}$ ranges over labels used to label each branch of the $n$-ary branching construct, $\mathtt{t}$ ranges over protocol variables for recursive*

*protocol definitions, and* $n \in \mathcal{N}$ *ranges over names of logical atoms used by assertions. The sets of actions* $\mathcal{P}$*, labels* $\mathcal{L}$*, and names* $\mathcal{N}$ *are parameters to the language and thus can be freely chosen. Furthermore* $+$ *ranges over a set of operators* $\mathcal{O}$ *used to represent branching choice and thus can also be instantiated.*

The prefixing action provides sequential composition (in the style of process calculi). Branching is $n$-ary, taking the form of a set of protocol choices with a label $\mathrm{l_i}$ for each choice. Looping behaviour is captured via the recursive protocol variable binding $\mu\mathtt{t}$, which respects the usual rules of binders, and recursion variables $\mathtt{t}$. Protocols can be annotated with assertions to introduce guarantees $\mathsf{assert}(n)$, requirements $\mathsf{require}(n)$, and linear requirements $\mathsf{consume}(n)$: $\mathsf{assert}(n)$ introduces a true logical atom $n$ into the scope of the following protocol, $\mathsf{require}(n)$ allows the protocol to proceed only if $n$ is in the scope (basically $\mathsf{consume}(n)$ presupposes $\mathsf{require}(n)$), and $\mathsf{consume}(n)$ removes the truth of logical atom $n$ from the scope of the following protocol.

We assume variables to be guarded in the standard way (they only occur under actions or branching). To simplify the theory, we assume that: (1) nested recursions are guarded, ruling out protocols of the form $\mu\mathtt{t}.\mu\mathtt{t}'.S$, with no loss of generality since $\mu\mathtt{t}.\mu\mathtt{t}'.S$ is behaviourally equivalent to $\mu\mathtt{t}.S[\mathtt{t}/\mathtt{t}']$, and (2) in $\mu\mathtt{t}.S$ variable $\mathtt{t}$ occurs free at least once in $S$, with no loss of generality since e.g., $\mu\mathtt{t}.\mathtt{?pay.end}$ is behaviourally equivalent to `?pay.end`. Unless otherwise stated, we consider protocols to be closed with respect to these recursion variables.

**Remark 1 (Language instantiation)** *In the examples we often instantiate the prefixing actions* $\mathcal{P}$ *to sends* `!T` *and receives* `?T` *capturing interaction with some other concurrent program, i.e.,* $p \in \{\mathtt{!T}, \mathtt{?T}\}$ *where* $T$ *is a type (e.g., integers, strings), and instantiate choice* $+$ *to a pair of* polarised choice *operators:* $+ \in \{\oplus, \&\}$*, either offering of a choice* $\oplus$ *or selecting from amongst some choices* $\&$*. This yields a session types-like syntax similar to the one used by Dardha, Giachino and Sangiorgi. [18].*

Examples often colour assertions green and labels purple for readability.

### 2.1 Assertion Examples

Consider a payment process `?pay.end` that receives a payment and terminates, and a dispatch process `!item.end` that sends a product link and terminates. We can interleave these two protocols in two ways: `?pay.!item.end` (payment first) or `!item.?pay.end` (dispatch first). By using assertions, we can require that payment happens before dispatch: below, $I_1$ asserts the logical atom *paid* as a post-condition to receiving payment while in $I_2$ the sending action depends on the logical atom *paid* as a pre-condition, and in doing so consumes it.

$$I_1 = \mathtt{?pay}.\mathsf{assert}(\mathit{paid}).\mathtt{end} \qquad I_2 = \mathsf{consume}(\mathit{paid}).\mathtt{!item.end}$$

The only interleaving composition of $I_1$ and $I_2$ that satisfies the constraints posed by the assertions is: $\mathtt{?pay}.\mathsf{assert}(\mathit{paid}).\mathsf{consume}(\mathit{paid}).\mathtt{!item.end}$.

Linear constraint $\mathsf{consume}(n)$ models a guarantee that can be used once, whereas non-linear constraint $\mathsf{require}(n)$ does not consume $n$. Using a mix of linear and non-linear constraints, we can model a prepaid buffet scenario where a payment remains valid (hungry) for several iterations until the meal ends (end):

$$\mu\mathtt{t}.\&\{\mathsf{hungry} : \mathsf{require}(paid).!\mathtt{food}.\mathtt{t},\ \mathsf{end} : \mathsf{consume}(paid).\mathtt{end}\}$$

**Example 1 (Asserted banking and PIN/TAN)** *The informal requirement on the banking and PIN/TAN example discussed in the introduction can be modelled using assertions. An asserted version of the banking protocol, given below as $S'_B$, uses $\mathsf{require}(pin)$ to ensure a successful PIN authentication before accessing the banking menu; $\mathsf{consume}(tan)$ to require one successful TAN authentication for each iteration involving a payment; and $\mathsf{consume}(pin)$ to remove the PIN guarantee when logging out. Assertions $\mathsf{assert}(pay)$ and $\mathsf{consume}(pay)$ ensure TAN authentication only happens in case of payment.*

$$S'_B = \mathsf{require}(pin).\mu\mathtt{t}.\&\left\{\begin{array}{ll} \mathsf{statement}: & !\mathtt{statement}.\mathtt{t} \\ \mathsf{payment}: & \mathsf{assert}(pay).\mathsf{consume}(tan).?\mathtt{details}.\mathtt{t} \\ \mathsf{logout}: & \mathsf{consume}(pin).\mathtt{end} \end{array}\right\}$$

*In the asserted authentication protocol $S'_A$ below, $\mathsf{assert}(pin)$ and $\mathsf{assert}(tan)$ provide guarantees of successful PIN and TAN authentication, respectively:*

$$S'_A = \ ?\mathtt{pin}.\oplus\left\{\begin{array}{ll} \mathsf{ok}: & \mathsf{assert}(pin).\mu\mathtt{r}.\mathsf{consume}(pay).!\mathtt{id}.?\mathtt{tan}.\oplus\left\{\begin{array}{ll} \mathsf{ok}: & \mathsf{assert}(tan).\mathtt{r} \\ \mathsf{fail}: & \mathtt{r} \end{array}\right\} \\ \mathsf{fail}: & \mathtt{end} \end{array}\right\}$$

### 2.2 Protocol Semantics

The semantics of a protocol is given in Definition 2 in terms of an environment that keeps track of guarantees, and lets protocols progress only if stated guarantees can be met by the environment. The semantics is up to the structural equivalence rules given below, where $S[\mu\mathtt{t}.S/\mathtt{t}]$ is the one-time unfolding of $\mu\mathtt{t}.S$.

$$\mu\mathtt{t}.S \equiv S \ (\textit{where } \mathtt{t} \notin \mathsf{fv}(S)) \qquad \mu\mathtt{t}.S \equiv S[\mu\mathtt{t}.S/\mathtt{t}]$$

**Definition 2 (Operational semantics)** *The semantics of protocols is defined by a labelled transition system (LTS) over configurations of the form $(A, S)$ where $A$ ranges over environments $A \subseteq \mathcal{N}$ (sets of logical atoms), with transition labels $\ell ::= p \mid +\mathrm{l} \mid \mathsf{assert}(n) \mid \mathsf{require}(n) \mid \mathsf{consume}(n)$ and the transition rules below:*

$$\begin{array}{lll}
(A, p.S) \xrightarrow{p} (A, S) & & \langle\mathtt{Inter}\rangle \\
(A, +\{\mathrm{l}_i : S_i\}_{i\in I}) \xrightarrow{+\mathrm{l}_j} (A, S_j) & (j \in I) & \langle\mathtt{Branch}\rangle \\
(A, \mathsf{assert}(n).S) \xrightarrow{\mathsf{assert}(n)} (A \cup \{n\}, S) & & \langle\mathtt{Assert}\rangle \\
(A, \mathsf{require}(n).S) \xrightarrow{\mathsf{require}(n)} (A, S) & (n \in A) & \langle\mathtt{Require}\rangle \\
(A, \mathsf{consume}(n).S) \xrightarrow{\mathsf{consume}(n)} (A \setminus \{n\}, S) & (n \in A) & \langle\mathtt{Consume}\rangle \\
\dfrac{(A, S) \xrightarrow{\ell} (A', S')}{(A, \mu\mathtt{t}.S) \xrightarrow{\ell} (A', S'[\mu\mathtt{t}.S/\mathtt{t}])} & & \langle\mathtt{Rec}\rangle
\end{array}$$

Rules ⟨Inter⟩ and ⟨Branch⟩ always allow a protocol to proceed with some action, resulting in the appropriate continuation, without any effect to the environment. Rule ⟨Assert⟩ adds atom $n$ to the environment. Rules ⟨Require⟩ and ⟨Consume⟩ both require the presence of atom $n$ in the environment for the protocol to continue. Although ⟨Require⟩ leaves the environment unchanged, ⟨Consume⟩ consumes the atom $n$ from the environment. In ⟨Rec⟩, $S'[\mu\mathtt{t}.S/\mathtt{t}]$ means that the recursive protocol is unfolded by substituting $\mu\mathtt{t}.S$ for $\mathtt{t}$ in $S'$.

We write: $(A,S)\not\rightarrow$ if $(A,S)\xrightarrow{\ell}(A',S')$ for no $\ell,A',S'$; $(A,S)\xrightarrow{\vec{\ell}}(A',S')$ for a vector $\vec{\ell}=\ell_1,\dots,\ell_n$ if $(A,S)\xrightarrow{\ell_1}\dots\xrightarrow{\ell_n}(A',S')$. We say that $(A',S')$ is *reachable* from $(A,S)$ if $(A,S)=(A',S')$ or $(A,S)\xrightarrow{\vec{\ell}}(A',S')$ for a vector $\vec{\ell}$. We omit labels and target states where immaterial.

**Definition 3 (Stuck state & progress)** *State $(A,S)$ is* stuck *if $S\not\equiv\mathtt{end}$ and $(A,S)\not\rightarrow$. A protocol $S$ enjoys* progress *if every state $(A',S')$ reachable from $(\emptyset,S)$ is not stuck.*

A protocol may reach a stuck state when it does not have sufficient pre-conditions in its environment $A$. In Example 1, $S'_B$ does not enjoy progress because the pre-condition expressed by require(*pin*) cannot be met; similarly, $S'_A$ does not enjoy progress because of unmet pre-condition consume(*pay*).

### 2.3 Well-Assertedness

Assertions are key to generating meaningful compositions of protocols. Following the labelled transitions semantics, we define a judgement which captures the pre- and post-conditions of a protocol implied by its assertions. We use the notation $A\ \{S\}\, A'$ reminiscent of a Hoare triple where $A$ and $A'$ are pre- and post-conditions of $S$.

**Definition 4 (Well-assertedness)** *Let $A$ be a set of names.* Well-assertedness *of a protocol $S$ with respect to $A$ is defined below, as an inference system on judgements of the form $A\ \{S\}\, A'$, where $A'$ is the set of names (logical atoms) resulting after the execution of $S$ given the set of names $A$.*

$$\frac{A\ \{S\}\,A'}{A\ \{p.S\}\,A'}\,[\text{act}]\qquad \frac{\forall i\in I.\ A\ \{S_i\}\,A_i}{A\ \{+\{\mathrm{l}_i:S_i\}_{i\in I}\}\ \bigcap_{i\in I}A_i}\,[\text{bra}]\qquad \frac{A\cup\{n\}\ \{S\}\,A'}{A\ \{\mathsf{assert}(n).S\}\,A'}\,[\text{assert}]$$

$$\frac{A\cup\{n\}\ \{S\}\,A'}{A\cup\{n\}\ \{\mathsf{require}(n).S\}\,A'}\,[\text{require}]\qquad \frac{A\setminus\{n\}\ \{S\}\,A' \qquad n\in A}{A\ \{\mathsf{consume}(n).S\}\,A'}\,[\text{consume}]$$

$$\frac{A\ \{S\}\,A\cup A'}{A\ \{\mu\mathtt{t}.S\}\,A\cup A'}\,[\text{rec}]\qquad \frac{-}{A\ \{\mathtt{end}\}\,A}\,[\text{end}]\qquad \frac{-}{A\ \{\mathtt{t}\}\,A}\,[\text{call}]$$

*We write $A\ \{S\}$ when $A\ \{S\}\,A'$ for some $A'$ (i.e., when the post-condition is not of interest). We say that $S$ is* very-well-asserted *if $\emptyset\ \{S\}$. We say that a state $(A,S)$ is well-asserted if $S$ is well-asserted with respect to $A$.*

Protocols $S'_A$ and $S'_B$ in Example 1 are not very-well-asserted but they are well-asserted with respect to $\{pin, tan\}$ and $\{pay\}$, respectively.

We now consider some properties of well-asserted protocols. Proofs are in Appendix D. Firstly, protocols that do not contain assertions are very-well-asserted:

**Proposition 1 (Very-well-assertedness)** *If $S$ is generated by the grammar in Definition 1 without the assertion fragment then it is very-well-asserted.*

Next, well-asserted protocols can have their environment weakened, akin to precondition weakening in Hoare logic:

**Proposition 2 (Environment weakening)** *If $A\ \{S\}$ and $A \subseteq A'$ then $A'\ \{S\}$. Hence, $\emptyset\ \{S\}$ implies $A\ \{S\}$ for all $A$.*

Next, Lemma 1 states that the redux of a well-asserted state is well-asserted, moreover the postconditions are not weakened by reduction:

**Lemma 1 (Reduction preserves well-assertedness)** *If $A\ \{S\} A'$ and there is a reduction $(A, S) \xrightarrow{\ell} (A'', S')$ then $\exists A''' \supseteq A'. A''\ \{S'\} A'''$.*

**Lemma 2 (Well-asserted protocols are not stuck)** *If $A\ \{S\}$ and $S$ is closed with respect to recursion variables* $(\mathsf{fv}(S) = \emptyset)$ *then* $(A, S)$ *is not stuck.*

Next, Lemma 3 shows that if a protocol "gets stuck", this is because it does not have enough preconditions to proceed. Thus, the protocol needs assumptions that may be provided by other protocols it could be composed with. Lemma 3 follows by induction on the length of a protocol's execution, combined with Lemmas 1 and 2.

**Lemma 3 (Progress of very-well-asserted protocols)** *If $S$ is very-well-asserted (i.e., $\emptyset\ \{S\}$) and closed then it exhibits* progress.

We next introduce protocol composition, which produces protocols that are meaningful with respect to their assertions (i.e., that exhibit progress).

## 3 Interleaving Compositions

We compose protocols by computing syntactic interleavings. We derive the 'interleaving composition' (IC) of two protocols $S_1$ and $S_2$ via a relation with judgements of the form: $T_L; T_R; A \vdash S_1 \circ S_2 \triangleright S$ where $S$ is the resulting composed protocol, and $A$ is the set of names (i.e., assertions) provided by the environment to $S$. We let $T$ range over recursion environments, defined as possibly empty lists of *distinct* protocol variables $\mathtt{t}$. Lists are concatenated via the , (comma) operator, which is overloaded to extend a list with a single element, e.g., written $T, \mathtt{t}$. In the judgements, we use two recursion environments $T_L$ and $T_R$ to keep track of the free protocol variables in $S_1$ and $S_2$ respectively in order to handle composition of recursive protocols. We use an underlining annotation $\underline{\mathtt{t}}$ to denote variables that were used to merge two recursive protocols into one recursive IC, and predicate $\mathtt{unused}(T)$ that is true if all variables in $T$ are not used (i.e., not underlined), and false otherwise. The 'used' annotation $\underline{\mathtt{t}}$ is instrumental in handling composition of nested recursions, as explained later.

$$\frac{T_L;\, T_R;\, A \vdash S_1 \circ S_2 \triangleright S}{T_L;\, T_R;\, A \vdash p.S_1 \circ S_2 \triangleright p.S} \qquad \frac{T_R;\, T_L;\, A \vdash S_2 \circ S_1 \triangleright S}{T_L;\, T_R;\, A \vdash S_1 \circ S_2 \triangleright S} \quad \text{[act/sym]}$$

$$\frac{T_L;\, T_R;\, A \cup \{n\} \vdash S_1 \circ S_2 \triangleright S}{T_L;\, T_R;\, A \cup \{n\} \vdash \mathsf{require}(n).S_1 \circ S_2 \triangleright \mathsf{require}(n).S} \quad \text{[require]}$$

$$\frac{T_L;\, T_R;\, A \setminus \{n\} \vdash S_1 \circ S_2 \triangleright S \qquad n \in A}{T_L;\, T_R;\, A \vdash \mathsf{consume}(n).S_1 \circ S_2 \triangleright \mathsf{consume}(n).S} \quad \text{[consume]}$$

$$\frac{T_L;\, T_R;\, A \cup \{n\} \vdash S_1 \circ S_2 \triangleright S}{T_L;\, T_R;\, A \vdash \mathsf{assert}(n).S_1 \circ S_2 \triangleright \mathsf{assert}(n).S} \quad \text{[assert]}$$

$$\frac{\forall i \in I \quad T_L;\, T_R;\, A \vdash S_i \circ S_2 \triangleright S'_i}{T_L;\, T_R;\, A \vdash +\{\mathrm{l}_i : S_i\}_{i\in I} \circ S_2 \triangleright +\{\mathrm{l}_i : S'_i\}_{i\in I}} \quad \text{[bra]}$$

$$\frac{T_L, \mathtt{t}_1;\, T_R;\, A \vdash S_1 \circ \mu\mathtt{t}_2.S_2 \triangleright S \quad A\ \{\mu\mathtt{t}_1.S\}}{T_L;\, T_R;\, A \vdash \mu\mathtt{t}_1.S_1 \circ \mu\mathtt{t}_2.S_2 \triangleright \mu\mathtt{t}_1.S} \qquad \frac{A\ \{\mu\mathtt{t}.S\} \quad \mathsf{fv}(\mu\mathtt{t}.S) = \emptyset}{T_L;\, T_R;\, A \vdash \mu\mathtt{t}.S \circ \mathtt{end} \triangleright \mu\mathtt{t}.S} \quad \text{[rec1/rec3]}$$

$$\frac{T_L;\, T_1, \underline{\mathtt{t}}, T_2;\, A \vdash S_1[\mathtt{t}/\mathtt{t}_1] \circ S_2 \triangleright S \qquad \mathtt{unused}(T_2)}{T_L;\, T_1, \mathtt{t}, T_2;\, A \vdash \mu\mathtt{t}_1.S_1 \circ S_2 \triangleright S} \quad \text{[rec2]}$$

$$\frac{\underline{\mathtt{t}} \in T_L \vee \underline{\mathtt{t}} \in T_R}{T_L;\, T_R;\, A \vdash \mathtt{t} \circ \mathtt{t} \triangleright \mathtt{t}} \qquad \frac{-}{T_L;\, T_R;\, A \vdash \mathtt{end} \circ \mathtt{end} \triangleright \mathtt{end}} \quad \text{[call/end]}$$

**Figure 2** Rules for iterleaving composition of protocols

**Definition 5 (Interleaving composition)** *IC is defined by the judgements in Figure 2.*

In Figure 2, rule [act] is for prefixes, [sym] is the commutativity rule, and [end] handles a terminated protocol. By combining [act] and [sym] one can obtain all interleavings of two sequences of actions.

Rule [require] includes the continuation of a protocol only if a required assertion $n$ is provided by the environment. Rule [consume] is similar except the assertion is removed in the precondition's environment. Conversely, [assert] adds assertion $n$ to the environment of the precondition. Rules [require], [assume], and [consume] may enforce a particular order in actions of an interleaving. For example, the reader can verify that the composition of $?\mathtt{pay}.\mathsf{assert}(p).\mathtt{end}$ and $\mathsf{consume}(p).!\mathtt{item}.\mathtt{end}$ produces (only) one interleaving $?\mathtt{pay}.\mathsf{assert}(p).\mathsf{consume}(p).!\mathtt{item}.\mathtt{end}$ that is obtained by applying [act], [assert], [sym], [consume], [act], and [end].

Rule [bra] is similar to [act] but the continuations are composed with each branch. For example the composition $+\{\mathrm{l}_1 : \mathtt{end},\ \mathrm{l}_2 : \mathtt{end}\} \circ\ !\mathtt{Int}.\mathtt{end}$ with initially empty environment produces the following two interleavings:

$+\{\mathrm{l}_1 :!\mathtt{Int}.\mathtt{end},\ \mathrm{l}_2 :!\mathtt{Int}.\mathtt{end}\}$ (applying [bra], [sym], [act], [end])
$!\mathtt{Int}.+\{\mathrm{l}_1 : \mathtt{end},\ \mathrm{l}_2 : \mathtt{end}\}$ (applying [sym], [act], [act], [sym], [bra], [end])

Rules [rec1] and [rec2] allow two recursive protocols to be composed. The composition of two recursive protocols, say $\mu\mathtt{t}_1.S_1$ and $\mu\mathtt{t}_2.S_2$, yields a recursive protocol where the recursion body is the composition of the two recursion bodies, and only one of the two protocol variables is used, either $\mathtt{t}_1$ or $\mathtt{t}_2$. For example, the composition of $\mu\mathtt{t}_1.!\mathtt{p}_1.\mathtt{t}_1$ and $\mu\mathtt{t}_2.!\mathtt{p}_2.\mathtt{t}_2$ yields e.g.,

$\mu\mathtt{t}_1.!\mathtt{p}_1.!\mathtt{p}_2.\mathtt{t}_1$ (applying [rec1], [act], [sym], [rec2], [act], [call])
$\mu\mathtt{t}_2.!\mathtt{p}_2.!\mathtt{p}_1.\mathtt{t}_2$ (applying [sym], and proceeding as above)

Rule [rec1] picks $\mathtt{t}_1$ as name for the interleaving composition, records $\mathtt{t}_1$ as the end of the $T_L$ list and continues with the composition of the recursion body $S_1$ with $\mu\mathtt{t}_2.S_2$. The premise $A\ \{\mu\mathtt{t}_1.S\}$ ensures well-assertedness of the *arbitrary repetition* of $S$, that is $\mu\mathtt{t}_1.S$ (the composition rules only check that $S$ is well-asserted). Rule [rec2] completes the merge of two recursions, with calls to $\mathtt{t}_2$ in this instance being redirected to $\mathtt{t}_1$ (via a substitution). Variable $\mathtt{t}_1$ is in the right recursion environment $T_1, \mathtt{t}_1, T_2$, namely a list of protocol variables, followed by unused $\mathtt{t}_1$, followed by a list of unused protocol variables $T_2$, yielding a protocol with just one recursion. In the premise of [rec2], $\mathtt{t}_1$ in this instance becomes used.

In [rec2], condition $\mathtt{unused}(T_2)$ prevents erroneous 'flattening' of nested recursions. For instance, in the composition of $S_1 = \mu\mathtt{t}.p.\mathtt{t}$ and $S_2 = \mu\mathtt{t}_1.q.\mu\mathtt{t}_2. + \{\mathrm{l}_1 : \mathtt{t}_1, \mathrm{l}_2 : \mathtt{t}_2\}$, merging $\mathtt{t}$ with both $\mathtt{t}_1$ and $\mathtt{t}_2$ would yield the undesirable derivation $S = \mu\mathtt{t}.p.q.+\{\mathrm{l}_1 : \mathtt{t}, \mathrm{l}_2.\mathtt{t}\}$ where $S$ does not preserve the behaviour of $S_2$. Behaviour preservation is formally defined later on; for now, observe that $S_2$ permits successive choices of the label $\mathrm{l}_2$ without any intervening actions, whereas $S$ requires an intervening $q$ action (and $p$ action) between any successive choices of label $\mathrm{l}_2$. See Example 7 in Appendix A for some derivations of interleaving compositions of $S_1$ and $S_2$. The requirement that $\mathtt{t}$ precedes only unused variables $T_2$ (captured by predicate $\mathtt{unused}(T_2)$) also prevents 'criss-cross' substitutions when composing two protocols with nested recursions which can also violate behaviour preservation in similar ways to the case observed above.

Consider now the composition of a recursive protocol with a non-recursive one e.g., $S_1 = \mu\mathtt{t}.p_1.\mathtt{t}$ with $S_2 = p_2.\mathtt{end}$. We do *not* want to derive the following protocol: $S = \mu\mathtt{t}.p_1.p_2.\mathtt{t}$ The problem with $S$ is that it allows execution $p_1, p_2, p_1, p_2, \ldots$ where action $p_2$ is repeatedly executed, while $S_2$ only prescribes one instance of $p_2$. Such a derivation would not preserve the behaviour of $S_2$. Our rules do not allow derivation of $S$ above because rule [call] checks that the component protocols share protocol variable $\mathtt{t}$ (i.e., they are both recursive and correctly merged).

Another undesirable composition of $S_1 = \mu\mathtt{t}.p_1.\mathtt{t}$ and $S_2 = p_2.\mathtt{end}$ is one where $S_1$ 'comes first' yielding $S' = \mu\mathtt{t}.p_1.\mathtt{t}$ which, morally, behaves as $S_2$ after an infinite loop. If this were a composition, it would violate a second property we discuss formally later, fairness, requiring each component protocol to be able to proceed until it terminates. $S'$ is not derivable thanks to [rec3], which only allows a recursive protocol to be introduced in an interleaving composition when the non-recursive component has already been all merged (i.e., it is $\mathtt{end}$). We can, e.g., derive the following composition of $S_1$ and $S_2$, where the terminating protocol $S_2$ comes first (hence satisfying fairness):

$p_2.\mu\mathtt{t}.p_1.\mathtt{t}$ (applying [act], [sym], [rec3]).

The premise $\mathsf{fv}(\mu\mathtt{t}.S) = \emptyset$ of [rec3] prevents it being used inappropriately in case of nested recursion, e.g., to prevent composition of $\mu\mathtt{t}_1.p_1.\mu\mathtt{t}_2.p_2.\mathtt{t}_1$ and $q.\mathtt{end}$ to produce (via [rec1], [act], [sym], [rec3]) $\mu \mathtt{t}_1.p_1.q.\mu \mathtt{t}_2.p_2.\mathtt{t}_1$, which violates behaviour preservation (discussed later) by repeating an action $q$ from a non-recursive context.

### 3.1 Variations on the Branching Rule

The branching rule of interleaving composition can be viewed as a *distributivity* property: sequential composition after a control-flow branch can be distributed inside the branches. Algebraically, we can informally describe this distributivity as follows, for a 2-way branch (*sans* labelling): $(S_1 + S_2) \circ T \equiv (S_1 \circ T) + (S_2 \circ T)$. Such a property is familiar in Kleene algebra models of programs and program reasoning [34] and monotone dataflow frameworks in static analysis [32]. Since interleaving composition generates a set of possible protocols it would be more accurate to express this property in terms of set membership rather than equality (for simplicity of the analogy, this elides the fact that each composition $\circ$ is itself a set):

$$(S_1 + S_2) \circ T \ni (S_1 \circ T) + (S_2 \circ T) \qquad \text{(distributivity)}$$

In this section we consider two variants of this distributive behaviour for composition called (1) 'weak branching' and (2) 'interchange branching' which can be summarised via the algebraic analogy as variants of distributivity, respectively:

$$(S_1 + S_2) \circ T \ni (S_1 \circ T) + S_2 \quad \wedge \quad (S_1 + S_2) \circ T \ni S_1 + (S_2 \circ T) \qquad \text{(weak)}$$
$$(S_1 + S_2) \circ (T_1 + T_2) \ni (S_1 \circ T_1) + (S_2 \circ T_2) \qquad \text{(interchange)}$$

In (weak), composition distributes inside one branch but not the other. In (interchange), composing branches with branches has a 'merging' effect on the branches rather than distributing within. (The 'interchange' terminology comes from similar properties in category theory [33]).

We motivate and discuss each variation from the protocol perspective. In the rest of this section we introduce two additional composition rules: [wbra] for weak branching, and [cbra] for interchange branching (which we will refer to as *correlating branching* as it better reflects the effects of the rule on the protocols). Note that these two variations grow the set of possible interleavings, rather than shrinking it: they provide more general composition behaviours but do not exclude the more specialised behaviours. For generality of the theory, the derivation of interleaving composition can apply any branching ([bra], [wbra], [cbra]). For practicality, our tool allows engineers to choose the kind of branching to use in any specific scenario (as shown in Section 5).

#### 3.1.1 Weak Branching for "Asymmetric" Guarantees

*Weak branching* allows partial execution of some protocols being composed even if there are not sufficient assertions to continue, as long as all protocols are completely executed in some execution path. For example, protocol $S_B$ below needs assertion $n$ to proceed. Assume we want to compose $S_B$ with a protocol $S_A$, which can provide $n$ in only one of its branches ok. Protocol $S_A$ may be an authentication server, granting or blocking access to $S_B$ depending on a password pwd. That is, for some $S'$:

$$S_A \;::=\; ?\mathtt{pwd}.\oplus\{\mathsf{ok} : \mathsf{assert}(n).\ \mathtt{end},\ \mathsf{ko} :\ \mathtt{end}\} \qquad S_B \;::=\; \mathsf{require}(n).S'$$

Since we want the actions of $S_B$ not to be executed after selection of label ko, we want interleaving composition to generate the following protocol:

$$S_{AB} = ?\texttt{pwd}.\oplus\{\mathsf{ok}: \ \mathsf{assert}(n).\mathsf{require}(n).S', \ \mathsf{ko}: \ \texttt{end}\}$$

Protocol $S_{AB}$ is not attainable using the rules of Definition 5: the derivation blocks composing $\mathsf{require}(n).S'$ with the second branch's end in the empty environment.[1] Instead, we introduce a 'weak branching' composition rule to allow asymmetric guarantees:

**Definition 6 (Weak branching)** Weak branching composition *of protocols is derived using the judgements in Definition 5 and the* additional *rule [wbra]:*

$$\frac{\begin{array}{c} I = I_A \cup I_B \qquad I_A \cap I_B = \emptyset \qquad I_A \neq \emptyset \\ \forall i \in I_A.\ T_L;\ T_R;\ A \vdash S_i \circ S \triangleright S_i' \qquad \forall i \in I_B.\ T_L;\ T_R;\ A \vdash S_i \circ S \not\triangleright \ \wedge\ A\ \{S_i\} \end{array}}{T_L;\ T_R;\ A \vdash +\{\mathsf{l}_i : S_i\}_{i\in I} \circ S \triangleright +\{\mathsf{l}_i : S_i'\}_{i\in I_A} \cup \{\mathsf{l}_i : S_i\}_{i \in I_B}}$$

Precondition $I_A \neq \emptyset$ ensures that each protocol's actions are executed in at least one execution path, and is key to the fairness property introduced in Definition 9. Hereafter we denote with $\triangleright_s$ derivations obtained using the judgements in Definition 5 only and $\triangleright_w$ for derivations with the additional rule [wbra].

**Example 2 (Weak IC of banking and PIN/TAN)** *Consider the banking and PIN/TAN protocols in Example 1 (p. 7). Interleaving composition of $S_A'$ and $S_B'$ using $\triangleright_s$ returns an empty set. When using $\triangleright_w$ instead, we can derive the following interleaving composition modelling a banking/authentication protocol that satisfies the requirements specified in Section 1.1.*

$$S_{BA} = ?\texttt{pin}.\oplus\left\{\begin{array}{l} \mathsf{ok}: \ \mathsf{assert}(pin).\mathsf{require}(pin).\mu\texttt{r}.\&\left\{\begin{array}{l} \mathsf{payment}: \ S_{TAN}, \\ \mathsf{statement}: !\texttt{statement}.\texttt{r}, \\ \mathsf{logout}: \ \mathsf{consume}(pin).\,\texttt{end} \end{array}\right\} \\ \mathsf{fail}: \ \texttt{end} \end{array}\right\}$$

$$S_{TAN} = \mathsf{assert}(pay).\mathsf{consume}(pay).!\texttt{id}.?\texttt{tan}.\oplus\left\{\begin{array}{l} \mathsf{ok}: \mathsf{assert}(tan).\,\mathsf{consume}(tan). \\ \qquad ?\texttt{details}.\texttt{r}, \\ \mathsf{fail}: \texttt{r} \end{array}\right\}$$

### 3.1.2 Correlating Branching

*Correlating branching* allows two protocols to be composed by 'correlating' each branch of one with at least one branch of the other.

Consider two branching protocols: $S_1$ offering two services s1 and s2, and $S_2$ offering two kinds of payment p1 and p2. When composing $S_1$ and $S_2$, we can correlate s1 with p1, and s2 with p2, using assertions:

$$\begin{array}{rcl} S_1 & = & \oplus\{\mathsf{s1} : \mathsf{assert}(one).\texttt{end},\ \ \mathsf{s2} : \mathsf{assert}(two).\texttt{end}\} \\ S_2 & = & \oplus\{\mathsf{p1} : \mathsf{consume}(one).\texttt{end},\ \ \mathsf{p2} : \mathsf{consume}(two).\texttt{end}\} \end{array}$$

[1] If we start from a non-empty environment $\{n\}$ we can derive $?\texttt{pwd}.\oplus\{\mathsf{ok} : \mathsf{assert}(n).\mathsf{require}(n).S', \mathsf{ko} : \mathsf{require}(n).S'\}$. However, initial assumption $\{n\}$ means that access to $S_B$ is granted regardless of the authentication outcome.

We would like to obtain the following composition:

$$S_{12} = \oplus \left\{ \begin{array}{l} \mathsf{s1} : \oplus\{\mathsf{p1} : \mathsf{assert}(one).\mathsf{consume}(one).\mathtt{end}\}, \\ \mathsf{s2} : \oplus\{\mathsf{p2} : \mathsf{assert}(two).\mathsf{consume}(two).\mathtt{end}\} \end{array} \right\}$$

Composition rule [bra] is too strict and returns an empty set for $S_1$ and $S_2$. Weak branching [wbra] is also not useful in this case, producing the interleaving below, which does not capture the intended correlation:

$$\oplus \left\{ \begin{array}{l} \mathsf{p1} : \oplus \left\{ \ \mathsf{s1} : \mathsf{assert}(one).\mathsf{consume}(one).\mathtt{end}, \mathsf{s2} : \mathsf{assert}(two).\mathtt{end} \ \right\}, \\ \mathsf{p2} : \oplus \left\{ \ \mathsf{s1} : \mathsf{assert}(one).\mathtt{end}, \mathsf{s2} : \mathsf{assert}(two).\mathsf{consume}(two).\mathtt{end} \ \right\} \end{array} \right\}$$

Definition 7 introduces a further rule [cbra], to allow for correlating compositions.

**Definition 7 (Correlating branching)** Correlating branching composition *is derived using the judgement in Definition 5 with the addition of rule [cbra] below:*

$$\frac{\begin{array}{c} \forall i \in I.\, J_i \neq \emptyset \ \wedge \bigcup_{i\in I} J_i = J \\ \forall j \in J_i \quad T_L; T_R; A \vdash S_i \circ S'_j \rhd S_{ij} \qquad \forall j \in J \setminus J_i \quad T_L; T_R; A \vdash S_i \circ S'_j \not\rhd \end{array}}{T_L; T_R; A \vdash +\{\mathrm{l_i} : S_i\}_{i\in I} \circ +'\{\mathrm{l'_j} : S'_j\}_{j\in J} \rhd +\{\mathrm{l_i} : +'\{\mathrm{l'_j} : S_{ij}\}_{j\in J_i}\}_{i\in I}}$$

The first premise requires that: (1) each branch of the first protocol can be correlated with at least one branch of the second protocol ($J_i \neq \emptyset$), and (2) each branch of the second protocol can be correlated with at least one branch of the first protocol ($\bigcup_{i\in I} J_i = J$). This precondition is critical to ensure the fairness property we introduce in Section 4 (Definition 9). Rule [cbra] allows us to obtain $S_{12}$ as the interleaving composition of $S_1$ and $S_2$ above, modelling the intended correlation.

Hereafter we denote with $\rhd_c$ (resp. $\rhd_{wc}$) derivations obtained using the judgements in Definition 7 with the addition of rule [cbra] (resp. [cbra] and [wbra]). The inclusion relation between the different kinds of judgement is shown on the right (with $\rhd_s$ and $\rhd_{wc}$ being the most and least strict, respectively).

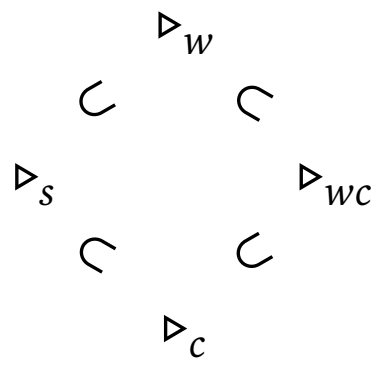

## 4 Properties of Interleaving Composition

In this section, we give the main properties of interleaving compositions. First, we give some general properties of well-assertedness and algebraic/scoping properties (i.e., sanity checks). Then, we give behaviour preservation and fairness, both formulated using a semantics of 'protocol ensembles' (a semantic counterpart of syntactic composition). Hereafter, we will denote with $\rhd$ any kind of judgement in $\{\rhd_s, \rhd_w, \rhd_c, \rhd_{wc}\}$.

### 4.1 Well-Assertedness of Compositions

Critical for the validity of our approach is that interleaving compositions preserve the constraints of assertions:

**Proposition 3 (Validity)** *If* $T_L; T_R; \emptyset \vdash S_1 \circ S_2 \triangleright S$ *then* $S$ *is very-well-asserted.*

Appendix E details the proof. A corollary of Proposition 3 and Lemma 3 (progress of very-well-asserted protocols) is that interleaving compositions enjoy progress:

**Corollary 1 (Progress)** *If* $T_L; T_R; \emptyset \vdash S_1 \circ S_2 \triangleright S$ *then* $S$ *enjoys progress.*

### 4.2 Algebraic and Scoping Properties

We consider algebraic properties and notions of open and closed protocol with respect to recursion variables. Appendix F details the proofs of these results.

Composing closed recursive protocols yields closed protocols. This property is a corollary of a more general property, that free variables are preserved by composition:

**Proposition 4** *If* $T_L; T_R; A \vdash S_1 \circ S_2 \triangleright S$ *then* $\mathsf{fv}(S_1) \cup \mathsf{fv}(S_2) = \mathsf{fv}(S)$.

That is, the free variables of a composed protocol are exactly the union of the free variables of the protocols being composed.

**Corollary 2 (Composition preserves closedness)** *For all* $A, S$ *and closed protocols* $S_1, S_2$, *if* $T_L; T_R; A \vdash S_1 \circ S_2 \triangleright S$ *then* $S$ *is a closed protocol.*

A useful algebraic property is that composition has `end` protocols as units:

**Proposition 5 (Interleaving composition has left- and right-units)** *For a protocol* $S$ *where* $A\ \{S\} \wedge \mathsf{fv}(S) = \emptyset$ *then* $T_L; T_R; A \vdash S \circ \mathtt{end} \triangleright S$ *and* $T_L; T_R; A \vdash \mathtt{end} \circ S \triangleright S$.

### 4.3 Behaviour Preservation and Fairness of Protocol Ensembles

In Section 3, we gave a syntactic definition of *interleaving composition*, which enacts the dependencies implied by assertions in protocols, and provides a blue-print of an implementation. In this section, we consider ‘protocol ensembles’, which can be understood as the *semantic compositions* of two asserted protocols. Semantic compositions have a behaviour that is similar to parallel composition (e.g., as in CCS), but unlike parallel composition the two asserted protocols cannot communicate with each other, i.e., there are no internal $\tau$ actions. All interactions in a semantic composition are directed towards other endpoints. Semantic composition provides a more general and somewhat familiar notion of composition, which we will use as a reference to analyse the properties of interleaving compositions.

Protocol ensembles, ranged over by $C$, are defined as follows:

$$C \quad ::= \quad S \quad | \quad S \,||\, S$$

By defining $C$ as either asserted protocols $S$ (which may be interleaving compositions) or semantic compositions $S \,||\, S$, we obtain a common LTS for comparing the behaviour of interleaving and semantic compositions. For simplicity we limit the theory to the composition of two protocols. The extension to $n$ protocols is straightforward e.g., based on labelling each protocol and its actions with a unique identifier.

The LTS for protocol ensembles extends the LTS for asserted protocols: it is defined over states of the form $(A, C)$, transition labels $\ell$ (as for asserted protocols), and by the rules in Definition 2 plus the following two rules:

$$\frac{(A, S_1) \xrightarrow{\ell} (A', S_1')}{(A, S_1 \,||\, S_2) \xrightarrow{\ell} (A', S_1' \,||\, S_2)} \langle \texttt{Com1} \rangle \qquad \frac{(A, S_2) \xrightarrow{\ell} (A', S_2')}{(A, S_1 \,||\, S_2) \xrightarrow{\ell} (A', S_1 \,||\, S_2')} \langle \texttt{Com2} \rangle$$

We write $(A, C) \rightarrow$ if $(A, C) \xrightarrow{\ell} (A', C')$ for some $\ell, A', C'$. Protocols in $C$ do not communicate internally, but may affect each other by changing or checking $A$.

**Behaviour preservation** Fix an LTS for protocol ensembles $(Q, L, \longrightarrow)$ defined on the set $Q$ of states $\mathbf{s}$ of the form $(A, C)$ and labels $L$. We use the standard notion of *simulation* [41] to compare protocols of interleaving compositions and protocol ensembles, using protocol ensembles as a correct general model to which interleaving compositions need to adhere.

**Definition 8 (Simulation)** *A (strong)* simulation *is a relation $\mathscr{R} \subseteq Q \times Q$ such that, whenever $\mathbf{s}_1 \mathscr{R} \mathbf{s}_2$: $\forall \ell \in L, \mathbf{s}_1' : \mathbf{s}_1 \xrightarrow{\ell} \mathbf{s}_1'$ implies $\exists \mathbf{s}_2' : \mathbf{s}_2 \xrightarrow{\ell} \mathbf{s}_2'$ and $\mathbf{s}_1' \,\mathscr{R}\, \mathbf{s}_2'$.*

We call 'similarity' the largest simulation relation. We write $\mathbf{s}_1 \lesssim \mathbf{s}_2$ when there exists a simulation $\mathscr{R}$ such that $\mathbf{s}_1 \mathscr{R} \mathbf{s}_2$. We say that $C_1$ preserves the behaviour of $C_2$ with respect to $A$ if $(A, C_1) \lesssim (A, C_2)$.

**Theorem 1 (Behaviour preservation of compositions - closed)**

$$\emptyset; \emptyset; A \vdash S_1 \circ S_2 \triangleright S \quad \Rightarrow \quad (A, S) \lesssim (A, S_1 \,||\, S_2)$$

Therefore, interleaving compositions will only show behaviour that would be allowed by a protocol ensemble. Clearly, protocol ensembles allow more possible executions than an interleaving composition, which is only one of the possible interleavings. The proof of Theorem 1 is by induction on the derivation of $S$ and, although the statement assumes closed protocols, some inductive hypotheses in the proof (e.g., premises of [rec1] or [rec2]) require reasoning about open protocols. The proof hence relies on a property (Lemma G.7 – Appendix G) on open protocols: (roughly) given two protocols and one of their interleaving compositions, any action of the interleaving composition is matched by an action of the ensemble of the two protocols, and this property is preserved upon transition. Note that, while environments $T_L$ and $T_R$ are trivially empty in Theorem 1 (closed protocols), they have a key role in proving Lemma G.7 (open protocols): they include the variables of each component protocol that have been bound in a derivation, and give critical information of the scope and structure of the original component protocols in that derivation. Appendix G details the proof.

**Fairness** Fix an ensemble of two protocols $S_0 \,||\, S_1$ and any of their interleaving compositions $S$. By fairness, each action of $S_0$ (resp. $S_1$) can be observed in at least one execution of $S$, possibly after a finite sequence of other actions by $S_1$ (resp. $S_0$). In the following, we write $(_, S)$ to denote $(A, S)$ when $A$ is immaterial.

**Definition 9 (Fairness)** *$S$ is* fair *w.r.t. $S_0$ and $S_1$ on $A$, if $\forall i \in \{0,1\}$ and any transition $(_, S_i) \xrightarrow{\ell} (_, S_i')$ there exists $\vec{r}$ such that: **1)** $(A, S_{|1-i|}) \xrightarrow{\vec{r}} (_, S'_{|1-i|})$, **2)** $(A, S) \xrightarrow{\vec{r}\ell} (A', S')$, and **3)** $S'$ is fair with respect to $S_i'$ and $S'_{|1-i|}$ on $A'$.*

**Theorem 2 (Fairness of compositions)** *If $\emptyset; \emptyset; A \vdash S_0 \circ S_1 \triangleright S$ then $S$ is* fair *w.r.t. $S_0$ and $S_1$ on $A$.*

A key aspect of fairness (Definition 9) is that it fixes $\ell$ and then requires at least one execution in which $\ell$ is eventually executed by $S$. This implies that although not all possible future branches include all parts of the protocols being composed, some will.

**Definition 10 (Strong fairness)** *$S$ is* strongly fair *w.r.t. $S_0$ and $S_1$ on $A$, if any $i \in \{0,1\}$ and all transitions $(_, S_i) \xrightarrow{\ell} (_, S_i')$ and $(A, S_{|1-i|}) \xrightarrow{\vec{r}}$, there exist $\vec{r'}$, $\vec{r''}$ with $(A, S_{|1-i|}) \xrightarrow{\vec{r'}} (_, S'_{|1-i|})$ and either:*

***1)*** *$\vec{r'}\vec{r''} = \vec{r}$ (i.e., $\vec{r'}$ is a prefix of $\vec{r}$), or*

***2)*** *$\vec{r'} = \vec{r}\vec{r''}$ (i.e., $\vec{r}$ is an ex prefix of $\vec{r'}$)*

*such that $(A, S) \xrightarrow{\vec{r'}\ell} (A', S')$ and $S'$ is strongly fair w.r.t. $S_i'$ and $S'_{|1-i|}$ on $A'$.*

By Definition 10, any action of a composition can be matched by an action of the protocols being composed, and this property is preserved by transition. Vectors $\vec{r}$, $\vec{r'}$, and $\vec{r''}$ are used to universally quantify on $\vec{r}$ and yet allow for the cases where $\ell$ comes before (1) or after (2) $\vec{r}$ in the composition. It follows a stronger fairness result for compositions using only [bra] that only holds for $\triangleright_s$ judgements.

**Theorem 3 (Strong fairness of compositions with $\triangleright_s$)** *If $\emptyset; \emptyset; A \vdash S_0 \circ S_1 \triangleright_s S$ then $S$ is* strongly fair *with respect to $S_0$ and $S_1$ on $A$.*

Appendix H details the proofs.

**Example 3 (Fairness and weak branching)** *Consider a simpler variant of the protocols in Section 3.1.1 (omitting password exchange and continuation):*

$$\begin{array}{rcl} S_A &=& \oplus\{\mathsf{ok} : \mathsf{assert}(n).\mathsf{end},\ \mathsf{ko} : \mathsf{end}\} \qquad S_B = \mathsf{require}(n).\ \mathsf{end} \\ S_{AB} &=& \oplus\{\mathsf{ok} : \mathsf{assert}(n).\mathsf{require}(n).\mathsf{end},\ \mathsf{ko} : \mathsf{end}\} \end{array}$$

*Observe $\emptyset; \emptyset; \emptyset \vdash S_A \circ S_B \not\triangleright_s S_{AB}$ and $\emptyset; \emptyset; \emptyset \vdash S_A \circ S_B \triangleright_w S_{AB}$. We show that $S_{AB}$ is a fair composition w.r.t. $S_A$ and $S_B$ on $\emptyset$, but it is not a strongly fair one.*

*First focus on fairness. $S_A$ can move with either label $\oplus\mathsf{ok}$ or $\oplus\mathsf{ko}$. In either case $(\emptyset, S_{AB})$ can immediately make a corresponding step with $\vec{r}$ empty. If $S_B$ moves, that is by label $\mathsf{require}(n)$, then for some environment $\{n\}$:*

$$(\{n\}, S_B) \xrightarrow{\mathsf{require}(n)} (\emptyset, \mathsf{end}) \tag{1}$$

*There exists a sequence of transitions with labels $\vec{r} = \oplus\mathsf{ok}, \mathsf{assert}(n)$ such that*

$$(\emptyset, S_B) \xrightarrow{\oplus\mathsf{ok},\mathsf{assert}(n)} (\{n\}, \mathsf{end}) \qquad (\emptyset, S) \xrightarrow{\oplus\mathsf{ok},\mathsf{assert}(n)} (\{n\}, \mathsf{require}(n).\mathsf{end}) \xrightarrow{\mathsf{require}(n)} (\emptyset, \mathsf{end})$$

*and* $\emptyset; \emptyset; \emptyset \vdash \texttt{end} \circ \texttt{end} \triangleright_w \texttt{end}$. *In the case above, we could select a 'good' path of* $S_A$ *and* $S_{AB}$ *that allows the transition with label* $\mathsf{require}(n)$ *to happen.*

*Focus now on strong fairness and again, consider the step in Equation* (1) *by* $S_B$. *Now we can pick an arbitrary* $\vec{r}$, *say,* $\oplus\mathsf{ok}$, *such that* $(\emptyset, S_B) \xrightarrow{\oplus \mathsf{ko}} (\emptyset, \texttt{end})$. *Looking at* $S_{AB}$, *there is no prefix nor extension of* $\vec{r} = \oplus\mathsf{ok}$ *that allows a* $\mathsf{require}(n)$ *step by* $S_{AB}$ *once the branch* ko *is taken. Therefore,* $S_{AB}$ *is not strongly fair with respect to* $S_A$ *and* $S_B$ *on* $\emptyset$.

### 4.4 Completeness

We discuss completeness of our composition rules: for every 'good' execution of $S_1 \,||\, S_2$ (i.e., non-terminating or reaching state $\texttt{end} \,||\, \texttt{end}$), can we obtain an interleaving composition of $S_1$ and $S_2$ that yields that execution? At present the answer is negative. For example, $S_a$ and $S_b$ below produce no interleavings (not even with $\triangleright_w$)

$$\begin{array}{lcl} S_a & = & \texttt{?pwd}.\mathsf{assert}(login).\texttt{?quit}.\mathsf{assert}(n).\mathsf{consume}(login).\texttt{end} \\ S_b & = & \mu \texttt{t}.\&\{\mathsf{balance} : \mathsf{require}(login).\texttt{!bal}.\texttt{t},\ \mathsf{finish} : \mathsf{consume}(n).\texttt{end}\} \end{array}$$

while it may be desirable to obtain:

$$\texttt{?pwd}.\mathsf{assert}(login).\mu\texttt{t}.\&\begin{cases} \mathsf{balance} : \mathsf{require}(login).\texttt{!bal}.\texttt{t}, \\ \mathsf{finish} : \texttt{?quit}.\mathsf{assert}(n).\mathsf{consume}(n).\mathsf{consume}(login).\texttt{end} \end{cases}$$

The IC above cannot be derived because [rec1] prevents composition of recursive with non-recursive protocols. A simplistic modification of [rec1] to allow composition of $\texttt{t}_1.S_1$ and $S_2$ (with $\texttt{Top}(S_2) = \emptyset$) would produce $\mu\texttt{t}.\texttt{?pwd}.\mathsf{assert}(login).\&\{\ldots\}$ which is not behaviour preserving (the password request is repeated). Similar tweaks to [rec2] have the same problem. With more complex rules, we may possibly allow weak composition of $S_a$ with $S_b$ only for syntactic subterms of $S_b$ that terminate (e.g, after the finish branch). Extending our rules in this direction, and investigating completeness, is future work. At present using $\triangleright_c$ we can still compose $S_b$ with a modified $S_a$, e.g.

$$\texttt{?pwd}.\mathsf{assert}(login).\mu\texttt{t}_a.\&\{\mathsf{void} : \texttt{t}_a, \mathsf{quit} : \texttt{?quit}.\mathsf{assert}(n).\mathsf{consume}(login).\texttt{end}\}$$

## 5 Implementation

To illustrate the proposed approach, we have implemented a tool for Erlang that offers *interleaving composition of protocols*, *code generation*, and *protocol extraction*.

*Interleaving composition* is defined as a function producing zero or more protocol compositions, giving an algorithmic implementation of the relation in Definition 5. Following the variations on the branching rule, the tool offers strong, weak, correlating, and weak/correlating (denoted All in the table) composition. The user can select the kind of branching they wish to use. Looking at Example 1, the strong composition of banking and authentication protocols returns an empty set as expected. When opting for weak composition instead, the tool outputs one IC, equivalent to Example 2:

**Table 1** Number of compositions for branching rule variations; running example in grey.

| Protocols | Strong | Weak | Correlating | All |
|---|---|---|---|---|
| 1) service(), login() | 0 | 1 | 0 | 1 |
| 2) s1(), s2() (Section 3.1.2) | 0 | 1 | 2 | 3 |
| 3) i1(), i2() (Section 2.1) | 1 | 1 | 1 | 1 |
| 4) http(), aws_auth() (from [28]) | 0 | 6 | 0 | 6 |
| 5) login(), booking() | 0 | 1 | 0 | 1 |
| 6) pin(), tan() | 0 | 1 | 0 | 1 |
| 7) pintan(), bank() | 0 | 1 | 0 | 1 |
| 8) resource(), server() | 1 | 1 | 1 | 2 |
| 9) userAgent(), agentInstrument() | 0 | 0 | 2 | 2 |
| 10) bankauthsimple(), keycard() | 0 | 1 | 0 | 1 |
| 11) auth_two_step(), email() | 0 | 9 | 0 | 9 |
| 12) sa(), sb() (Section 4.4) | 0 | 12 | 2 | 14 |

**Listing 1** PIN/TAN Banking Protocol rendered in our Erlang AST for protocols

```
bank_pintan() ->
  {act,r_pin, {branch, [{ok, {assert, pin, {require, pin, {rec, "r",
                              {branch, [{payment, {assert, pay, {consume, pay, {act,s_id, {act,r_tan,
                                 {branch, [{ok, {assert, tan, {consume, tan, {act, r_details, {rvar, "r"}}}}},
                                          {fail, {rvar, r}}]}}}}}}},
                                          {statement, {act, s_statement, {rvar, "r"}}},
                                          {logout, {consume, pin, endP}}]}}}}},
                        {fail, endP}]}}
```

Offering all four composition options (corresponding to $\triangleright_s$, $\triangleright_w$, $\triangleright_c$, $\triangleright_{wc}$ in the theory) instead of offering only the less restrictive weak/correlating branching $\triangleright_{wc}$, may improve the relevance of compositions returned. As observed in Section 3.1.2, using [wbra] in a context where we need to correlate branches likely returns irrelevant compositions (e.g. row 12). One way to reduce the number of irrelevant compositions, is to introduce more assertions. In fact, one of the aims of the tool is to support step-wise understanding of the protocol via progressive use of assertions. An alternative would be to annotate branching instances with the different options, which would further increase relevance of the returned results. This is left for future work. Table 1 shows the number of interleaving compositions obtained for each variation of the branching rule for a suite of examples. The suite includes: ad-hoc examples to validate the theory (rows 1 - 3, 7, 12), examples from literature, such as the HTTP example from [28] (row 4), and other examples inspired from real-world applications such as Gmail's two-steps authentication (row 11). By appropriately selecting composition options and assertions, the tool returns a small number of interleaving compositions. The number of compositions increases in examples with recursions, especially nested recursions as can be seen in rows 4, 11, which would require some additional assertions to choose among the interleavings.

*Code generation* takes a protocol definition and produces an Erlang stub. Protocol structures (action, sequence, choice) can be represented as a directed graph and then as finite state machines that transition based on the messages received. The finite state machines are used to generate a stub that uses the Erlang/OTP gen_statem [1], a generic abstraction which supports the implementation of finite state machine modules. Not only is it convenient to represent the protocol as a state machine, but `gen_statem` offers some useful features. Internal events from the state machine to itself are a good way to represent branches that make a selection among some choices. 'Postponing events' and timeouts provide functionality for further implementation of the generated code stubs. *Actions* and *branches* are represented as events that trigger a state transition. We use function declarations to represent incoming events, and function applications to represent outgoing events. Each state has its own handler function used to send an event to the state machine. When the event is received the corresponding state function is called and the transition to the next state is made. The default generated event is an asynchronous communication (called a 'cast' in Erlang/OTP parlance). For sending actions and selecting branches, the event type is internal, an event from your state machine to itself. *End* is represented by the terminate function of a `gen_statem` module, whilst the *fixed-point* and the *recursive variables* dictate the control flow of the state machine. State variables must be declared by including them in a record definition — `Data`. Following Frama-C [17], we represent *assertions* as specially formatted comments. For example: `{assert, pay}` is represented as an Erlang comment `%assert pay`. These comments are positioned before code that implements the state to which this assertion acts as a pre-condition in the protocol. Listing 2 shows an excerpt of the code generated for the PIN/TAN Banking protocol, `bank_pintan()`, containing the states generated for the first action and branch.

**Listing 2** PIN/TAN Banking State Machine

```
state1(cast, Pin, Data) -> {next_state, state2, Data}.
%assert pin
%require pin
state2(cast, ok, Data) -> {next_state, state3, Data};
state2(cast, fail, Data) -> {stop, normal, Data}.
%assert pay
%consume pay
state3(cast, payment, Data) -> {next_state, state4, Data};
state3(cast, statement,Data) ->{next_state, state10,Data};
%consume pin
state3(cast, logout, Data) -> {stop, normal, Data}.
```

*Protocol extraction and migration.* Protocol extraction generates protocols from code via a static analysis of Erlang modules implementing state machines using either `gen_statem`, or `gen_fsm` behaviour. When assertions are expressed using the comments illustrated above, they are also extracted. The obtained protocol can be annotated with extra assertions as necessary and composed with another to obtain a more complex protocol. The extraction option preserves local code that can be migrated when generating a new stub. For example, starting out from an existing implementation of banking, we can use the tool to extract the protocol $S_B$, obtain a composition with $S_A$, and generate a new module where pre-existing code for banking can be migrated.

*Re-engineering.* To extend the banking/authentication server with a keycard authentication option, we can compose the PIN/TAN Banking Protocol with e.g., the keycard protocol below. Assertions ensure that the branching for TAN or keycard authentication is plugged in (using assertion `keyp`) to the payment option of the PIN/TAN protocol, and that TAN authentication in PIN/TAN protocol is plugged only in the `tan` branch of the keycard protocol (using assertion `otp`):

**Listing 3** Keycard Option Protocol

```
keycard() -> {rec, "y", {require, keyp, {branch, [{tan, {assert, otp, {rvar, "y"}}},
                                                   {keycard, {rvar, "y"}}]}}}.
```

By adding an assertion of keyp and a consume of otp at the beginning of the branch payment of the PIN/TAN Protocol one would obtain the desired extension using the weak composition option. Our tool can then be used to generate a stub for the extended protocol and migrate reusable code from the implementation of the PIN/TAN Banking Protocol to the new implementation. These features satisfy the requirements laid out in Section 1.1: supporting re-engineering driven by the composition of protocols. The tool generates stubs from ICs, extracts protocols, and reuse code upon composition with different protocols. See our artifact for the complete benchmark [12].

## 6 Related Work and Conclusion

There is a vast literature on protocol specification (both theory and practice, e.g. see the survey of Lai [36]). Most techniques provide a monolithic view of protocols. We studied protocol composition using 'assertions' to specify contact points and constraints between the protocols. We have given correctness in terms of behaviour preservation, fairness and well-assertedness, and shown that all compositions enjoy it. There are three main lines of research that relate to our work.

Firstly, *software adaptors* give typed protocol interfaces between software components [45]. The idea is similar to the structured view of communication in session types [26], with the notion of *duality* capturing when opposite sides of a protocol are compatible. Composition in these works is about sound composition of protocol implementations, whereas we address the (upfront) creation of composite protocols.

Secondly, protocol composition has been studied as the run-time 'weaving' of component actions. Barbanera et al. study such a composition in the setting of communicating finite state machines [2, 3]. Participants in two communicating systems can be transformed into coupled 'gateways', forming a composite system. A compatibility relation is based on dual behaviour of the two gateways. Safety of the resulting system is by this compatibility, along with conditions of 'no mixed states' and determinism for sends and receives. Building from this idea, later work studies synchronous CFSMs, and replaces the two coupled gateways with a single one [5], and composition/decomposition on global types in the setting of multiparty session types [4]. Montesi and Yoshida study composition in the setting of choreographies [40]. Their composition relies on the use of partial choreographies, which can mix global descriptions with communication among external peers. Inspired by aspect-oriented programming,the work in [43] supports protocol extensions with 'aspectual' session types, that allow messages in session types to be matched and consequently introduce new behaviour in addition to, or in place of, the matched messages. Unlike the above approaches, we focus on a syntactic, statically derivable notion of composition. We use process calculi to model protocols as simple (i.e., mono-thread) objects that can be used by humans to reason about the desired application logic and generate/engineer modular code. The work in [35] looks at composition of aspects and modular verification of aspect-oriented programs, focussing on maintaining a relationship between models and aspects. Various works look at composition of features into coherent software systems [15, 24, 31, 46], focussing on resolving conflict stemming from feature interactions. Instead, we focus on establishing primitives for humans to reason on what a composite protocol should be, and support code generation.

The third pertinent thread in the literature defines syntactic compositions in the form of Team Automata [6, 7, 20] or related calculi [7]. These works define different ways of composing machines, primarily based on synchronising machines via common actions. In contrast, our means of composition is via assertions (orthogonal to actions) which express directional (i.e., rely-guarantee-style) dependencies. Our use of assertions aims to reflect programming practice. Assertions are kept in our generated code and can be used to enable protocol extraction and re-engineering, and as code documentation. Our composition cannot capture the whole range of synchronisations offered by Team Automata. Conversely, Team Automata cannot capture the range of compositions in our approach. One can encode some interleaving compositions as Team Automata, by modelling each assert($n$)-require($n$) or assert($n$)-consume($n$) pair as a common synchronisation action. However, the options offered by Team Automata (e.g., 'free', 'state indispensable', or 'action indispensable') do not capture our requirement that synchronisation (i) always happens on assertion-actions and (ii) never happens on communication actions (these are a separate syntactic and semantic entity). Furthermore, our assertions do not imply immediate synchronisation: an assert($n$) can occur in a protocol some way before a require($n$). Thus an attempted encoding of Team Automata into our protocols, encoding synchronisation actions as unique assert($n$)-consume($n$) pairs, would not preserve the behaviour of Team Automata for all possible compositions (just those where 'annihilating' pairs appeared contiguously). Thus, Team Automata and our approach overlap in some synchronising behaviours, but not all. A formal study of such overlap is further work.

Unlike approaches to safe communication discussed above, we do not focus on communication safety, which is an orthogonal concern. As discussed in Appendix B, our parameterisable language allows us to inherit communication-safety properties from session types by instantiating our protocol language to a session type syntax (e.g. that in [18]), with asynchrony [16] and multiparty sessions [10].

We are working on a factorisation function that decomposes protocols, as a kind of algebraic inverse to composition. This would allow us to 'close the loop', factorizing protocols into simple components for later (re)composition. We plan to extend recursion to model quantified recursion and assertion environments as multisets.

**Acknowledgements** This work has been partially supported by the BehAPI project funded by the EU H2020 RISE under the Marie Sklodowska-Curie action (No: 778233), EPSRC project EP/T014512/1 (STARDUST), and EPSRC project EP/T013516/1 (Granule). We thank Christian Kissig for contributing with ideas and discussions, and the anonymous reviewers for their feedback. Orchard is also supported by the generosity of Eric and Wendy Schmidt by recommendation of the Schmidt Futures program.

## A Illustrative Examples of IC Derivations

**Example 4 (Composition with** [act] and [sym]**)** *Consider the protocols* !Int.end *and* !String.end*. By combining [act] and [sym] one can obtain all interleavings of two sequences of actions,* !Int.!String.end *and* !String.!Int.end*, as shown with the two example derivations below:*

$$
\dfrac{\dfrac{\dfrac{\dfrac{-}{\emptyset;\emptyset;\emptyset \vdash \mathtt{end} \circ \mathtt{end} \triangleright \mathtt{end}}\,\textit{[end]}}{\emptyset;\emptyset;\emptyset \vdash !\mathtt{String.end} \circ \mathtt{end} \triangleright !\mathtt{String.end}}\,\textit{[act]}}{\emptyset;\emptyset;\emptyset \vdash \mathtt{end} \circ !\mathtt{String.end} \triangleright !\mathtt{String.end}}\,\textit{[sym]}}{\emptyset;\emptyset;\emptyset \vdash !\mathtt{Int.end} \circ !\mathtt{String.end} \triangleright !\mathtt{Int}.!\mathtt{String.end}}\,\textit{[act]}
$$

$$
\dfrac{\dfrac{\dfrac{\dfrac{\dfrac{-}{\emptyset;\emptyset;\emptyset \vdash \mathtt{end} \circ \mathtt{end} \triangleright \mathtt{end}}\,\textit{[end]}}{\emptyset;\emptyset;\emptyset \vdash !\mathtt{Int.end} \circ \mathtt{end} \triangleright !\mathtt{Int.end}}\,\textit{[act]}}{\emptyset;\emptyset;\emptyset \vdash \mathtt{end} \circ !\mathtt{Int.end} \triangleright !\mathtt{Int.end}}\,\textit{[sym]}}{\emptyset;\emptyset;\emptyset \vdash !\mathtt{String.end} \circ !\mathtt{Int.end} \triangleright !\mathtt{String}.!\mathtt{Int.end}}\,\textit{[act]}}{\emptyset;\emptyset;\emptyset \vdash !\mathtt{Int.end} \circ !\mathtt{String.end} \triangleright !\mathtt{String}.!\mathtt{Int.end}}\,\textit{[sym]}
$$

**Example 5 (Composition with assertions)** *Let* $I_2 = \mathsf{consume}(p).!\mathtt{item.end}$.

$$
\dfrac{\dfrac{\dfrac{\dfrac{\dfrac{\dfrac{-}{\emptyset;\emptyset;\emptyset \vdash \mathtt{end} \circ \mathtt{end} \triangleright \mathtt{end}}\,\textit{[end]}}{\emptyset;\emptyset;\emptyset \vdash !\mathtt{item.end} \circ \mathtt{end} \triangleright !\mathtt{item.end}}\,\textit{[act]}}{\emptyset;\emptyset;\{p\} \vdash \mathsf{consume}(p).!\mathtt{item.end} \circ \mathtt{end} \triangleright \mathsf{consume}(p).!\mathtt{item.end}}\,\textit{[consume]}}{\emptyset;\emptyset;\{p\} \vdash \mathtt{end} \circ I_2 \triangleright \mathsf{consume}(p).!\mathtt{item.end}}\,\textit{[sym]}}{\emptyset;\emptyset;\emptyset \vdash \mathsf{assert}(p).\mathtt{end} \circ I_2 \triangleright \mathsf{assert}(p).\mathsf{consume}(p).!\mathtt{item.end}}\,\textit{[assert]}}{\emptyset;\emptyset;\emptyset \vdash ?\mathtt{pay}.\mathsf{assert}(p).\mathtt{end} \circ I_2 \triangleright ?\mathtt{pay}.\mathsf{assert}(p).\mathsf{consume}(p).!\mathtt{item.end}}\,\textit{[act]}
$$

**Example 6 (Composition with** [rec1] and [rec2] **)**

$$
\dfrac{\dfrac{\dfrac{\dfrac{\dfrac{\dfrac{\underline{\mathtt{t}_1} \in \underline{\mathtt{t}_1}}{\emptyset;\underline{\mathtt{t}_1};\emptyset \vdash \mathtt{t}_1 \circ \mathtt{t}_1 \triangleright \mathtt{t}_1}\,\textit{[call]}}{\emptyset;\underline{\mathtt{t}_1};\emptyset \vdash !\mathtt{p}_2.\mathtt{t}_1 \circ \mathtt{t}_1 \triangleright \mathtt{l}_2 : \mathtt{t}_1}\,\textit{[act]}}{\emptyset;\mathtt{t}_1;\emptyset \vdash \mu\mathtt{t}_2.!\mathtt{p}_2.\mathtt{t}_2 \circ \mathtt{t}_1 \triangleright \mathtt{l}_2 : \mathtt{t}_1}\,\textit{[rec2]}}{\mathtt{t}_1;\emptyset;\emptyset \vdash \mathtt{t}_1 \circ \mu\mathtt{t}_2.!\mathtt{p}_2.\mathtt{t}_2 \triangleright \mathtt{l}_2 : \mathtt{t}_1}\,\textit{[sym]}}{\mathtt{t}_1;\emptyset;\emptyset \vdash \mathtt{q}.\mathtt{t}_1 \circ \mu\mathtt{t}_2.!\mathtt{p}_2.\mathtt{t}_2 \triangleright \mathtt{q}.\mathtt{l}_2 : \mathtt{t}_1}\,\textit{[act]}}{\emptyset;\emptyset;\emptyset \vdash \mu\mathtt{t}_1.\mathtt{q}.\mathtt{t}_1 \circ \mu\mathtt{t}_2.!\mathtt{p}_2.\mathtt{t}_2 \triangleright \mu\mathtt{t}_1.\mathtt{q}.\mathtt{l}_2 : \mathtt{t}_1}\,\textit{[rec1]}
$$

**Example 7 (Composition of nested recursion with weak branching)** *The following are just two or the possible derivations, given as illustration. They use weak branching introduced in Section 3.1.1. In the second derivation below, note that* $\mathtt{t}$ *could be substituted to* $\mathtt{t}_1$ *instead of* $\mathtt{t}_2$ *(yielding a different derivation).*

$$
\begin{array}{cl}
\underline{\mathtt{t}} \in \underline{\mathtt{t}} & \\ \hline
\mathtt{t}_2;\, \underline{\mathtt{t}};\, \emptyset \vdash \mathtt{t} \circ \mathtt{t} \triangleright \mathtt{t} \text{ (only the first branch derives)} & [\textit{call}] \\ \hline
\mathtt{t}_2;\, \underline{\mathtt{t}};\, \emptyset \vdash +\{\mathtt{l}_1 : \mathtt{t}, \mathtt{l}_2 : \mathtt{t}_2\} \circ \mathtt{t} \triangleright +\{\mathtt{l}_1 : \mathtt{t}, \mathtt{l}_2 : \mathtt{t}_2\} & [\textit{wbra}] \\ \hline
\mathtt{t}_2;\, \underline{\mathtt{t}};\, \emptyset \vdash +\{\mathtt{l}_1 : \mathtt{t}, \mathtt{l}_2 : \mathtt{t}_2\} \circ \mathtt{t} \triangleright +\{\mathtt{l}_1 : \mathtt{t}, \mathtt{l}_2 : \mathtt{t}_2\} & [\textit{act}] \\ \hline
\emptyset;\, \underline{\mathtt{t}};\, \emptyset \vdash \mu\mathtt{t}_2.+\{\mathtt{l}_1 : \mathtt{t}, \mathtt{l}_2 : \mathtt{t}_2\} \circ \mathtt{t} \triangleright \mu\mathtt{t}_2.+\{\mathtt{l}_1 : \mathtt{t}, \mathtt{l}_2 : \mathtt{t}_2\} & [\textit{rec1}] \\ \hline
\emptyset;\, \underline{\mathtt{t}};\, \emptyset \vdash q.\mu\mathtt{t}_2.+\{\mathtt{l}_1 : \mathtt{t}, \mathtt{l}_2 : \mathtt{t}_2\} \circ \mathtt{t} \triangleright q.\mu\mathtt{t}_2.+\{\mathtt{l}_1 : \mathtt{t}, \mathtt{l}_2 : \mathtt{t}_2\} & [\textit{act}] \\ \hline
\emptyset;\, \mathtt{t};\, \emptyset \vdash \mu\mathtt{t}_1.q.\mu\mathtt{t}_2.!\mathtt{p}_2.+\{\mathtt{l}_1 : \mathtt{t}_1, \mathtt{l}_2 : \mathtt{t}_2\} \circ \mathtt{t} \triangleright q.\mu\mathtt{t}_2.+\{\mathtt{l}_1 : \mathtt{t}, \mathtt{l}_2 : \mathtt{t}_2\} & [\textit{rec2}] \\ \hline
\mathtt{t};\, \emptyset;\, \emptyset \vdash \mathtt{t} \circ \mu\mathtt{t}_1.q.\mu\mathtt{t}_2.+\{\mathtt{l}_1 : \mathtt{t}_1, \mathtt{l}_2 : \mathtt{t}_2\} \triangleright q.\mu\mathtt{t}_2.+\{\mathtt{l}_1 : \mathtt{t}, \mathtt{l}_2 : \mathtt{t}_2\} & [\textit{sym}] \\ \hline
\mathtt{t};\, \emptyset;\, \emptyset \vdash p.\mathtt{t} \circ \mu\mathtt{t}_1.q.\mu\mathtt{t}_2.+\{\mathtt{l}_1 : \mathtt{t}_1, \mathtt{l}_2 : \mathtt{t}_2\} \triangleright p.q.\mu\mathtt{t}_2.+\{\mathtt{l}_1 : \mathtt{t}, \mathtt{l}_2 : \mathtt{t}_2\} & [\textit{act}] \\ \hline
\emptyset;\, \emptyset;\, \emptyset \vdash \mu\mathtt{t}.p.\mathtt{t} \circ \mu\mathtt{t}_1.q.\mu\mathtt{t}_2.+\{\mathtt{l}_1 : \mathtt{t}_1,\, \mathtt{l}_2.\mathtt{t}_2\} \triangleright \mu\mathtt{t}.p.q.\mu\mathtt{t}_2.+\{\mathtt{l}_1 : \mathtt{t}, \mathtt{l}_2 : \mathtt{t}_2\} & [\textit{rec1}]
\end{array}
$$

$$
\begin{array}{cl}
\underline{\mathtt{t}_2} \in \mathtt{t}_1, \underline{\mathtt{t}_2} & \\ \hline
\mathtt{t}_1, \underline{\mathtt{t}_2};\, \emptyset;\, \emptyset \vdash \mathtt{t}_2 \circ \mathtt{t}_2 \triangleright \mathtt{t}_2 \text{ (only the second branch derives)} & [\textit{call}] \\ \hline
\mathtt{t}_1, \underline{\mathtt{t}_2};\, \emptyset;\, \emptyset \vdash +\{\mathtt{l}_1 : \mathtt{t}_1, \mathtt{l}_2 : \mathtt{t}_2\} \circ \mathtt{t}_2 \triangleright +\{\mathtt{l}_1 : \mathtt{t}_1,\, \mathtt{l}_2 : \mathtt{t}_2\} & [\textit{wbra}] \\ \hline
\emptyset;\, \mathtt{t}_1, \underline{\mathtt{t}_2};\, \emptyset \vdash \mathtt{t}_2 \circ +\{\mathtt{l}_1 : \mathtt{t}_1, \mathtt{l}_2 : \mathtt{t}_2\} \triangleright +\{\mathtt{l}_1 : \mathtt{t}_1,\, \mathtt{l}_2 : \mathtt{t}_2\} & [\textit{sym}] \\ \hline
\emptyset;\, \mathtt{t}_1, \underline{\mathtt{t}_2};\, \emptyset \vdash p.\mathtt{t}_2 \circ +\{\mathtt{l}_1 : \mathtt{t}_1, \mathtt{l}_2 : \mathtt{t}_2\} \triangleright p.+\{\mathtt{l}_1 : \mathtt{t}_1,\, \mathtt{l}_2 : \mathtt{t}_2\} & [\textit{act}] \\ \hline
\emptyset;\, \mathtt{t}_1, \mathtt{t}_2;\, \emptyset \vdash \mu\mathtt{t}.p.\mathtt{t} \circ +\{\mathtt{l}_1 : \mathtt{t}_1, \mathtt{l}_2 : \mathtt{t}_2\} \triangleright p.+\{\mathtt{l}_1 : \mathtt{t}_1,\, \mathtt{l}_2 : \mathtt{t}_2\} & [\textit{rec2}] \\ \hline
\mathtt{t}_1, \mathtt{t}_2;\, \emptyset;\, \emptyset \vdash +\{\mathtt{l}_1 : \mathtt{t}_1, \mathtt{l}_2 : \mathtt{t}_2\} \circ \mu\mathtt{t}.p.\mathtt{t} \triangleright p.+\{\mathtt{l}_1 : \mathtt{t}_1,\, \mathtt{l}_2 : \mathtt{t}_2\} & [\textit{sym}] \\ \hline
\mathtt{t};\, \emptyset;\, \emptyset \vdash \mu\mathtt{t}_2.+\{\mathtt{l}_1 : \mathtt{t}, \mathtt{l}_2 : \mathtt{t}_2\} \circ \mu\mathtt{t}.p.\mathtt{t} \triangleright \mu\mathtt{t}_2.p.+\{\mathtt{l}_1 : \mathtt{t}_1,\, \mathtt{l}_2 : \mathtt{t}_2\} & [\textit{rec1}] \\ \hline
\mathtt{t}_1;\, \emptyset;\, \emptyset \vdash q.\mu\mathtt{t}_2.+\{\mathtt{l}_1 : \mathtt{t}_1, \mathtt{l}_2 : \mathtt{t}_2\} \circ \mu\mathtt{t}.p.\mathtt{t} \triangleright q.\mu\mathtt{t}_2.p.+\{\mathtt{l}_1 : \mathtt{t}_1, \mathtt{l}_2 : \mathtt{t}_2\} & [\textit{act}] \\ \hline
\emptyset;\, \emptyset;\, \emptyset \vdash \mu\mathtt{t}_1.q.\mu\mathtt{t}_2.+\{\mathtt{l}_1 : \mathtt{t}_1, \mathtt{l}_2 : \mathtt{t}_2\} \circ \mu\mathtt{t}.p.\mathtt{t} \triangleright \mu\mathtt{t}_1.q.\mu\mathtt{t}_2.p.+\{\mathtt{l}_1 : \mathtt{t}_1, \mathtt{l}_2 : \mathtt{t}_2\} & [\textit{rec1}]
\end{array}
$$

# B Examples with Alternative Instantiations of the Protocol Language

The protocol language given in Section 2 is parametric on the set of action/branching prefixes and branching labels and our results hold regardless of the specific instantiation used. Instantiation allows us to apply our framework to different scenarios. Many of the examples in this paper are in the context of concurrency or distribution, yet protocols are pervasive in monolithic sequential code, e.g., for interacting with an operating system or libraries. A classic example is the stateful protocol of file handling in which files must be opened and closed, and only read and written to according to their permissions between those two actions. Our protocol language can easily model such situations, with interleaving composition providing a range of choices to a developer about how to combine and interact multiple stateful protocols. In this section we provide a more concrete discussion of a few use cases focussing on interaction and communication, providing hints at possible synergies between our framework and techniques already studied for specific formalisms.

## B.1 Protocols for Communicating Processes

Interaction structures of a communicating process can be characterised using process calculi such as CCS [38]. For example, a CCS process $S = com.S + \overline{out}.0$ can be understood as a protocol prescribing the interactions supported by a server: repeatedly receive commands (*com* is a receive action) or decide to logout (*out* is a send action) and terminate. The CCS notation can be expressed in our protocol language by instantiating

$$\mathscr{P} = \{a, \overline{a} \mid a \in \mathit{Names}\} \quad + \in \{+\} \quad \mathscr{L} = \mathscr{P}$$

where $a$ (resp. $\overline{a}$) models a receive (resp. send) action on channel $a$, and *Names* is a set of channel names. Hence, $S$ above can be represented in our framework as the process below $\mu\mathtt{t}.+\{\mathit{com} : \mathtt{t}, \overline{\mathit{out}} : \mathtt{end}\}$.

### B.2 Protocols as Session Types

As mentioned in Remark 1 we often used a session types-like syntax in our examples. In this section we show how to use our framework in combination with session types techniques. In Example 2 we have used interleaving composition to generate a banking/authentication protocol $S_{BA}$ using a session types-like syntax, and in Section 5 we have shown that our tool can generate a stub implementation of $S_{BA}$, which one can then extend with local (i.e., non communication-related) behaviour. Assume this implementation, say `bank_pt.erl`, is published as a web API. $S_{BA}$ can be published as part of the API specification (as a *behavioural* API [8]).

So far, our framework has provided assurances on the relationship between component protocols and their interleaving composition (which pertains to the engineering within one node in a distributed system – in this case the banking/authentication server). Session types serve an orthogonal purpose: to provide assurances about the *inter-relations* of the protocols implemented in different nodes, e.g., given a well-defined banking/authentication server, how to derive a *suitable* client?

Anyone willing to develop a client for the banking/authentication service can use $S_{BA}$ to algorithmically derive a client protocol by using the notion of *duality* of Session Types [25, 26, 44]. The dual of a protocol is obtained algorithmically, by swapping the 'direction' of each action and branching: ! with ?, ⊕ with &, and vice-versa. We let $\overline{S_{BA}}$ denote the dual of $S_{BA}$, omitting the assertions for readability.

$$\overline{S_{BA}} = !\mathtt{pin}.\&\left\{\begin{array}{ll} \mathsf{ok}: & \mu\mathtt{r}.\oplus\left\{\begin{array}{ll} \mathsf{payment}: & ?\mathtt{id}.!\mathtt{tan}.\&\left\{\begin{array}{ll}\mathsf{ok}: & !\mathtt{details}.\mathtt{r}\\ \mathsf{fail}: & \mathtt{r}\end{array}\right\}\\ \mathsf{statement}: & ?\mathtt{statement}.\mathtt{r}\\ \mathsf{logout}: & \mathtt{end}\end{array}\right\}\\ \mathsf{fail}: & \mathtt{end}\\ \mathsf{logout}: & \mathsf{consume}(\mathit{pin}).\mathtt{end}\end{array}\right\}$$

Our tool can be used again to generate a stub for $\overline{S_{BA}}$ (e.g., file `co_bank_pt.erl`). Session types duality and communication structuring (e.g., determinism, no mixed choices) yields a *safe* distributed system, resulting from, say, the concurrent and possibly distributed execution of `bank_pt.erl` and `co_bank_pt.erl`. In this context, *safe* means no deadlock and no communication mismatches even when communications are asynchronous [16] as in Erlang.

Ideally, we could model higher-order messages by instantiating prefixes to incorporate the entire protocol language itself, preventing use of delegated channels using assertions.

### B.3 Interleaving Composition and Multiparty Session Types

Consider a protocol, modelled using the session types syntax, that specifies the possible interactions between a user $U$ and a remote instrument $I$ (e.g., a camera). Below, $S_I$ is the protocol specified from the perspective of $I$, which offers a menu with two choices: set or get. In case of set, $I$ receives coordinates to update its own state, and in case of get, $I$ sends a picture from the current coordinates. Protocol $\overline{S_U}$ is the dual of $S_I$ (i.e., specified from the perspective of $U$).

$$S_I = \mu\mathtt{t}.\&\{\mathsf{set}\colon ?\mathtt{coord}.\mathtt{t}, \mathsf{get}\colon !\mathtt{snap}.\mathtt{t}\} \qquad \overline{S_U} = \mu\mathtt{t}.+\{\mathsf{set}\colon !\mathtt{coord}.\mathtt{t}, \mathsf{get}\colon ?\mathtt{snap}.\mathtt{t}\}$$

The protocols above may have been designed top-down or extracted using our tool out of an existing system. Assume we need to modify the scenario above by introducing a proxy agent $A$, so that $U$ and $I$ will only interact via $A$. We need to: (1) express the protocols so that it is clear which roles are involved in each interaction, and (2) define the protocol for $A$. We address (1) by using a different instantiation of the protocol language to make roles explicit. For example, by fixing a set *Roles* of protocol roles, a set $\mathscr{T}$ of datatypes, and letting $\mathscr{L} \subset \mathbb{N}$ and:

$$\mathscr{P} = \{ab\#\mathtt{T} \mid a, b \in Roles, \# \in \{!, ?\}, T \in \mathscr{T}\} \qquad + \in \{ab\# \mid a, b \in Roles, \# \in \{+, \&\}\}$$

Then we obtain protocols in a multiparty session type syntax (e.g., local types in [10]), where $ab!$ (resp. $ab?$) denotes a send action by $a$ (resp. receive action by $b$) in an asynchronous interaction from $a$ to $b$. Similarly for branching and selection. This instantiation allows us to model the following multiparty versions of $S_I$ and $S_U$, respectively:

$$S_{AI} = \mu\mathtt{t}.AI\,\&\begin{cases}\mathsf{set}: AI\,?\mathtt{coord}.\mathtt{t},\\ \mathsf{get}: IA\,!\mathtt{snap}.\mathtt{t}\end{cases} \qquad S_{UA} = \mu\mathtt{t}.UA\,+\begin{cases}\mathsf{set}: UA\,!\mathtt{coord}.\mathtt{t},\\ \mathsf{get}: AU\,?\mathtt{snap}.\mathtt{t}\end{cases}$$

We would like $A$ to act as a server for $U$ and as a client for $I$. Concretely, we could generate the protocol for $A$ as a specific 'forwarding' interleaving of the dual of $S_{AI}$ and the dual of $S_{UA}$.

We use assertions to ensure that $A$ sends $I$ a menu choice only after having received one from $U$ (and this must be the same choice received by $U$), and then $A$ must reflect the behaviour of $I$ following the given choice. The asserted protocols to be composed into the protocol for $A$ are given below, where $\mathsf{assert}(set)/\mathsf{consume}(set)$ and $\mathsf{assert}(get)/\mathsf{consume}(get)$ express the desired correlation between branches, and $\mathsf{assert}(f)/\mathsf{consume}(f)$ model the forwarding pattern:

$$\mu\mathtt{t}.AI\,+\left\{\begin{array}{l}\mathsf{set}: \mathsf{consume}(set).\,\mathsf{consume}(f).\,AI\,!\mathtt{coord}.\mathtt{t},\\ \mathsf{get}: \ \mathsf{consume}(get).\,AI\,?\mathtt{snap}.\mathsf{assert}(f).\,\mathtt{t}\end{array}\right\}$$

$$\mu\mathtt{t}.\,UA\,\&\left\{\begin{array}{l}\mathsf{set}: \mathsf{assert}(set).\,UA?\mathtt{coord}.\mathsf{assert}(f).\,\mathtt{t},\\ \mathsf{get}: \mathsf{assert}(get).\,\mathsf{consume}(f).AU\,!\mathtt{snap}.\mathtt{t}\end{array}\right\}$$

Using interleaving composition with correlating branching (Section 3.1.2) on the two protocols above we obtain the following interleaving composition specifying the protocol for $A$, where we omit the assertions for readability:

$$S_{UAI} = \mu \texttt{t}.\, UA \,\&\, \begin{cases} \text{set} : AI \;+\{\text{set} : UA?\texttt{coord}.AI\,!\texttt{coord}.\texttt{t}\}, \\ \text{set} : AI \;+\{\text{get} : IA?\texttt{snap}.AU\,!\texttt{snap}.\texttt{t}\} \end{cases}$$

Now that we have the protocols of the extended multiparty system we can use our tool to generate code for $S_{UAI}$, $S_{AI}$, and $S_{UA}$. Thanks to the extraction/migration functionality of our tool, pre-existing local code for $U$ and $I$ can be reused in the new code for $U$ and $I$, where the specific endpoints for communication can be added manually. The tool does not yet support syntax for specifying roles, so in the case of an agent communicating with several parties such as $S_{UAI}$ the direction of the communication would need to be specified manually. In the multiparty scenario, correctness of the interaction structures can be checked using multiparty compatibility [19] and verification techniques [37].

## C Basic Properties of Protocols

We first define some additional results in this appendix which are used for some of the key lemmas of this section. We have the following set of inversion lemmas on well-assertedness:

**Lemma C.1 (Prefix well-asserted inversion)** $\forall A, A', S.\ A\ \{p.S\}\, A' \implies A\ \{S\}\, A'$.

**Proof 1** *There is only one rule that provides the well-assertedness of prefixing:*

$$\frac{A\ \{S\}\, A'}{A\ \{p.S\}\, A'}$$

**Lemma C.2 (Branch well-asserted inversion)** $\forall A, A', I, \{S_i\}_{i\in I}$.

$$A\ \{+\{\mathsf{l}_i : S_i\}_{i\in I}\}\, A' \implies A' \equiv \exists \{A_i\}_{i\in I}.\bigcap_{i\in I} A_i \ \wedge\ \forall i \in I.A\ \{S_i\}\, A_i$$

**Proof 2** *There is only one rule that provides the well-assertedness of branching:*

$$\frac{\forall i \in I.\ B\ \{S_i\}\, B_i}{B\ \{+\{\mathsf{l}_i : S_i\}_{i\in I}\}\ \bigcap_{i\in I} B_i)}$$

*Given the antecedent of this lemma, we then have that* $\{A_i\}_{i\in I} = \{B_i\}_{i\in I}$ *and* $A' = \bigcap_{i\in I} B_i$ *then the premise provides the consequent of the lemma,* $\forall i \in I.A\ \{S_i\}\, A_i$.

**Lemma C.3 (Assert well-asserted inversion)** $\forall A, A', n, S.\ A\ \{\text{assert}(n).S\}\, A' \implies A \cup \{n\}\ \{S\}\, A'$.

**Proof 3** *There is only one rule that provides the well-assertedness of assert:*

$$\frac{A \cup \{n\}\ \{S\}\, A'}{A\ \{\mathsf{assert}(n).S\}\, A'}$$

**Lemma C.4 (Require well-asserted inversion)** $\forall A, A', n, S.\ A\ \{\mathsf{require}(n).S\}\, A' \implies A\ \{S\}\, A' \wedge n \in A.$

**Proof 4** *There is only one rule that provides the well-assertedness of require:*

$$\frac{B \cup \{n\}\ \{S\}\, B'}{B \cup \{n\}\ \{\mathsf{require}(n).S\}\, B'}$$

*Thus we have that* $A = B \cup \{n\}$ *and so* $n \in A$ *and* $A' = B'$ *thus the premise provides the consequent of this lemma.*

**Lemma C.5 (Consume well-asserted inversion)** $\forall A, A', n, S. A\, \{\mathsf{consume}(n).S\}\, A' \implies n \in A \wedge A \setminus \{n\}\ \{S\}\, A'.$

**Proof 5** *There is only one rule that provides the well-assertedness of consume:*

$$\frac{B\ \{S\}\, B' \qquad n \in (B \cup \{n\})}{B \cup \{n\}\ \{\mathsf{consume}(n).S\}\, B'}$$

*Thus let* $A = B \cup \{n\}$ *and* $A' = B'$ *and therefore* $n \in A$*. The (first) premise of this rule then provides the consequent of this lemma.*

**Lemma C.6 (Recursion well-asserted inversion)** $\forall A, A', n, S.\ A\, \{\mu \mathtt{t}.S\}\, A' \implies A\, \{S\}\, A' \wedge A \subseteq A'.$

**Proof 6** *There is only one rule that provides the well-assertedness of recursion:*

$$\frac{B\ \{S\}\, B \cup B'}{B\ \{\mu \mathtt{t}.S\}\, B \cup B'}$$

*Thus we let* $A = B$ *and* $A' = B \cup B'$ *yielding* ($A \subseteq A'$) *and then premise provides the consequent of this lemma.*

**Lemma C.7 (Well-asserted unfolding extension)** *For all*

$$A\ \{S[\mu \mathtt{t}.S/\mathtt{t}]\}\, A' \Rightarrow A\ \{S[\mu \mathtt{t}.e.S/\mathtt{t}]\}\, A'$$

*where* $e$ *ranges over* $p$, $\mathsf{require}(n), \mathsf{assert}(n), \mathsf{consume}(n), \mu \mathtt{t}'$ *(in the last case then* . *becomesa scoping rather than a prefixing, by overloading).*

**Proof 7** ▪ *(act)* $S = p.S'$. *Assume* $A\ \{p.S'[\mu\mathtt{t}.p.S'/\mathtt{t}]\}A'$ *then by inversion (Lemma C.1) this yields* $A\ \{S'[\mu\mathtt{t}.p.S'/\mathtt{t}]\}A'$.
*By induction then* $A\ \{S'[\mu\mathtt{t}.e.p.S'/\mathtt{t}]\}A'$*, which then allows us to derive:*

$$\frac{A\ \{S'[\mu\mathtt{t}.e.p.S'/\mathtt{t}]\}A'}{A\ \{p.S'[\mu\mathtt{t}.e.p.S'/\mathtt{t}]\}A'}\text{[act]}$$

*which equals our goal*

$$A\ \{(p.S')[\mu\mathtt{t}.e.p.S'/\mathtt{t}]\}A'$$

*by the definition of syntactic substitution.*

- *(bra)* $S = +\{\mathrm{l}_i : S_i\}_{i\in I}$ *then by inversion (Lemma C.2) this yields* $A\ \{S_i\}A_i$ *for all* $i \in I$*. with* $A' \equiv \exists\{A_i\}_{i\in I}.\bigcap_{i\in I} A_i$.
*By induction on each then we have* $A\ \{S_i[\mu\mathtt{t}.e.+\{\mathrm{l}_i : S_i\}_{i\in I}/\mathtt{t}]\}A'_i$ *allowing us to re-derive branching well-assertedness:*

$$\frac{A\ \{S_i[\mu\mathtt{t}.e.+\{\mathrm{l}_i : S_i\}_{i\in I}/\mathtt{t}]\}A'_i}{A\ \{+\{\mathrm{l}_i : S_i[\mu\mathtt{t}.e.+\{\mathrm{l}_i : S_i\}_{i\in I}/\mathtt{t}]\}_{i\in I}\}A'}\text{[bra]}$$

*which equals our goal by the definition of syntactic substitution:*

$$A\ \{(+\{\mathrm{l}_i : S_i\}_{i\in I})[\mu\mathtt{t}.e.+\{\mathrm{l}_i : S_i\}_{i\in I}/\mathtt{t}]\}A'$$

- *(assert)* $S = \mathsf{assert}(n).S'$. *Assuming* $A\ \{\mathsf{assert}(n).S'[\mu\mathtt{t}.\mathsf{assert}(n).S'/\mathtt{t}]\}A'$ *then by inversion (Lemma C.3) this yields* $A \cup \{n\}\ \{S'[\mu\mathtt{t}.\mathsf{assert}(n).S'/\mathtt{t}]\}A'$.
*By induction then* $A \cup \{n\}\ \{S'[\mu\mathtt{t}.e.\mathsf{assert}(n).S'/\mathtt{t}]\}A'$*, which then allows us to derive:*

$$\frac{A \cup \{n\}\ \{S'[\mu\mathtt{t}.e.\mathsf{assert}(n).S'/\mathtt{t}]\}A'}{A\ \{\mathsf{assert}(n).S'[\mu\mathtt{t}.e..S'/\mathtt{t}]\}A'}\text{[assert]}$$

*which equals our goal*

$$A\ \{(\mathsf{assert}(n).S')[\mu\mathtt{t}.e.\mathsf{assert}(n).S'/\mathtt{t}]\}A'$$

*by the definition of syntactic substitution.*

- *(require)* $S = \mathsf{require}(n).S'$. *Assuming* $A\ \{\mathsf{require}(n).S'[\mu\mathtt{t}.\mathsf{require}(n).S'/\mathtt{t}]\}A'$ *then by inversion (Lemma C.4) this yields* $A\ \{S'[\mu\mathtt{t}.\mathsf{require}(n).S'/\mathtt{t}]\}A'$ *(with* $n \in A$*).*
*By induction then* $A\ \{S'[\mu\mathtt{t}.e.\mathsf{require}(n).S'/\mathtt{t}]\}A'$*, which then allows us to derive:*

$$\frac{A\ \{S'[\mu\mathtt{t}.e.\mathsf{require}(n).S'/\mathtt{t}]\}A' \qquad n \in A}{A\ \{\mathsf{require}(n).S'[\mu\mathtt{t}.e.\mathsf{require}(n).S'/\mathtt{t}]\}A'}\text{[require]}$$

*which equals our goal*

$$A\ \{(\mathsf{require}(n).S')[\mu\mathtt{t}.e.\mathsf{require}(n).S'/\mathtt{t}]\}A'$$

*by the definition of syntactic substitution.*

- *(consume)* $S = \mathsf{consume}(n).S'$. *Assume* $A\ \{\mathsf{consume}(n).S'[\mu\mathtt{t}.\mathsf{consume}(n).S'/\mathtt{t}]\}A'$ *then by inversion (Lemma C.5) this yields* $A \setminus \{n\}\ \{S'[\mu\mathtt{t}.\mathsf{consume}(n).S'/\mathtt{t}]\}A'$ *(with* $n \in A$*).*
  *By induction then* $A \setminus \{n\}\ \{S'[\mu\mathtt{t}.e.\mathsf{consume}(n).S'/\mathtt{t}]\}A'$*, which then allows us to derive:*

$$\frac{A \setminus \{n\}\ \{S'[\mu\mathtt{t}.e.\mathsf{consume}(n).S'/\mathtt{t}]\}A' \qquad n \in A}{A\ \{\mathsf{consume}(n).S'[\mu\mathtt{t}.e.\mathsf{consume}(n).S'/\mathtt{t}]\}A'}[consume]$$

  *which equals our goal*

$$A\ \{(\mathsf{consume}(n).S')[\mu\mathtt{t}.e.\mathsf{consume}(n).S'/\mathtt{t}]\}A'$$

  *by the definition of syntactic substitution.*
- *(rec)* $S = \mu\mathtt{t}'.S'$ *Assume* $A\ \{(\mu\mathtt{t}'.S')[\mu\mathtt{t}.(\mu\mathtt{t}'.S')/\mathtt{t}]\}A'$ *then by inversion (Lemma C.6) this yields* $A\ \{S'[\mu\mathtt{t}.(\mu\mathtt{t}'.S')/\mathtt{t}]\}A'$ *(with* $A \subseteq A'$*).*
  *By induction then* $A\ \{S'[\mu\mathtt{t}.e.(\mu\mathtt{t}'.S')/\mathtt{t}]\}A'$*, which then allows us to derive:*

$$\frac{A\ \{S'[\mu\mathtt{t}.e.(\mu\mathtt{t}'.S')/\mathtt{t}]\}A'}{A\ \{\mu\mathtt{t}'.S'[\mu\mathtt{t}.\mu\mathtt{t}'.S'/\mathtt{t}]\}A'}[rec]$$

  *(note the post-condition here satisfies* $\exists A''.A \cup A'' = A'$ *since* $A \subseteq A'$*).*
  *The conclusion equals our goal*

$$A\ \{(\mu\mathtt{t}'.S')[\mu\mathtt{t}.\mu\mathtt{t}'.S'/\mathtt{t}]\}A'$$

  *by the definition of syntactic substitution and uniquness of binders property.*
- *(end)* $S = \mathtt{end}$ *Assuming* $A\ \{\mathtt{end}[\mu\mathtt{t}.\mathtt{end}/\mathtt{t}]\}A'$.
  *Since* $\mathtt{end}[\mu\mathtt{t}.\mathtt{end}/\mathtt{t}] = [\mu\mathtt{t}.e.\mathtt{end}/\mathtt{t}]$ *then this holds trivially from the assumption.*
- *(call)* $S = \mathtt{t}'$*. Assuming* $A\ \{\mathtt{t}'[\mu\mathtt{t}.\mathtt{t}'/\mathtt{t}]\}A'$.
  - $\mathtt{t} = \mathtt{t}'$ *then* $\mathtt{t}'[\mu\mathtt{t}.\mathtt{t}'/\mathtt{t}] = \mathtt{t}[\mu\mathtt{t}.\mathtt{t}/\mathtt{t}] = \mu\mathtt{t}.\mathtt{t}$ *Such a protocol is not allowed by the syntactic guardness requirement, so this case is trivial (ex falso quodlibet).*
  - $\mathtt{t} \neq \mathtt{t}'$ *then* $\mathtt{t}'[\mu\mathtt{t}.\mathtt{t}'/\mathtt{t}] = \mathtt{t}'$ *Then the goal hoalds trivially here as from the assumption we get* $A\ \{\mathtt{t}'[\mu\mathtt{t}.e.\mathtt{t}'/\mathtt{t}]\}A'$.

**Lemma C.8 (Well-asserted unfolding extension under branch)** *For all*

$$A\ \{S_i\}A' \wedge A\ \{+\{\mathrm{l}_i : S_i\}_{i\in I}\}A' \Rightarrow A\ \{S_i[+\{\mathrm{l}_i : \mu\mathtt{t}.S_i\}_{i\in I}/\mathtt{t}]\}A'$$

**Proof 8** *By induction over the structure of* $S_i$.

*In each case, we proceed by induction, rebuilding well-assertedness (exactly as in the proof of Lemma C.7). The key case is when we have a recursion variable that we are substituting into,* $\mathtt{t}'$.

- $\mathtt{t}' \equiv \mathtt{t}$ *then we substitute here:* $\mathtt{t}'[+\{\mathrm{l}_i : \mu\mathtt{t}.S_i\}_{i\in I}/\mathtt{t}] = +\{\mathrm{l}_i : \mu\mathtt{t}.S_i\}_{i\in I}$ *and so well-assertedness holds by the second conunct of the premise.*
- $\mathtt{t}' \neq \mathtt{t}'$ *then trivially* $A\ \{\mathtt{t}'\}A'$ *since* $\mathtt{t}'[+\{\mathrm{l}_i : \mu\mathtt{t}.S_i\}_{i\in I}/\mathtt{t}] = \mathtt{t}'$

**Lemma C.9 (Well-asserted unfolding)** *For all sets of names $A$,$A'$ and protocols $S$, then:*

$$A\ \{S\} A' \implies A\ \{S[\mu t.S/t]\} A'$$

**Proof 9** *By induction on the structure of terms $S$:*

- *(act) $S = p.S'$ with assumption $A\ \{p.S'\} A'$. By Lemma C.1 (inversion) we then have $A\ \{S'\} A'$.*
  *By induction on this judgment we have that $A\ \{S'[\mu t.S'/t]\} A'$.*
  *By Lemma C.7 then this gives us $A\ \{S'[\mu t.p.S'/t]\} A'$ which we can use to build the well-assertedness derivation:*

$$\frac{A\ \{S'[\mu t.p.S'/t]\} A'}{A\ \{p.S'[\mu t.p.S'/t]\} A'}\textit{[act]}$$

  *which yields our goal by the definition of syntactic substitution.*

- *(bra) $S = +\{l_i : S_i\}_{i\in I}$ with assumption $A\ \{+\{l_i : S_i\}_{i\in I}\} A'$.*
  *By inversion (Lemma C.2) this yields $A\ \{S_i\} A_i$ for all $i \in I$. with $A' \equiv \exists\{A_i\}_{i\in I}.\bigcap_{i\in I} A_i$.*
  *By induction on each $i \in I$ then we have that $A\ \{S_i[\mu t.S_i/t]\} A'$. Applying Lemma C.8 on each then this give us: $A\ \{S_i[+\{l_i : \mu t.S_i\}_{i\in I}/t]\} A'$ which we can then use to build the well-assertedness derivation:*

$$\frac{\forall i \in I.\ A\ \{S_i[\mu t. + \{l_i : S_i\}_{i\in I}/t]\} A_i}{A\ \{+\{l_i : S_i[\mu t. + \{l_i : S_i\}_{i\in I}/t]\}_{i\in I}\} \bigcap_{i\in I} A_i}$$

  *which by the definition of syntactic substitution yields the goal:*

$$A\ \{(+\{l_i : S_i\}_{i\in I})[\mu t. + \{l_i : S_i\}_{i\in I}/t]\} \bigcap_{i\in I} A_i\ \textit{[bra]}$$

- *(require) $S = \mathsf{require}(n).S'$ with assumption $A\ \{\mathsf{require}(n).S'\} A'$.*
  *By Lemma C.4 (inversion) we then have $A\ \{S'\} A'$ and $n \in A$.*
  *By induction on this judgment we have that $A\ \{S'[\mu t.S'/t]\} A'$.*
  *By Lemma C.7 then this gives us $A\ \{S'[\mu t.\mathsf{require}(n).S'/t]\} A'$ which we can use to build the well-assertedness derivation:*

$$\frac{A\ \{S'[\mu t.\mathsf{require}(n).S'/t]\} A'}{A\ \{\mathsf{require}(n).S'[\mu t.\mathsf{require}(n).S'/t]\} A'}\textit{[require]}$$

  *(where $\exists A''.A = A'' \cup \{n\}$ sinch $n \in A$). which yields our goal by the definition of syntactic substitution.*

- *(consume) $S = \mathsf{consume}(n).S'$ with assumption $A\ \{\mathsf{consume}(n).S'\} A'$.*
  *By Lemma C.5 (inversion) we then have $A \setminus \{n\}\ \{S'\} A'$ and $n \in A$.*
  *By induction on this judgment we have that $A \setminus \{n\}\ \{S'[\mu t.S'/t]\} A'$.*

*By Lemma C.7 then this gives us* $A \setminus \{n\}\ \{S'[\mu\mathtt{t}.\mathsf{consume}(n).S'/\mathtt{t}]\}A'$ *which we can use to build the well-assertedness derivation:*

$$\frac{A \setminus \{n\}\ \{S'[\mu\mathtt{t}.\mathsf{consume}(n).S'/\mathtt{t}]\}A' \qquad n \in A}{A\ \{\mathsf{consume}(n).S'[\mu\mathtt{t}.\mathsf{consume}(n).S'/\mathtt{t}]\}A'}\textit{[consume]}$$

*which yields our goal by the definition of syntactic substitution.*

- *(assert)* $S = \mathsf{assert}(n).S'$ *with assumption* $A\ \{\mathsf{assert}(n).S'\}A'$.
  *By Lemma C.3 (inversion) we then have* $A \setminus \{n\}\ \{S'\}A'$ *and* $n \in A$.
  *By induction on this judgment we have that* $A \cup \{n\}\ \{S'[\mu\mathtt{t}.S'/\mathtt{t}]\}A'$.
  *By Lemma C.7 then this gives us* $A \cup \{n\}\ \{S'[\mu\mathtt{t}.\mathsf{assert}(n).S'/\mathtt{t}]\}A'$ *which we can use to build the well-assertedness derivation:*

$$\frac{A \setminus \{n\}\ \{S'[\mu\mathtt{t}.\mathsf{assert}(n).S'/\mathtt{t}]\}A'}{A\ \{\mathsf{assert}(n).S'[\mu\mathtt{t}.\mathsf{assert}(n).S'/\mathtt{t}]\}A'}\textit{[assert]}$$

  *which yields our goal by the definition of syntactic substitution.*

- *(rec)* $S = \mu\mathtt{t}'.S'$ *with assumption* $A\ \{\mu\mathtt{t}'.S'\}A'$.
  *By Lemma C.6 (inversion) we then have* $A\ \{S'\}A'$ *and* $A \subseteq A'$.
  *By induction on this judgment we have that* $A\ \{S'[\mu\mathtt{t}.S'/\mathtt{t}]\}A'$.
  *By Lemma C.7 then this gives us* $A\ \{S'[(\mu\mathtt{t}.(\mu\mathtt{t}'.S'))/\mathtt{t}]\}A'$ *which we can use to build the well-assertedness derivation:*

$$\frac{A\ \{S'[\mu\mathtt{t}.\mu\mathtt{t}'.S'/\mathtt{t}]\}A'}{A\ \{\mu\mathtt{t}'.S'[\mu\mathtt{t}.\mu\mathtt{t}'.S'/\mathtt{t}]\}A'}\textit{[rec]}$$

  *(leveraging* $A \subseteq A'$*) which yields our goal by the definition of syntactic substitution.*

- *(end)* $\dfrac{-}{A\ \{\mathsf{end}\}A}$ *Trivial since* $\mathsf{end}[\mu\mathtt{t}.\mathsf{end}/\mathtt{t}] = \mathsf{end}$.

- *(call)* $S = \mathtt{t}'$ *with assumption* $A\ \{\mathtt{t}'\}A'$ *thus* $A \equiv A'$.
  *Case*
  - $\mathtt{t} = \mathtt{t}'$ *thus* $\mathtt{t}'[\mu\mathtt{t}.\mathtt{t}'/\mathtt{t}] = \mu\mathtt{t}.\mathtt{t}$. *Then we can apply construct well-assertedness by the derivation:*

$$\frac{A\ \{\mathtt{t}\}A}{A\ \{\mu\mathtt{t}.\mathtt{t}\}A}\textit{[rec]}$$

  - $\mathtt{t} \neq \mathtt{t}'$ *then* $\mathtt{t}'[\mu\mathtt{t}.\mathtt{t}'/\mathtt{t}] = \mathtt{t}'$ *therefore using the assumption:* $A\ \{\mathtt{t}[\mu\mathtt{t}.\mathtt{t}'/\mathtt{t}]\}A'$.

## D Proof of Lemmas 1, 2, 3 on Well-Assertedness and Progress

**Lemma 1 (Reduction preserves well-assertedness)** *If* $A\ \{S\}A'$ *and there is a reduction* $(A, S) \xrightarrow{\ell} (A'', S')$ *then* $\exists A''' \supseteq A'.A''\ \{S'\}A'''$.

**Proof 10** *By induction on the structure of* $A\ \{S\} A'$ .

- *(act)*

$$\frac{A\ \{S\} A'}{A\ \{p.S\} A'}$$

*Then the only possible reduction is* $\langle\mathtt{Inter}\rangle$*:*

$$(A, p.S) \xrightarrow{p} (A, S)$$

*Therefore we can conclude with the premise of* $A\ \{S\} A'$ *which shows that* $S$ *is well-asserted (and trivially* $A \supseteq A$*).*

- *(bra)*

$$\frac{\forall i \in I.\ A\ \{S_i\} A_i}{A\ \{+\{\mathrm{l}_i : S_i\}_{i\in I}\} \bigcap_{i\in I} A_i}$$

*Then the only possible reduction is* $\langle\mathtt{Branch}\rangle$*:*

$$(A, +\{\mathrm{l_i} : S_i\}_{i\in I}) \xrightarrow{+\mathrm{l_j}} (A, S_j)\ \ (j \in I)$$

*Therefore we can conclude with the premise* $A\ \{S_j\} A_j$ *which shows that* $S_j$ *is well-asserted (and* $A_j \supseteq \bigcap_{i\in I} A_i$ *since* $j \in I$*).*

- *(require)*

$$\frac{A \cup \{n\}\ \{S\} A'}{A \cup \{n\}\ \{\mathsf{require}(n).S\} A'}$$

*Then the only possible reduction is* $\langle\mathtt{Require}\rangle$ *with:*

$$(A \cup \{n\}, \mathsf{require}(n).S) \xrightarrow{\mathsf{require}(n)} (A \cup \{n\}, S)\ \ (n \in (A \cup \{n\}))$$

*(note the trivial satisfaction of the side condition here). Therefore we can conclude with the premise* $A \cup \{n\}\ \{S\} A'$ *which shows that* $S$ *is well-asserted (and trivially* $A' \supseteq A'$*).*

- *(consume)*

$$\frac{A\ \{S\} A'}{A \cup \{n\}\ \{\mathsf{consume}(n).S\} A'}$$

*Then the only possible reduction is* $\langle\mathtt{Consume}\rangle$*:*

$$(A \cup \{n\}, \mathsf{consume}(n).S) \xrightarrow{\mathsf{consume}(n)} ((A \cup \{n\}) \setminus \{n\}, S)\ \ (n \in (A \cup \{n\}))$$

*Thus since* $(A \cup \{n\}) \setminus \{n\} = A$ *we can conclude with the premise* $A\ \{S\} A'$ *showing that* $S$ *is well-asserted (and trivially* $A' \supseteq A'$*).*

- *(assert)*

$$\frac{A \cup \{n\}\ \{S\} A'}{A\ \{\mathsf{assert}(n).S\} A'}$$

*Then the only possible reduction is* $\langle\mathtt{Assert}\rangle$*:*

$$(A, \mathsf{assert}(n).S) \xrightarrow{\mathsf{assert}(n)} (A \cup \{n\}, S)$$

*Thus we can conclude with the premise* $A \cup \{n\}\ \{S\} A'$ *showing that* $S$ *is well-asserted (and trivially* $A' \supseteq A'$*).*

- *(rec)*

$$\frac{A\ \{S\} A \cup A'}{A\ \{\mu t.S\} A \cup A'}$$

*Then the only possible reduction is* $\langle \texttt{Rec} \rangle$*:*

$$\frac{(A, S) \xrightarrow{\ell} (A'', S')}{(A, \mu t.S) \xrightarrow{\ell} (A'', S'[\mu t.S/t])} \qquad (*)$$

*We now proceed by an inner induction on the structure of* $S$ *to prove that* $\exists A''' \supseteq A \cup A'.A''\ \{S'[\mu t.S/t]\} A'''$.

- *(prefix)* $S = p.S_1$ *thus* $A\ \{p.S_1\} A \cup A'$, *and thus* $(*)$ *must be the reduction:*

$$\frac{\dfrac{}{(A, p.S_1) \xrightarrow{p} (A, S_1)}\langle \texttt{Inter} \rangle}{(A, \mu t.p.S_1) \xrightarrow{p} (A, S_1[\mu t.p.S_1/t])}\langle \texttt{Rec} \rangle$$

*thus* $A'' = A$.
*By lemma C.9 on* $A\ \{p.S_1\} A \cup A'$ *then* $A\ \{p.S_1[\mu t.p.S_1/t]\} A \cup A'$ *(**)*
*Then by the definition of substitution and lemma C.1 (inversion of prefix well-assertedness) on (**) we get:* $A\ \{S_1[\mu t.p.S_1/t]\} A \cup A'$ *providing the goal with* $A''' = A \cup A'$.

- *(branch)* $S = +\{\mathrm{l_i} : S_i\}_{i \in I}$ *thus* $A\ \{+\{\mathrm{l_i} : S_i\}_{i \in I}\} A \cup A'$, *and thus* $(*)$ *must be the reduction:*

$$\frac{\dfrac{}{(A, +\{\mathrm{l_i} : S_i\}_{i \in I}) \xrightarrow{+\mathrm{l_j}} (A, S_j) \qquad (j \in I)}\langle \texttt{Branch} \rangle}{(A, \mu t.p. + \{\mathrm{l_i} : S_i\}_{i \in I}) \xrightarrow{p} (A, S_j[\mu t. + \{\mathrm{l_i} : S_i\}_{i \in I}/t])}\langle \texttt{Rec} \rangle$$

*thus* $A'' = A$.
*By Lemma C.9 on* $A\ \{+\{\mathrm{l_i} : S_i\}_{i \in I}\} A \cup A'$ *and unfolding the definition of syntactic substitution then* $A\ \{+\{\mathrm{l_i} : S_i[\mu t. + \{\mathrm{l_i} : S_i\}_{i \in I}/t]\}_{i \in I}\} A \cup A'$ *(**).*
*Then by the definition of substitution and Lemma C.2 (inversion of branch well-formendess) on (**) we get:* $\exists \{A_i\}_{i \in I}.A \cup A' \equiv \bigcap_{i \in I} A_i\ \wedge\ \forall i \in I.A\ \{S_i[\mu t. + \{\mathrm{l_i} : S_i\}_{i \in I}/t]\} A_i$.

*Then taking* $i = j$ *we get* $A\ \{S_j[\mu t. + \{\mathrm{l_i} : S_i\}_{i \in I}/t]\} A_j$ *providing the goal of this lemma with* $A''' = A_j$ *and* $A_j \supseteq A \cup A' = \bigcap_{i \in I} A_i$ *since* $j \in I$.

- *(assert)* $S = \mathsf{assert}(n).S_1$ *thus* $A\ \{\mathsf{assert}(n).S_1\} A \cup A'$, *and thus* $(*)$ *must be the reduction:*

$$\frac{\dfrac{}{(A, \mathsf{assert}(n).S_1) \xrightarrow{\mathsf{assert}(n)} (A \cup \{n\}, S_1)}\langle \texttt{Assert} \rangle}{(A, \mu t.\mathsf{assert}(n).S_1) \xrightarrow{\mathsf{assert}(n)} (A \cup \{n\}, S_1[\mu t.\mathsf{assert}(n).S_1/t])}\langle \texttt{Rec} \rangle$$

*thus* $A'' = A \cup \{n\}$.

*By lemma C.9 on* $A\ \{\mathsf{assert}(n).S_1\}A \cup A'$ *then (unfolding substitution)* $A\ \{\mathsf{assert}(n).S_1[\mu\mathtt{t}.\mathsf{assert}(n).S_1/\mathtt{t}]\}A \cup A'$ *(**)*
*Then by the definition of substitution and lemma C.3 (inversion of assert well-assertedness) on (**) we get:* $A \cup \{n\}\ \{S_1[\mu\mathtt{t}.\mathsf{assert}(n).S_1/\mathtt{t}]\}A \cup A'$ *providing the goal with* $A''' = A \cup A'$.

- *(require)* $S = \mathsf{require}(n).S_1$ *thus* $A\ \{\mathsf{require}(n).S_1\}A \cup A'$, *and thus* $(*)$ *must be the reduction:*

$$\dfrac{\dfrac{}{(A, \mathsf{require}(n).S_1) \xrightarrow{\mathsf{require}(n)} (A, S_1)}\langle\mathtt{Require}\rangle\ (n \in A)}{(A, \mu\mathtt{t}.\mathsf{require}(n).S_1) \xrightarrow{\mathsf{require}(n)} (A, S_1[\mu\mathtt{t}.\mathsf{require}(n).S_1/\mathtt{t}])}\langle\mathtt{Rec}\rangle$$

*and thus* $A'' = A$.
*By lemma C.9 on* $A\ \{\mathsf{require}(n).S_1\}A \cup A'$ *then (unfolding substitution)* $A\ \{\mathsf{require}(n).S_1[\mu\mathtt{t}.\mathsf{require}(n).S_1/\mathtt{t}]\}A \cup A'$ *(**)*
*Then by the definition of substitution and lemma C.4 (inversion of require well-assertedness) on (**) with* $n \in A$ *we get:*
$A\ \{S_1[\mu\mathtt{t}.\mathsf{require}(n).S_1/\mathtt{t}]\}A \cup A'$ *matching the goal with* $A''' = A \cup A'$.

- *(consume)* $S = \mathsf{consume}(n).S_1$ *thus* $A\ \{\mathsf{consume}(n).S_1\}A \cup A'$, *and thus* $(*)$ *must be the reduction:*

$$\dfrac{\dfrac{}{(A, \mathsf{consume}(n).S_1) \xrightarrow{\mathsf{consume}(n)} (A \setminus \{n\}, S_1)}\langle\mathtt{Consume}\rangle\ (n \in A)}{(A, \mu\mathtt{t}.\mathsf{consume}(n).S_1) \xrightarrow{\mathsf{consume}(n)} (A \setminus \{n\}, S_1[\mu\mathtt{t}.\mathsf{consume}(n).S_1/\mathtt{t}])}\langle\mathtt{Rec}\rangle$$

*and thus* $A'' = A \setminus \{n\}$.
*By lemma C.9 on* $A\ \{\mathsf{consume}(n).S_1\}A \cup A'$ *then (unfolding substitution)* $A\ \{\mathsf{consume}(n).S_1[\mu\mathtt{t}.\mathsf{consume}(n).S_1/\mathtt{t}]\}A \cup A'$ *(**)*
*Then by the definition of substitution and lemma C.5 (inversion of consume well-assertedness) on (**) with* $n \in A$ *we get:*
$A \setminus \{n\}\ \{S_1[\mu\mathtt{t}.\mathsf{consume}(n).S_1/\mathtt{t}]\}A \cup A'$ *matching the goal with* $A''' = A \cup A'$.

- *(rec)* $S = \mu\mathtt{t}_1.S_1$ *thus* $A\ \{\mu\mathtt{t}_1.S_1\}A \cup A'$, *and thus* $(*)$ *must be the reduction:*

$$\dfrac{\dfrac{(A, S_1) \xrightarrow{\ell} (A_1, S_1')}{(A, \mu\mathtt{t}_1.S_1) \xrightarrow{\ell} (A_1, S_1'[\mu\mathtt{t}_1.S_1/\mathtt{t}_1])}}{(A, \mu\mathtt{t}.\mu\mathtt{t}_1.S_1) \xrightarrow{\mathsf{consume}(n)} (A_1, S_1'[\mu\mathtt{t}_1.S_1/\mathtt{t}_1][\mu\mathtt{t}.\mu\mathtt{t}_1.S_1/\mathtt{t}])}\langle\mathtt{Rec}\rangle$$

*and thus* $A'' = A_1$.
*By lemma C.9 on* $A\ \{\mu\mathtt{t}_1.S_1\}A \cup A'$ *then (unfolding substitution and since* $\mathtt{t} \neq \mathtt{t}_1$*)* $A\ \{\mu\mathtt{t}_1.S_1[\mu\mathtt{t}.\mu\mathtt{t}_1.S_1/\mathtt{t}]\}A \cup A'$ *(**)*
*Then by the definition of substitution and lemma C.6 (inversion of consume well-assertedness) on (**) with* $A \subseteq A \cup A'$ *by usual set theory laws, then we get:*
$A\ \{S_1[\mu\mathtt{t}.\mu\mathtt{t}_1.S_1/\mathtt{t}]\}A \cup A'$ *matching the goal with* $A''' = A \cup A'$.

*Thus by this sublemma we conclude that* $\exists A''' \supseteq A \cup A'.A''\ \{S'[\mu\mathtt{t}.S/\mathtt{t}]\}A'''$.

- *(call)*

$$\frac{-}{A\ \{\mathtt{t}\}\,A}$$

*A recursive variable on its own cannot reduce thus this case holds trivially since the premise is false.*

- *(end)*

$$\frac{-}{A\ \{\mathtt{end}\}\,A}$$

*The terminated process* end *cannot reduce thus this case holds trivially since the premise is false.*

**Lemma 2 (Well-asserted protocols are not stuck)** *If $A\ \{S\}$ and $S$ is closed with respect to recursion variables ($\mathsf{fv}(S) = \emptyset$) then $(A, S)$ is not stuck.*

**Proof 11** *We proceed by structural induction on the derivation of well-assertedness $A\ \{S\}\,A'$ (and thus simultaneously on the structure of $S$ since every syntactic construction has one well-assertedness rule):*

- *$S =$ end there are no further reductions possible and the thesis trivially holds: we conclude with the first conjunct of the definition 3.*
- *$S = \mathtt{t}$ then progress is trivial since the premise is that the $S$ is closed.*
- *$S = p.S'$ with:*

$$\frac{A\ \{S'\}\,A'}{A\ \{p.S'\}\,A'} \quad [\text{act}]$$

*Thus $S$ can reduce by* ⟨Inter⟩ *as $(A, p'.S') \xrightarrow{p} (A, S')$*

- *$S = +\{\mathsf{l}_i : S_i\}_{i\in I}$ with:*

$$\frac{\forall i \in I.\ A\ \{S_i\}\,A_i}{A\ \{+\{\mathsf{l}_i : S_i\}_{i\in I}\}\ \bigcap_{i\in I} A_i)} \quad (2)$$

*where $A' = \bigcap_{i\in I} A_i$. Thus, $S$ can reduce by* ⟨Branch⟩ *to $(A, S_j)$ for any $j \in I$.*

- *$S = \mathsf{assert}(n).S'$ with*

$$\frac{A \cup \{n\}\ \{S'\}\,A'}{A\ \{\mathsf{assert}(n).S'\}\,A'}$$

*Thus $S$ can reduce by* ⟨Assert⟩ *to $(A \cup \{n\}, S')$*

- *$S = \mathsf{consume}(n).S'$ with*

$$\frac{A \setminus \{n\}\ \{S'\}\,A' \qquad n \in A}{A\ \{\mathsf{consume}(n).S'\}\,A'}$$

*Thus $S$ can reduce by* ⟨Consume⟩ *to $(A \setminus \{n\}, S')$ since well-assertedness gives us that $n \in A$. to give us the side condition of this reduction rule.*

- $S = \mathsf{require}(n).S'$ *with*

$$\frac{A \cup \{n\}\ \{S'\} A'}{A \cup \{n\}\ \{\mathsf{require}(n).S'\} A'}$$

*Thus* $S$ *can reduce by* $\langle\texttt{Require}\rangle$ *to* $(A \cup \{n\}, S')$ *since well-assertedness gives us that* $n \in (A \cup \{n\})$ *to give us the side condition of this reduction rule.*

- $S = \mu\mathtt{t}.S'$ *with:*

$$\frac{A\ \{S'\} A'}{A\ \{\mu\mathtt{t}.S'\} A'} \tag{3}$$

*By induction on the first premise we get that* $S' = \mathsf{end}$ *or* $(A, S') \rightarrow (A''', S'')$.

  - *In the case of* $S' = \mathsf{end}$ *then we have* $S = \mu\mathtt{t}.\mathsf{end}$ *which by structural congruence (Section 2.2) then means* $S = \mathsf{end}$.
  - *In the case of* $(A, S') \rightarrow (A''', S'')$ *this provides the premise of the* $\langle\texttt{Rec}\rangle$ *rule such that* $S$ *can reduce to* $(A''', S''[\mu\mathtt{t}.S'/\mathtt{t}])$.

Below we write:

$$(A, S) \rightarrow^n (A', S') \quad \text{if} \quad \begin{cases} n = 1 & \exists \ell.(A, S) \xrightarrow{\ell} (A', S') \\ n > 1 & \exists \ell.(A, S) \xrightarrow{\ell} (A'', S'') \wedge (A'', S'') \rightarrow^{n-1} (A', S') \end{cases}$$

**Lemma 3 (Progress of very-well-asserted protocols)** *If* $S$ *is very-well-asserted (i.e.,* $\emptyset\ \{S\}$*) and closed then it exhibits* progress.

**Proof 12** *We proceed by induction on the length of the reduction sequence* $n$ *and prove a stronger lemma:*

*If* $S$ *is* closed $(\mathsf{fv}(S) = \emptyset)$ *and* very-well-asserted $(\exists A'.\emptyset\ \{S\} A')$ *then* $S$ *has* progress*, i.e.,* $\forall A, n, S'$ *if* $(\emptyset, S) \rightarrow^n (A, S')$ *then* $(S' = \mathsf{end} \ \vee\ (\exists A'', S''.(A, S') \rightarrow (A'', S'')$ ***and*** $\exists A'''.A''\ \{S''\} A''')$.

- $n = 0$.
*Thus,* $A = \emptyset$ *and* $S' = S$.
*By lemma 2, with* $\emptyset\ \{S\} A'$ *then we get that* $S = \mathsf{end} \vee \exists A'', S''.(A, S) \rightarrow (A'', S'')$.
*In the latter case (of a reduction), we then apply lemma 1 to get that* $\exists A'''.A''\ \{S''\} A'''$.
- $n = k + 1$
*Then we have the assumption that* $(\emptyset, S) \rightarrow^{k+1} (A_{k+1}, S'_{k+1})$ *and thus there exists* $(\emptyset, S) \rightarrow^k (A_k, S'_k) \rightarrow (A_{k+1}, S'_{k+1})$.
*We can apply the lemma inductively on* $(\emptyset, S) \rightarrow^k (A_k, S'_k)$ *(i.e., the* $n = k$ *case) to get that* $S'_k = \mathsf{end}$ *(not possible because of the* $k+1$ *reduction here) or* $\exists A'', S'.(A_k, S'_k) \rightarrow (A'_{k+1}, S''_{k+1})$ *and that* $\exists A_1.A'_{k+1}\ \{S''_{k+1}\} A_1$.
*Since reduction is deterministic we have that:* $(A_{k+1}, S'_{k+1}) = (A'_{k+1}, S''_{k+1})$.
*The goal here is to show that either* $S'_{k+1} = \mathsf{end}$ *or that* $\exists A_{k+2}, S'_{k+2}.(A_{k+1}, S'_{k+1}) \rightarrow (A_{k+2}, S'_{k+2})$ *where* $\exists A_2.A_{k+2}\ \{S'_{k+2}\} A_2$.

*By lemma 2 (local progress) with* $A'_{k+1}\ \{S'_{k+1}\} A_1$ *then we have that* $(S'_{k+1} = \mathtt{end}) \vee \exists A_{k+2}, S'_{k+2}.(A_{k+1}, S'_{k+1}) \rightarrow (A_{k+2}, S'_{k+2})$*. In the case of the left disjunct we are done. In the case of the right disjunct, we then have the remaining piece of evidence via lemma 1 (reduction preserves well-assertedness) with the* $A'_{k+1}\ \{S'_{k+1}\} A_1$ *and* $(A_{k+1}, S'_{k+1}) \rightarrow (A_{k+2}, S'_{k+2})$ *which gives us that* $\exists A_2. A_{k+2}\ \{S'_{k+2}\} A_2$.
*Thus we are done.*

## E Proof of Proposition 3 on Validity of Composition

**Proposition 3 (Validity)** *If* $T_L; T_R; \emptyset \vdash S_1 \circ S_2 \triangleright S$ *then* $S$ *is very-well-asserted.*

**Proof 13** *Proposition 3 follows immediately from Lemma 5, given in this section, setting* $A = \emptyset$*. The proof of Lemma 5 relies on an auxiliary lemma, Lemma 4, given below. Lemma 4 makes use of environment weakening (Proposition Proposition 2) given in earlier sections.*

**Lemma 4** *If* $A_0\ \{S\} A$ *and* $A\ \{S'\}$ *then* $A_0\ \{S[S'/\mathtt{end}]\}$ .

**Proof 14** *The proof is by induction on the size of* $S$*, proceeding by case analysis on the syntax of* $S$.

**Base cases** *There are two base cases:* $S = \mathtt{end}$ *and* $S = \mathtt{t}$*. If* $S = \mathtt{end}$*, by [end]* $A_0 = A$*. The thesis follows then immediately from the hypothesis* $A\ \{S'\}$ *since* $\mathtt{end}[S'/\mathtt{end}] = S'$
*If* $S = \mathtt{t}$ *the thesis follows immediately by hypothesis* $A_0\ \{S\} A$ *since* $S[S'/\mathtt{end}] = S$.

**Inductive cases** *The inductive cases are analyzed below:*

- *Case* $S = p.S''$*. By [act] on hypothesis* $A_0\ \{p.S''\} A$ *follows (as premise)*

  $$A_0\ \{S''\} A \tag{4}$$

  *By induction, Equation (4) and hypothesis* $A\ \{S'\}$ *follows*

  $$A_0\ \{S''[S'/\mathtt{end}]\} \tag{5}$$

  *By applying rule [act] to Equation (5) obtain* $A_0\ \{p.S''[S'/\mathtt{end}]\}$ *as required.*
- *Case* $S = \mathsf{require}(n).S''$*. By [assume] on hypothesis* $A_0\ \{\mathsf{require}(n).S''\} A$ *follows (as premise)*

  $$A_0\ \{S''\} A \tag{6}$$

  *with* $n \in A_0$*. By induction, Equation (6) and hypothesis* $A\ \{S'\}$ *follows*

  $$A_0\ \{S''[S'/\mathtt{end}]\} \tag{7}$$

  *By applying rule [assume] to Equation (7) – observe that* $n \in A_0$ *– obtain*

  $$A_0\ \{\mathsf{require}(n).S''[S'/\mathtt{end}]\}$$

  *as required.*

- *Case* $S = \mathsf{consume}(n).S''$. *By [consume] on hypothesis* $A_0\ \{\mathsf{consume}(n).S''\} A$ *follows (as premise)* $n \in A_0$ *and*

$$A_0 \setminus \{n\}\ \{S''\} A \tag{8}$$

*By induction, Equation* (8) *and hypothesis* $A\ \{S'\}$ *follows*

$$A_0 \setminus \{n\}\ \{S''[S'/\mathtt{end}]\} \tag{9}$$

*By applying rule [consume] to Equation* (9) *obtain* $A_0\ \{\mathsf{consume}(n).S''[S'/\mathtt{end}]\}$ *as required.*

- *Case* $S = \mathsf{assert}(n).S''$. *By [assert] on hypothesis* $A_0\ \{\mathsf{assert}(n).S''\} A$ *follows (as premise)*

$$A_0 \cup \{n\}\ \{S''\} A \tag{10}$$

*By induction, Equation* (10) *and hypothesis* $A\ \{S'\}$ *follows*

$$A_0 \cup \{n\}\ \{S''[S'/\mathtt{end}]\} \tag{11}$$

*By applying rule [assert] to Equation* (11) *obtain* $A_0\ \{\mathsf{assert}(n).S''[S'/\mathtt{end}]\}$ *as required.*

- *Case* $S = \mu\mathtt{t}.S''$. *By [rec] on hypothesis* $A_0\ \{\mu\mathtt{t}.S''\} A$ *follows (as premise)*

$$A_0\ \{S''\} A \tag{12}$$

*By induction, Equation* (12) *and hypothesis* $A\ \{S'\}$ *follows*

$$A_0\ \{S''[S'/\mathtt{end}]\} \tag{13}$$

*By applying rule [rec] to Equation* (13) *obtain* $A_0\ \{\mu.S''[S'/\mathtt{end}]\}$ *as required.*

- *Case* $S = +\{\mathrm{l}_i : S_i\}_{i\in I}$. *By [bra] on hypothesis* $A_0\ \{+\{\mathrm{l}_i : S_i\}_{i\in I}\} A$ *follows (as premise)*

$$\forall i \in I.\ A_0\ \{S_i\} A_i \tag{14}$$

*for some* $\{A_i\}_{i\in I}$. *Since by [bra] applied in Equation* (14) $\bigcap_{i\in I} A_i = A$ *then*

$$\forall i \in I.\ A \subseteq A_i \tag{15}$$

*By Proposition 1 (environment weakening), Equation* (14), *and Equation* (15), *follows*

$$A_i\ \{S_i\}\ \textit{for all } i \in I. \tag{16}$$

*By induction, Equation* (14) *and Equation* (16) *give*

$$\forall i \in I.\ A_0\ \{S_i[S'/\mathtt{end}]\} \tag{17}$$

*By applying Equation* (17) *as a premise of [bra] obtain* $A_0\ \{+\{\mathrm{l}_i : S_i\}_{i\in I}[S'/\mathtt{end}]\}$ *as required.*

**Lemma 5** *If* $T_L; T_R; A \vdash S_1 \circ S_2 \triangleright S$ *then* $A\{S\}$.

**Proof 15** *The proof is by induction on the derivation, proceeding by case analysis on the last rule (Definition 5) applied.*

**Base cases.** *There are two base cases: the last application was rule [end] or [call]. If the last (and only) rule applied was [end] in Definition 5 then* $S = \texttt{end}$ *and by rule [end] in Definition 4* $A\ \{\texttt{end}\}$ *as required. If the last (and only) rule applied was [call] in Definition 5 then* $S = \texttt{t}$ *and by rule [call] in Definition 4* $A\ \{\texttt{t}\}$ *as required.*

**Inductive cases.** *We show below the inductive cases.*

- *Case last rule is [act]. The conclusion is on the form* $T_L; T_R; A \vdash p.S_1' \circ S_2 \triangleright p.S'$ *with premise*

$$T_L; T_R; A \vdash S_1' \circ S_2 \triangleright S' \tag{18}$$

  *By induction, from Equation (18) it follows:*

$$A\ \{S'\} \tag{19}$$

  *By applying rule [act] in Definition 4 to Equation (19) it follows* $A\ \{p.S'\}$ *as required.*
- *Case last rule is [sym]. The conclusion is on the form* $T_L; T_R; A \vdash S_1 \circ S_2 \triangleright S$ *with premise*

$$T_L; T_R; A \vdash S_2 \circ S_1 \triangleright S \tag{20}$$

  *By induction, from Equation (20) it follows that* $A\{S\}$ *as required.*
- *Case last rule is [require]. The conclusion is on the form*

$$T_L; T_R; \{n\} \cup A' \vdash \mathsf{require}(n).S_1' \circ S_2 \triangleright \mathsf{require}(n).S'$$

  *with premise*

$$T_L; T_R; \{n\} \cup A' \vdash S_1' \circ S_2 \triangleright S' \tag{21}$$

  *By induction, from Equation (21) follows* $\{n\} \cup A'\ \{S'\}$ *. By using* $\{n\} \cup A'\ \{S'\}$ *as a premise for rule [require] in Definition 4 we obtain* $\{n\} \cup A'\ \{\mathsf{require}(n).S'\}$ *as required.*
- *Case last rule is [consume]. The conclusion is on the form*

$$T_L; T_R; \{n\} \cup A' \vdash \mathsf{consume}(n).S_1' \circ S_2 \triangleright \mathsf{consume}(n).S'$$

  *with premise*

$$T_L; T_R; A' \vdash S_1' \circ S_2 \triangleright S' \tag{22}$$

  *By induction, from Equation (22) it follows*

$$A'\ \{S'\} \tag{23}$$

  *By applying Equation (23) as a premise for rule [consume] in Definition 4 obtain* $\{n\} \cup A'\ \{\mathsf{consume}(n).S'\}$ *as required.*

- *Case last rule is [assert]. The conclusion is on the form*

  $$T_L; T_R; A \vdash \mathsf{assert}(n).S_1' \circ S_2 \triangleright \mathsf{assert}(n).S'$$

  *with premise*

  $$T_L; T_R; A \cup \{n\} \vdash S_1' \circ S_2 \triangleright S' \tag{24}$$

  *By induction, from Equation* (24) *it follows*

  $$A \cup \{n\} \; \{S'\} \tag{25}$$

  *By applying rule [act] in Definition 4 to Equation* (19) *it follows* $A \; \{\mathsf{assert}(n).S'\}$ *as required.*
- *Case last rule is [bra] (without weakening). The conclusion is on the form*

  $$T_L; T_R; A \vdash +\{\mathrm{l}_i : S_i\}_{i\in I} \circ S_2 \triangleright +\{\mathrm{l}_i : S_i'\}_{i\in I}$$

  *with premise*

  $$\forall i \in I.\ T_L; T_R; A \vdash S_i \circ S_2 \triangleright S_i' \tag{26}$$

  *By induction, from Equation* (26) *it follows:*

  $$\forall i \in I.\ A \; \{S_i'\} \tag{27}$$

  *By applying Equation* (27) *as premise of [bra] in Definition 4 it follows* $A \; \{+\{\mathrm{l}_i : S_i'\}_{i\in I}\}$ *as required.*
- *Case last rule is [bra] (with weakening). The conclusion is on the form*

  $$T_L; T_R; A \vdash +\{\mathrm{l}_i : S_i\}_{i\in I} \circ S_2 \triangleright +\{\mathrm{l}_i : S_i'\}_{i\in I_A} \cup +\{\mathrm{l}_i : S_i\}_{i\in I_B}$$

  *with premises* $I_A \cup I_B = I$, $I_A \cap I_B = \emptyset$ *and* $I_A = \emptyset$, *and*

  $$\forall i \in I_A.\ T_L; T_R; A \vdash S_i \circ S_2 \triangleright S_i' \tag{28}$$

  $$\forall i \in I_B.\ A \; \{S_i\} \tag{29}$$

  *By induction, from Equation* (28) *it follows:*

  $$\forall i \in I_A.\ A \; \{S_i'\} \tag{30}$$

  *By applying Equation* (29) *and Equation* (30) *as premise of [bra] in Definition 4 it follows* $A \; \{+\{\mathrm{l}_i : S_i'\}_{i\in I_A} \cup +\{\mathrm{l}_i : S_i\}_{i\in I_B}\}$ *as required.*
- *Case last rule is [rec1]. The conclusion is of the form*

  $$T_L; T_R; A \vdash \mu\mathtt{t}_1.S_1' \circ \mu\mathtt{t}_2.S_2' \triangleright \mu\mathtt{t}_1.S$$

  *with premise*

  $$T_L, \mathtt{t}_1; T_R; A \vdash S_1' \circ \mu\mathtt{t}_2.S_2' \triangleright S \quad A \; \{\mu\mathtt{t}_1.S\}$$

  *The thesis follows by condition* $A \; \{\mu\mathtt{t}_1.S\}$ *in the premise above.*

- *Case last rule is [rec2]. The conclusion is of the form*

  $$T_L; T_1, \mathtt{t}, T_2; A \vdash \mu\mathtt{t}_1.S_1' \circ S_2' \triangleright S$$

  *with premise*

  $$T_L; T_1, \underline{\mathtt{t}}, T_2; A \vdash S_1'[\mathtt{t}/\mathtt{t}_1] \circ S_2' \triangleright S \quad \mathtt{unused}(T_2)$$

  *By induction* $A\ \{S_1\}$ *which is the thesis.*
- *Case last rule is [rec3]. Immediate by hypothesis.*

## F Proof of Algebraic and Scoping Properties

**Definition 11 (Substituting for** end**)** *Given two protocols* $S$ *and* $S'$ *then* $S[S'/\mathtt{end}]$ *is defined:*

$$
\begin{aligned}
(p.S)[S'/\mathtt{end}] &= p.S[S'/\mathtt{end}] \\
(+\{\mathsf{l_i} : S_i\}_{i\in I})[S'/\mathtt{end}] &= +\{\mathsf{l_i} : S_i[S'/\mathtt{end}]\}_{i\in I}) \\
(\mathsf{assert}(n).S)[S'/\mathtt{end}] &= \mathsf{assert}(n).S[S'/\mathtt{end}] \\
(\mathsf{consume}(n).S)[S'/\mathtt{end}] &= \mathsf{consume}(n).S[S'/\mathtt{end}] \\
(\mathsf{require}(n).S)[S'/\mathtt{end}] &= \mathsf{require}(n).S[S'/\mathtt{end}] \\
(\mu\mathtt{t}.S)[S'/\mathtt{end}] &= (\mu\mathtt{t}.S[S'/\mathtt{end}]) \\
\mathtt{t}[S'/\mathtt{end}] &= \mathtt{t} \\
\mathtt{end}[S'/\mathtt{end}] &= S'
\end{aligned}
$$

**Lemma F.1** *Given* $T_L; T_R; A \vdash S_1 \circ S_2 \triangleright S$ *where* $T_L \cap T_R = \emptyset$ *then the following hold:*

1. $\mathtt{t} \in T_L \implies \mathtt{t} \in fn(S_1) \vee (\mathtt{t} \notin fn(S_1) \wedge \mathtt{t} \notin fn(S))$
2. $\mathtt{t} \in T_R \implies \mathtt{t} \in fn(S_2) \vee (\mathtt{t} \notin fn(S_2) \wedge \mathtt{t} \notin fn(S))$

**Proof 16**

- *(act)*

$$\frac{T_L; T_R; A \vdash S_1' \circ S_2 \triangleright S}{T_L; T_R; A \vdash p.S_1' \circ S_2 \triangleright p.S}$$

*By induction and where* $\mathtt{t} \in fn(S_1')$ *implies* $\mathtt{t} \in fn(p.S_1')$ *for part (1) and* $(\mathtt{t} \notin fn(S_1') \wedge \mathtt{t} \notin fn(S))$ *implies* $(\mathtt{t} \notin fn(p.S_1') \wedge \mathtt{t} \notin fn(p.S))$.
*and part (2) follows directly.*

- *(sym)*

$$\frac{T_R; T_L; A \vdash S_2 \circ S_1 \triangleright S}{T_L; T_R; A \vdash S_1 \circ S_2 \triangleright S}$$

*By induction and swapping parts (1) and (2) in the induction to get the thesis.*

- *(require)*

$$\frac{T_L; T_R; A \cup \{n\} \vdash S_1 \circ S_2 \triangleright S}{T_L; T_R; A \cup \{n\} \vdash \mathsf{require}(n).S_1 \circ S_2 \triangleright \mathsf{require}(n).S}$$

*Similar to the case for (act) since* $\mathsf{require}(n)$ *does not affect recursion variables.*

- *(consume)*

$$\frac{T_L; T_R; A \setminus \{n\} \vdash S_1 \circ S_2 \triangleright S \qquad n \in A}{T_L; T_R; A \vdash \mathsf{consume}(n).S_1 \circ S_2 \triangleright \mathsf{consume}(n).S}$$

*Similar to the case for (act) since* $\mathsf{consume}(n)$ *does not affect recursion variables.*

- *(assert)*

$$\frac{T_L; T_R; A \cup \{n\} \vdash S_1 \circ S_2 \triangleright S}{T_L; T_R; A \vdash \mathsf{assert}(n).S_1 \circ S_2 \triangleright \mathsf{assert}(n).S}$$

*Similar to the case for (act) since* $\mathsf{assert}(n)$ *does not affect recursion variables.*

- *(bra)*

$$\frac{\forall i \in I \;\; T_L; T_R; A \vdash S_i \circ S_2 \triangleright S_i'}{T_L; T_R; A \vdash +\{\mathsf{l_i} : S_i\}_{i \in I} \circ S_2 \triangleright +\{\mathsf{l_i} : S_i'\}_{i \in I}}$$

*By induction (over each premise* $i \in I$*) where any* $\mathtt{t} \in fn(S_i)$ *implies that* $\mathtt{t} \in fn(+\{\mathsf{l_i} : S_i\}_{i \in I})$ *and otherwise if all* $\mathtt{t} \notin fn(S_i)$ *then* $\mathtt{t} \notin +\{\mathsf{l_i} : S_i'\}_{i \in I}$ *and part (2) follows directly since it is unchanged.*

- *(rec1)*

$$\frac{T_L, \mathtt{t}_1; T_R; A \vdash S_1 \circ \mu\mathtt{t}_2.S_2 \triangleright S \quad A\ \{\mu\mathtt{t}_1.S\}}{T_L; T_R; A \vdash \mu\mathtt{t}_1.S_1 \circ \mu\mathtt{t}_2.S_2 \triangleright \mu\mathtt{t}_1.S}$$

*For part (1), assuming* $\mathtt{t} \in T_L$ *from the implication, then we also have* $\mathtt{t} \in (T_L, \mathtt{t}_1)$ *and since all bound variables are fresh we have that* $(T_L, \mathtt{t}_1) \cap T_R = \emptyset$ *which we apply on the inductive argument to get* $\mathtt{t} \in fn(S_1) \vee (\mathtt{t} \notin fn(S_1) \wedge \mathtt{t} \notin fn(S))$ *which we case split on:*

- $\mathtt{t} \in fn(S_1)$ *thus* $\mathtt{t} \in fn(\mu\mathtt{t}_1.S_1)$ *following the assumption of all bound variables being initial fresh (and not overlapping* $T$*).*
- $(\mathtt{t} \notin fn(S_1) \wedge \mathtt{t} \notin fn(S))$ *which then implies* $(\mathtt{t} \notin fn(\mu\mathtt{t}_1.S_1) \wedge \mathtt{t} \notin fn(\mu\mathtt{t}_1.S))$.

*Part (2) follows directly by induction since that side of* $\circ$ *is unmodified in the premise.*

- *(rec2)*

$$\frac{T_L; T_1, \underline{\mathtt{t}'}, T_2; A \vdash S_1[\mathtt{t}'/\mathtt{t}_1] \circ S_2 \triangleright S \qquad \mathtt{unused}(T_2)}{T_L; T_1, \mathtt{t}', T_2; A \vdash \mu\mathtt{t}_1.S_1 \circ S_2 \triangleright S}$$

*Part (1). Assuming* $\mathtt{t} \in T_L$. *Case split on whether* $\mathtt{t} = \mathtt{t}'$ *or not:*

- $\mathtt{t} = \mathtt{t}'$ *which contradicts the premise that* $T_L \cap T_R = \emptyset$.
- $\mathtt{t} \neq \mathtt{t}'$ *then by induction on the premise we have that* $\mathtt{t} \in fn(S_1[\mathtt{t}'/\mathtt{t}_1]) \vee (\mathtt{t} \notin fn(S_1[\mathtt{t}'/\mathtt{t}_1]) \wedge \mathtt{t} \notin fn(S))$ *on which we case split:*
  * $\mathtt{t} \in fn(S_1[\mathtt{t}'/\mathtt{t}_1])$ *therefore* $\mathtt{t} \in fn(\mu\mathtt{t}_1.S_1)$ *since no variable capture is possible.*
  * $(\mathtt{t} \notin fn(S_1[\mathtt{t}'/\mathtt{t}_1]) \wedge \mathtt{t} \notin fn(S))$ *therefore* $(\mathtt{t} \notin fn(\mu\mathtt{t}_1.S_1) \wedge \mathtt{t} \notin fn(S))$.

*Part (2) follows directly by induction since* $S_2$ *remains an unchanged and there is only a chance of annotation in* $T_R$ *which does not affect the meaning.*

- *(rec3)*

$$\frac{A\ \{\mu\mathtt{t}'.S\} \quad \mathsf{fv}(\mu\mathtt{t}'.S) = \emptyset}{T_L; T_R; A \vdash \mu\mathtt{t}'.S \circ \mathtt{end} \triangleright \mu\mathtt{t}'.S}$$

*Part (1), for all* $\mathtt{t} \in T_L$ *the goal is* $\mathtt{t} \in fn(\mu\mathtt{t}'.S) \vee (\mathtt{t} \notin fn(\mu\mathtt{t}'.S) \wedge \mathtt{t} \notin fn(\mu\mathtt{t}'.S))$. *Either* $\mathtt{t} \in \mathsf{fv}(\mu\mathtt{t}'.S)$ *satisfying the goal or* $\mathtt{t} \notin \mathsf{fv}(\mu\mathtt{t}'.S)$ *also satisfying the goal here. Part (2), for all* $\mathtt{t} \in T_R$ *then* $\mathtt{t} \notin \mathsf{fv}(\mathtt{end})$ *and* $\mathtt{t} \notin \mathsf{fv}(\mu\mathtt{t}'.S)$ *by the side condition of the rule.*

- *(call)*

$$\frac{\underline{\mathtt{t}'} \in T_L \vee \underline{\mathtt{t}'} \in T_R}{T_L; T_R; A \vdash \mathtt{t}' \circ \mathtt{t}' \triangleright \mathtt{t}'}$$

*Part (1), case on whether* $\mathtt{t} = \mathtt{t}'$

- $\mathtt{t} = \mathtt{t}'$ *- then trivially* $\mathtt{t} \in fn(\mathtt{t}')$ *as* $\mathtt{t} = \mathtt{t}'$
- $\mathtt{t} \neq \mathtt{t}'$ *- then trivially* $\mathtt{t} \notin fn(\mathtt{t}')$ *(first conjunct) and* $\mathtt{t} \notin \mathsf{fv}(\mathtt{t}')$ *(second conjunct)*

*Part (2) is as above for Part (1) due to the symmetry in this rule.*

- *(end)*

$$\frac{-}{T_L; T_R; A \vdash \mathtt{end} \circ \mathtt{end} \triangleright \mathtt{end}}$$

*For both part (1) and (2) we trivially have that for all* $\mathtt{t} \in T_L$ *(and* $\mathtt{t} \in T_R$*) then* $\mathtt{t} \notin \mathsf{fv}(\mathtt{end})$ *satisfying the goal here.*

**Proposition 4** *If* $T_L; T_R; A \vdash S_1 \circ S_2 \triangleright S$ *then* $\mathsf{fv}(S_1) \cup \mathsf{fv}(S_2) = \mathsf{fv}(S)$.

**Proof 17** *By induction on* $T_L; T_R; R \vdash S_1 \circ S_2 \triangleright S$*:*

- *(act)*

$$\frac{T_L; T_R; A \vdash S_1' \circ S_2 \triangleright S}{T_L; T_R; A \vdash p.S_1' \circ S_2 \triangleright p.S}$$

*By induction and then that* $\mathsf{fv}(p.S) = \mathsf{fv}(S)$.

- *(sym)*

$$\frac{T_L; T_R; A \vdash S_2 \circ S_1 \triangleright S}{T_L; T_R; A \vdash S_1 \circ S_2 \triangleright S}$$

*By induction and commutativity of* $\cup$ *on sets.*

- *(require)*

$$\frac{T_L; T_R; A \cup \{n\} \vdash S_1' \circ S_2 \triangleright S}{T_L; T_R; A \cup \{n\} \vdash \mathsf{require}(n).S_1' \circ S_2 \triangleright \mathsf{require}(n).S}$$

*By induction and then* $\mathsf{fv}(\mathsf{require}(n).S) = \mathsf{fv}(S)$ *(recall free variables are with respect to recursion variables rather than assertion names).*

- *(consume)*

$$\frac{T_L; T_R; A \setminus \{n\} \vdash S_1' \circ S_2 \triangleright S \qquad n \notin A}{T_L; T_R; A \vdash \mathsf{consume}(n).S_1' \circ S_2 \triangleright \mathsf{consume}(n).S}$$

*By induction and then* $\mathsf{fv}(\mathsf{consume}(n).S) = \mathsf{fv}(S)$.

- *(assert)*

$$\frac{T_L; T_R; A \cup \{n\} \vdash S_1' \circ S_2 \triangleright S}{T_L; T_R; A \vdash \mathsf{assert}(n).S_1' \circ S_2 \triangleright \mathsf{assert}(n).S}$$

*By induction and then* $\mathsf{fv}(\mathsf{assert}(n).S) = \mathsf{fv}(S)$.

- *(bra)*

$$\frac{\forall i \in I \;\; T_L; T_R; A \vdash S_i \circ S_2 \triangleright S_i'}{T_L; T_R; A \vdash +\{l_i : S_i\}_{i\in I} \circ S_2 \triangleright +\{l_i : S_i'\}_{i\in I}}$$

*By induction we have that* $\mathsf{fv}(S_i) \cup \mathsf{fv}(S_2) \supseteq \mathsf{fv}(S_i')$ *then since* $\bigcup_{i\in I} \mathsf{fv}(S_i') = \mathsf{fv}(+\{l_i : S_i'\}_{i\in I})$ *and* $\bigcup_{i\in I} \mathsf{fv}(S_i) = \mathsf{fv}(+\{l_i : S_i\}_{i\in I})$ *we get that* $\mathsf{fv}(+\{l_i : S_i'\}_{i\in I}) \cup \mathsf{fv}(S_2) \supseteq \mathsf{fv}(+\{l_i : S_i'\}_{i\in I})$.

- *(rec1)*

$$\frac{T_L, \mathtt{t}_1; T_R; A \vdash S_1 \circ \mu\mathtt{t}_2.S_2 \triangleright S \qquad A \; \{\mu\mathtt{t}_1.S\}}{T_L; T_R; A \vdash \mu\mathtt{t}_1.S_1 \circ \mu\mathtt{t}_2.S_2 \triangleright \mu\mathtt{t}_1.S}$$

*By induction* $\mathsf{fv}(S_1) \cup \mathsf{fv}(S_2) = \mathsf{fv}(S)$. *Since* $\mathsf{fv}(\mu\mathtt{t}_1.S_1) = \mathsf{fv}(S_1) \setminus \{\mathtt{t}_1\}$ *and* $\mathsf{fv}(\mu\mathtt{t}_1.S) = \mathsf{fv}(S) \setminus \{\mathtt{t}_1\}$ *then* $\mathsf{fv}(\mu\mathtt{t}_1.S_1) \cup \mathsf{fv}(S_2) = \mathsf{fv}(\mu\mathtt{t}_1.S)$ *as desired.*

- *(rec2)*

$$\frac{T_L; T_1, \underline{\mathtt{t}}, T_2; A \vdash S_1[\mathtt{t}/\mathtt{t}_1] \circ S_2 \triangleright S \qquad \mathtt{unused}(T_2)}{T_L; T_1, \mathtt{t}, T_2; A \vdash \mu\mathtt{t}_1.S_1 \circ S_2 \triangleright S}$$

*By induction we have that* $\mathsf{fv}(S_1[\mathtt{t}/\mathtt{t}_1]) \cup \mathsf{fv}(S_2) = \mathsf{fv}(S)$.
*We case split on whether* $\mathtt{t}_1 \in \mathsf{fv}(S_1)$.

- *If* $\mathtt{t}_1 \notin \mathsf{fv}(S_1)$ *then* $S_1[\mathtt{t}/\mathtt{t}_1] = S_1$ *and thus* $\mathsf{fv}(\mu\mathtt{t}_1.S_1) = \mathsf{fv}(S_1) = \mathsf{fv}(S_1[\mathtt{t}/\mathtt{t}_1])$ *and so* $\mathsf{fv}(\mu\mathtt{t}_1.S_1) \cup \mathsf{fv}(S_2) = \mathsf{fv}(S)$*;*
- *If* $\mathtt{t}_1 \in \mathsf{fv}(S_1)$ *then we have* $\mathtt{t} \in \mathsf{fv}(S_1[\mathtt{t}/\mathtt{t}_1])$*.*
  *In this case, by binding semantics and definition of substitution, then* $\mathsf{fv}(\mu\mathtt{t}_1.S_1) = \mathsf{fv}(S_1[\mathtt{t}/\mathtt{t}_1]) \setminus \{\mathtt{t}\}$*.*
  *By lemma F.1, we have* $\mathtt{t} \in \mathsf{fv}(S_2)$ *hence we proceed by the following reasoning:*

$$\begin{array}{rcl} & & \mathsf{fv}(S) \\ \{i.h.\} & = & \mathsf{fv}(S_1[\mathtt{t}/\mathtt{t}_1]) \cup \mathsf{fv}(S_2) \\ \{\text{lemma F.1 on premise}\} & = & \mathsf{fv}(S_1[\mathtt{t}/\mathtt{t}_1]) \cup \{\mathtt{t}\} \cup \mathsf{fv}(S_2) \\ \{(A \setminus B) \cup B = A \cup B\} & = & (\mathsf{fv}(S_1[\mathtt{t}/\mathtt{t}_1]) \setminus \{\mathtt{t}\}) \cup \{\mathtt{t}\} \cup \mathsf{fv}(S_2) \\ \{\text{binding + substitution }\} & = & \mathsf{fv}(\mu\mathtt{t}_1.S_1) \cup \{\mathtt{t}\} \cup \mathsf{fv}(S_2) \\ \{\text{lemma F.1 on conclusion}\} & = & \mathsf{fv}(\mu\mathtt{t}_1.S_1) \cup \mathsf{fv}(S_2) \end{array}$$

  *QED.*

- *(rec3) Since* $S_1 = \mu t.S'$ *here, and the result of composition is* $S = \mu t.S'$*, then* $\mathsf{fv}(S_1) = \mathsf{fv}(S)$ *as required.*
- *(call)*

$$\frac{\underline{\mathtt{t}} \in T_L \vee \underline{\mathtt{t}} \in T_R}{T_L; T_R; A \vdash \mathtt{t} \circ \mathtt{t} \triangleright \mathtt{t}}$$

  *Thus* $\mathsf{fv}(\mathtt{t}) \cup \mathsf{fv}(\mathtt{t}) = \mathsf{fv}(\mathtt{t})$ *trivially.*
- *(end)*

$$\frac{-}{T_L; T_R; A \vdash \mathtt{end} \circ \mathtt{end} \triangleright \mathtt{end}}$$

  *Trivial since* $\mathsf{fv}(\mathtt{end}) = \emptyset$*.*

**Corollary 2 (Composition preserves closedness)** *For all* $A, S$ *and closed protocols* $S_1, S_2$*, if* $T_L; T_R; A \vdash S_1 \circ S_2 \triangleright S$ *then* $S$ *is a closed protocol.*

**Proof 18** *Simple corollary of proposition 4 since* $\emptyset = \mathsf{fv}(S)$*.*

**Proposition 5 (Interleaving composition has left- and right-units)** *For a protocol* $S$ *where* $A\ \{S\} \wedge \mathsf{fv}(S) = \emptyset$ *then* $T_L; T_R; A \vdash S \circ \mathtt{end} \triangleright S$ *and* $T_L; T_R; A \vdash \mathtt{end} \circ S \triangleright S$*.*

**Proof 19** *We split the proposition into two parts. First proving the right unit, then the left unit.*

*For the right unit, we proceed by induction on the derivation of* $A\ \{S\}$*:*

- $S = p.S'$

$$\frac{A\ \{S'\} A'}{A\ \{p.S'\} A'}$$

  *Then by induction on* $S'$ *we have that* $T_L; T_R; A \vdash S' \circ \mathtt{end} \triangleright S'$ *and thus by (act) we get* $T_L; T_R; A \vdash p.S' \circ \mathtt{end} \triangleright p.S'$

- $S = +\{l_i : S_i\}_{i \in I}$

$$\frac{\forall i \in I.\, A\ \{S_i\} A_i}{A\ \{+\{l_i : S_i\}_{i\in I}\} \bigcap_{i \in I} A_i}$$

*By induction on the premise for each $S_i$ we get that $T_L; T_R; A \vdash S_i \circ \mathtt{end} \triangleright S_i$. Applying all these as the premises of (bra), we get that:*

$$T_L; T_R; A \vdash +\{l_i : S_i\}_{i\in I} \circ \mathtt{end} \triangleright +\{l_i : S_i\}_{i\in I}$$

*Satisfying the goal.*

- $S = \mathsf{require}(n).S'$

$$\frac{A' \cup \{n\}\ \{S'\} A''}{A' \cup \{n\}\ \{\mathsf{require}(n).S'\} A''}$$

*thus $A = A' \cup \{n\}$*
*By induction on the premise with $S'$ then we have $T_L; T_R; A' \cup \{n\} \vdash S' \circ \mathtt{end} \triangleright S'$. Applying this to (require) for interleaving composition then gives us:*

$$T_L; T_R; A' \cup \{n\} \vdash \mathsf{require}(n).S' \circ \mathtt{end} \triangleright \mathsf{require}(n).S'$$

*Satisfying the goal.*

- $S = \mathsf{consume}(n).S'$

$$\frac{A \setminus \{n\}\ \{S'\} A' \qquad n \in A}{A\ \{\mathsf{consume}(n).S'\} A'}$$

*By induction on the premise with $S'$ then we have $T_L; T_R; A \setminus \{n\} \vdash S' \circ \mathtt{end} \triangleright S'$. Applying this to (consume) for interleaving composition (with the side condition of $n \in A$ from the well-assertedness rule) then gives us:*

$$T_L; T_R; A \vdash \mathsf{consume}(n).S' \circ \mathtt{end} \triangleright \mathsf{consume}(n).S'$$

*Satisfying the goal.*

- $S = \mathsf{assert}(n).S'$

$$\frac{A \cup \{n\}\ \{S'\} A'}{A\ \{\mathsf{assert}(n).S'\} A'}$$

*By induction on the premise with $S'$ then we have $T_L; T_R; A \cup \{n\} \vdash S' \circ \mathtt{end} \triangleright S'$. Applying this to (assert) for interleaving composition then gives us:*

$$T_L; T_R; A \vdash \mathsf{assert}(n).S' \circ \mathtt{end} \triangleright \mathsf{assert}(n).S'$$

*Satisfying the goal.*

- $S = \mu\mathtt{t}.S'$

$$\frac{A\ \{S'\} A \cup A'}{A\ \{\mu\mathtt{t}.S'\} A \cup A'}$$

*If $\mathsf{fv}(\mu\mathtt{t}.S) = \emptyset$ then we apply (rec3) and obtain the thesis. If $\mathsf{fv}(\mu\mathtt{t}.S) \neq \emptyset$ the hypothesis does not hold hence done.*

- $S = \mathtt{end}$

  $$\frac{-}{A\ \{\mathtt{end}\}\ A}$$

  *We can then conclude with our goal via the (end) rule of interleaving composition:* $T_L; T_R; A \vdash \mathtt{end} \circ \mathtt{end} \triangleright \mathtt{end}$.
- $S = \mathtt{t}$

  $$\frac{-}{A\ \{\mathtt{t}\}\ A}$$

  *In this case* $\mathsf{fv}(S) = \{\mathtt{t}\} \neq \emptyset$ *which contradicts the hypothesis hence done.*

*Thus we have proved the right unit property of interleaving composition: that for all* $S$ *we have* $T_L; T_R; A \vdash S \circ \mathtt{end} \triangleright S$.

*To prove the left unit property we can then compose the above result with the (sym) rule of interleaving composition:*

$$\frac{T_L; T_R; A \vdash S \circ \mathtt{end} \triangleright S}{T_L; T_R; A \vdash \mathtt{end} \circ S \triangleright S}\ [sym]$$

*giving the left-unit property.* ■

## G Proof of Behaviour Preservation (Theorem 1)

We recall the theorem for convenience.

**Theorem 1 (Behaviour preservation of compositions - closed)**

$$\emptyset; \emptyset; A \vdash S_1 \circ S_2 \triangleright S \quad \Rightarrow \quad (A, S) \lesssim (A, S_1 \,||\, S_2)$$

**Proof 20** *By Lemma 6 derivation ensures behaviour preservation of one transition step, and also ensures that derivability of* $S$ *is preserved by transition possibly upon unfolding of* $S_1$ *or* $S_2$*. The thesis follows by observing that* $(A, S_i)$ *for* $i \in \{1, 2\}$ *is behaviourally equivalent to its unfoldings.*

### G.1 Behaviour Preservation - Auxiliary Definitions and Lemmas

**Lemma G.1** *If* $(A, S_1) \xrightarrow{\ell} (A', S_1')$ *and* $\mathtt{t} \notin fn(S_1)$ *then* $\mathtt{t} \notin fn(S_1')$.

**Proof sketch.** The proof is by induction observing that no reduction rule adds free names.

**Lemma G.2** $(A, S_1[\mathtt{t}/\mathtt{t}_1]) \xrightarrow{\ell} (A', S_1') \wedge \mathtt{t} \notin fn(S_1) \Longrightarrow (A, S_1) \xrightarrow{\ell} (A', S_1'[\mathtt{t}_1/\mathtt{t}])$.

**Proof sketch.** The proof is by induction on the proof of $\xrightarrow{\ell}$. All cases are base cases (trivial) except ⟨rec⟩. For ⟨rec⟩ we consider two cases:

1. $S_1 = \mu\mathtt{t}_1.S_1$. In this case $\mu\mathtt{t}_1.S_1[\mathtt{t}/\mathtt{t}_1] = \mu\mathtt{t}_1.S_1$ and by ⟨rec⟩

$$\frac{(A, S_1) \xrightarrow{\ell} (A', S_1')}{\mu\mathtt{t}_1.S_1 \xrightarrow{\ell} (A', S_1'[\mu\mathtt{t}_1.S_1/\mathtt{t}_1])}$$

The thesis is by observing that $S_1'[\mu\mathtt{t}_1.S_1/\mathtt{t}_1] = S'[\mu\mathtt{t}_1.S_1/\mathtt{t}_1][\mathtt{t}_1/\mathtt{t}]$ since
   - $\mathtt{t} \notin fn(\mu\mathtt{t}_1.S_1)$ by hypothesis;
   - $\mathtt{t} \notin fn(S_1'[\mu\mathtt{t}_1.S_1/\mathtt{t}_1])$ by Lemma G.1.

2. $S_1 = \mu\mathtt{t}_2.S_1$. By hypothesis (and rule ⟨rec⟩)

$$\frac{(A, S_1[\mathtt{t}/\mathtt{t}_1]) \xrightarrow{\ell} (A', S_1')}{\mu\mathtt{t}_2.S_1[\mathtt{t}/\mathtt{t}_1] \xrightarrow{\ell} (A', S_1'[\mu\mathtt{t}_2.S_1[\mathtt{t}/\mathtt{t}_1]/\mathtt{t}_2])} \tag{31}$$

By induction using the premise of equation (31)

$$(A, S_1) \xrightarrow{\ell} (A', S_1'[\mathtt{t}_1/\mathtt{t}])$$

The above used as premise of ⟨rec⟩ gives

$$(A, \mu\mathtt{t}_2.S_1) \xrightarrow{\ell} (A', S_1'[\mathtt{t}_1/\mathtt{t}][\mu\mathtt{t}_2.S_1/\mathtt{t}_2]) \tag{32}$$

Looking at equation (31), we need to prove that $S_1'[\mu\mathtt{t}_2.S_1[\mathtt{t}/\mathtt{t}_1]/\mathtt{t}_2][\mathtt{t}_1/\mathtt{t}]$ is equal to $S_1'[\mathtt{t}_1/\mathtt{t}][\mu\mathtt{t}_2.S_1/\mathtt{t}_2]$ from equation (32). We show this below.

$$\begin{aligned} & S_1'[\mu\mathtt{t}_2.S_1[\mathtt{t}/\mathtt{t}_1]/\mathtt{t}_2][\mathtt{t}_1/\mathtt{t}] \\ & = S_1'[\mathtt{t}_1/\mathtt{t}][\mu\mathtt{t}_2.S_1[\mathtt{t}/\mathtt{t}_1][\mathtt{t}_1/\mathtt{t}]/\mathtt{t}_2] && \text{(distribution of substitution)} \\ & = S_1'[\mathtt{t}_1\mathtt{t}][\mu\mathtt{t}_2.S_1/\mathtt{t}_2] && \text{(since } \mathtt{t} \notin fn(S_1) \text{ then } S_1'[\mathtt{t}/\mathtt{t}_1][\mathtt{t}_1/\mathtt{t}] = S_1'\text{)} \end{aligned}$$

As desired.

**Lemma G.3**

$$A\{S\} \wedge A'\{S'\}A \Rightarrow A'\{S'[S/\mathtt{t}_1]\}$$

**Proof 21** *By induction on the syntax of $S'$*

**Base cases**

- *If $S' = \mathtt{t}$ then $A'\{\mathtt{t}\}A$. If $\mathtt{t} \neq \mathtt{t}_1$ then thesis is by hypothesis. If $\mathtt{t} = \mathtt{t}_1$ then $A'\{\mathtt{t}_1\}A$ and $A' = A$ by [call] so $A\{\mathtt{t}_1\}A$ hence hypothesis $A\{S_1\}$ yields the thesis.*
- *If $S' = \mathtt{end}$ then $A'\{\mathtt{end}\}A$ and the thesis is the hypothesis as $\mathtt{t}_1 \notin fn(\mathtt{end})$.*

**Inductive cases**

- *If* $S' = p.S''$ *then by well-formedness rule [act]*

$$\frac{A'\{S''\}A}{A'\{p.S''\}A}$$

*By induction* $A'\{S''[S/\mathtt{t}_1]\}A$ *which, by [act] gives* $A'\{p.S''[S/\mathtt{t}_1]\}A$ *as desired.*
- *If* $S' = \mathsf{consume}(n).S''$ *then by well-formedness rule [consume]*

$$\frac{A' \setminus \{n\}\{S''\}A}{A'\{\mathsf{consume}(n).S''\}A}$$

*By induction* $A' \setminus \{n\}\{S''[S/\mathtt{t}_1]\}A$ *which by [consume] gives*

$$A'\{\mathsf{consume}(n).S''[S/\mathtt{t}_1]\}A$$

*as desired.*
- *The cases for assert, assume, and branching are similar to consume.*
- *If* $S' = \mu\mathtt{t}.S''$ *then by well-formedness rule [rec]*

$$\frac{A'S''A \cup A''}{A'\mu\mathtt{t}.S''A \cup A''}$$

*We have two cases: if* $\mathtt{t} = \mathtt{t}_1$ *then* $A'\{\mu\mathtt{t}_1.S''[S/\mathtt{t}_1] = \mu\mathtt{t}_1.S''\}A \cup A''$ *hence done; if* $\mathtt{t} \neq \mathtt{t}_1$ *then by induction* $A'S''[S/\mathtt{t}_1]A \cup A''$ *which used as premise of [rec] gives* $A'\{\mu\mathtt{t}.S''[S/\mathtt{t}_1]\}A \cup A''$.

**Lemma G.4 (Environment Unfolding 1)**

$$\emptyset; \emptyset; \hat{A} \vdash \mu\mathtt{t}_1.\hat{S}_1 \circ \hat{S}_2 \triangleright \mu\mathtt{t}_1.\hat{S} \tag{33}$$

*and*

$$T_L; T_R; A \vdash S_1 \circ S_2 \triangleright S \qquad \underline{\mathtt{t}_1} \in T_L \tag{34}$$

*and*

$$A\{S\}\hat{A} \tag{35}$$

*imply*

$$T_L \setminus \underline{\mathtt{t}_1}; T_R; A \vdash S_1[\mu\mathtt{t}_1.\hat{S}_1/\mathtt{t}_1] \circ S_2[\hat{S}_2/\mathtt{t}_1] \triangleright S[\mu\mathtt{t}_1.\hat{S}/\mathtt{t}_1]$$

**Proof 22** *The proof focusses on proving* $\triangleright_{wc}$ *which is the most general case; the other cases can be obtained by simply omitting the inductive cases for rules not used by that kind of composition (e.g., for* $\triangleright_s$ *omit the [wbra] and [cbra] case). The proof by induction on the derivation of* $S$ *by case analysis on the last rule used.*

**[rec1] -** $S_1 = \mu\mathtt{t}.S_1'$ **and** $S = \mu\mathtt{t}.S'$. *By hypothesis (showing the last rule application)*

$$\frac{T_L, \mathtt{t}; T_R; A \vdash S_1' \circ S_2 \triangleright S' \quad A\{\mu\mathtt{t}.S'\}}{T_L; T_R; A \vdash \mu\mathtt{t}.S_1' \circ S_2 \triangleright \mu\mathtt{t}.S'} \tag{36}$$

*By induction, considering the premise of equation* (36)

$$T_L, \mathtt{t} \setminus \underline{\mathtt{t}_1}; T_R; A \vdash S_1'[\mu\mathtt{t}_1.\hat{S}_1/\mathtt{t}_1] \circ S_2[\hat{S}_2/\mathtt{t}_1] \triangleright S'[\mu\mathtt{t}_1.\hat{S}/\mathtt{t}_1] \tag{37}$$

*By hypothesis equation* (35) $A\{\mu\mathtt{t}.S'\}\hat{A}$ *and by hypothesis equation* (33) $\hat{A}\{S\}$, *hence by Lemma G.3*

$$A\{\mu\mathtt{t}.S'[\mu\mathtt{t}_1.\hat{S}/\mathtt{t}_1]\} \tag{38}$$

*Then I can use equation* (37) *and equation* (38) *as premise of [rec1] obtaining the thesis*

$$T_L \setminus \underline{\mathtt{t}_1}; T_R; A \vdash \mu\mathtt{t}.S_1'[\mu\mathtt{t}_1.\hat{S}_1/\mathtt{t}_1] \circ S_2[\hat{S}_2/\mathtt{t}_1] \triangleright \mu\mathtt{t}.S'[\mu\mathtt{t}_1.\hat{S}/\mathtt{t}_1]$$

*as required.*

**[rec2] -** $S_1 = \mu\mathtt{t}_2.S_1'$ *By hypothesis (showing the last rule application by [rec2])*

$$\frac{T_L; T_1, \underline{\mathtt{t}}, T_2; A \vdash S_1'[\mathtt{t}/\mathtt{t}_2] \circ S_2 \triangleright S \quad \mathtt{unused}(T_2)}{T_L; T_1, \mathtt{t}, T_2; A \vdash \mu\mathtt{t}_2.S_1' \circ S_2 \triangleright S} \tag{39}$$

*By induction*

$$T_L \setminus \underline{\mathtt{t}_1}; T_1, \underline{\mathtt{t}}, T_2; A \vdash S_1'[\mathtt{t}/\mathtt{t}_2][\mu\mathtt{t}_1.\hat{S}_1/\mathtt{t}_1] \circ S_2[\hat{S}_2/\mathtt{t}_1] \triangleright S[\mu\mathtt{t}_1.\hat{S}/\mathtt{t}_1] \quad \mathtt{unused}(T_2)$$

*Because [rec2] a variable in the right recursion environment can be used/substituted at most once, hence the occurrences of* $\mathtt{t}$ *in* $\mu\mathtt{t}_2.S_1'[\mu\mathtt{t}_1.\hat{S}_1/\mathtt{t}_1]$ *are all and only those occurrences that were formerly substituted occurrences of* $\mathtt{t}_2$. *There exists, therefore, a substitution of* $\mathtt{t}$ *with* $\mathtt{t}$ *that yields* $S_2$. *By applying the above as a premise of [rec2] we obtain*

$$T_L \setminus \underline{\mathtt{t}_1}; T_1, \mathtt{t}, T_2; A \vdash \mu\mathtt{t}_2.S_1'[\mu\mathtt{t}_1.\hat{S}_1/\mathtt{t}_1] \circ S_2[\hat{S}_2/\mathtt{t}_1] \triangleright S[\mu\mathtt{t}_1.\hat{S}/\mathtt{t}_1] \tag{40}$$

*as required.*

**[call]** *If* $S_1' = \mathtt{t}$ *the thesis is immediate as* $\mathtt{t}[\mu\mathtt{t}_1.\hat{S}_1/\mathtt{t}_1] = \mathtt{t}[\mu\mathtt{t}_1.\hat{S}/\mathtt{t}_1] = \mathtt{t}[S_1/\mathtt{t}_1] = \mathtt{t}$. *If* $S_1' = \mathtt{t}_1$ *then the thesis is equivalent to hypothesis.*

**[consume] -** $S_1 = \mathsf{consume}(n).S_1'$ *By hypothesis*

$$\frac{T_L; T_R; A \vdash S_1' \circ S_2 \triangleright S}{T_L; T_R; A \cup \{n\} \vdash \mathsf{consume}(n).S_1' \circ S_2 \triangleright S}$$

*By induction, considering the premise of the derivation above:*

$$T_L \setminus \underline{\mathtt{t}_1}; T_R; A \vdash S_1'[\mu\mathtt{t}_1.\hat{S}_1/\mathtt{t}_1] \circ S_2[\hat{S}_2/\mathtt{t}_1] \triangleright S[\mu\mathtt{t}_1.\hat{S}/\mathtt{t}_1]$$

*by using the above as a premise of [consume] we obtain*

$$T_L \setminus \underline{\mathtt{t}_1}; T_R; A \cup n \vdash \mathsf{consume}(n).S_1'[\mu\mathtt{t}_1.\hat{S}_1/\mathtt{t}_1] \circ S_2[\hat{S}_2/\mathtt{t}_1] \triangleright S[\mu\mathtt{t}_1.\hat{S}/\mathtt{t}_1]$$

*as desired.*

**[wbra] -** $S_1 = +\{l_i : S_i\}_{i\in I}$ *By hypothesis*

$$\frac{\begin{array}{ll}\forall i \in I_A & T_L; T_R; A \vdash S_i \circ S_2 \triangleright S_i' \\ \forall i \in I_B & A\{S_i\} \wedge T_L; T_R; A \vdash S_i \circ S_2 \not\triangleright S_i'\end{array}}{T_L; T_R; A \vdash +\{l_i : S_i\}_{i\in I} \circ S_2 \triangleright +\{l_i : S_i'\}_{i\in I_A} \cup \{l_i : S_i\}_{i\in I_B}} \tag{41}$$

*By induction:*

$$\forall i \in I_A \quad T_L \setminus \underline{\mathtt{t}_1}; T_R; A \vdash S_i[\mu\mathtt{t}_1.\hat{S}_1/\mathtt{t}_1] \circ S_2[\hat{S}_2/\mathtt{t}_1] \triangleright S_i'[\mu\mathtt{t}_1.\hat{S}/\mathtt{t}_1] \tag{42}$$

*and by second premise of equation (41), used as a prefix still preserves the negative result in the derivation below (it worth nothing if $\mathtt{t}_1$ is removed by the environment):*

$$\forall i \in I_B \quad T_L \setminus \underline{\mathtt{t}_1}; T_R; A \vdash S_i[\mu\mathtt{t}_1.\hat{S}_1/\mathtt{t}_1] \circ S_2[\hat{S}_2/\mathtt{t}_1] \not\triangleright \tag{43}$$

*By applying equation (42) and equation (43) as premise of [wbra] we obtain*

$$T_L \setminus \underline{\mathtt{t}_1}; T_R; A \vdash +\{l_i : S_i[\mu\mathtt{t}_1.\hat{S}_1/\mathtt{t}_1]\}_{i\in I} \circ S_2[\hat{S}_2/\mathtt{t}_1] \triangleright$$
$$+\{l_i : S_i'[\mu\mathtt{t}_1.\hat{S}/\mathtt{t}_1]\}_{i\in I_A} \cup \{l_i : S_i[\mu\mathtt{t}_1.\hat{S}/\mathtt{t}_1]\}_{i\in I_B}$$

*which by definition of substitution is equivalent to*

$$T_L \setminus \underline{\mathtt{t}_1}; T_R \setminus \{\mathtt{t}_1, \underline{\mathtt{t}_1}\}; A \vdash +\{l_i : S_i\}_{i\in I}[\mu\mathtt{t}_1.\hat{S}_1/\mathtt{t}_1] \circ S_2[\hat{S}_2/\mathtt{t}_1] \triangleright$$
$$(+\{l_i : S_i'\}_{i\in I_A} \cup \{l_i : S_i\}_{i\in I_B})[\mu\mathtt{t}_1.\hat{S}/\mathtt{t}_1]$$

*as desired.*

**[bra] -** $S_1 = +\{l_i : S_i\}_{i\in I}$ *This case is a special case of [wbra] above with $I_B = \emptyset$.*

We denote with $\mathtt{unfold}(S, \mathtt{t})$ the one-time unfolding of $S$ with respect to $\mathtt{t}$. Namely, if there exists term $\mu\mathtt{t}.\hat{S}$ occurring syntactically in $S$, $\mathtt{unfold}(S, \mathtt{t}) = S[\hat{S}.\mu\mathtt{t}.\hat{S}/\mu\mathtt{t}.\hat{S}]$, otherwise $\mathtt{unfold}(S, \mathtt{t}) = S$.

**Lemma G.5 (Environment Unfolding 2)**

$$\emptyset; \emptyset; \hat{A} \vdash \mu\mathtt{t}_1.\hat{S}_1 \circ S_2 \triangleright \mu\mathtt{t}_1.\hat{S} \tag{44}$$

*and*

$$T_L; T_R; A \vdash S_1 \circ S_2 \triangleright S \qquad \mathtt{t}_1 \in \mathtt{Top}(S_1) \cap \mathtt{Top}(S) \tag{45}$$

*and*

$$A\{S\}\hat{A} \tag{46}$$

*imply there exists $\mathtt{t}$ such that*

$$T_L; T_R; A \vdash S_1[\mu\mathtt{t}_1.\hat{S}_1/\mathtt{t}_1] \circ \mathtt{unfold}(S_2, \mathtt{t}) \triangleright S[\mu\mathtt{t}_1.\hat{S}/\mathtt{t}_1]$$

**Proof 23** *Proceeds similarly to the proof of Lemma G.4. The only report the interesting cases:*

**[rec1] -** $S_1 = \mu \mathtt{t}.S_1'$ **and** $S = \mu \mathtt{t}.S'$**.** *By hypothesis (showing the last rule application)*

$$\frac{T_L, \mathtt{t}; T_R; A \vdash S_1' \circ S_2 \triangleright S' \quad A\{\mu \mathtt{t}.S'\}}{T_L; T_R; A \vdash \mu \mathtt{t}.S_1' \circ S_2 \triangleright \mu \mathtt{t}.S'} \tag{47}$$

*By induction, considering the premise of equation* (36) *there exists* $\mathtt{t}$ *in* $S_2$ *such that*

$$T_L, \mathtt{t}; T_R; A \vdash S_1'[\mu \mathtt{t}_1.\hat{S}_1/\mathtt{t}_1] \circ \mathtt{unfold}(S_2, \mathtt{t}) \triangleright S'[\mu \mathtt{t}_1.\hat{S}/\mathtt{t}_1] \tag{48}$$

*By hypothesis equation* (46) $A\{\mu \mathtt{t}.S'\}\hat{A}$ *and by hypothesis equation* (44) $\hat{A}\{S\}$*, hence by Lemma G.3*

$$A\{\mu \mathtt{t}.S'[\mu \mathtt{t}_1.\hat{S}/\mathtt{t}_1]\} \tag{49}$$

*Then I can use equation* (48) *and equation* (49) *as premise of [rec1] obtaining the thesis.*

**[rec2] -** $S_1 = \mu \mathtt{t}_2.S_1'$ *This case is not possible since hypothesis* $\mathtt{Top}(S) = \mathtt{t}_2$ *cannot hold due to the assumption that the component protocols use different protocol variables, hence* $\mathtt{t}_2$ *does not appear in* $S_2$*, not it can appear in* $S_1[\mathtt{t}/\mathtt{t}_2]$*. By inspection of the rules, observe that* $S$ *only has protocol variables that occur in either component protocols.*

### G.2 On the Relation $R1$

**Definition 12 (Folding)**

$$\begin{aligned}
&\circlearrowleft(p.S, \mathtt{t}) = p.\circlearrowleft(S, \mathtt{t})\\
&\circlearrowleft(\mathsf{assert}(n).S_1, \mathtt{t}) = \mathsf{assert}(n).\circlearrowleft(S, \mathtt{t})\\
&\circlearrowleft(\mathsf{consume}(n).S_1, \mathtt{t}) = \mathsf{consume}(n).\circlearrowleft(S, \mathtt{t})\\
&\circlearrowleft(\mathsf{require}(n).S_1, \mathtt{t}) = \mathsf{require}(n).\circlearrowleft(S, \mathtt{t})\\
&\circlearrowleft(+\{\mathrm{li} : S_i\}_{i\in I}, \mathtt{t}) = +\{\mathrm{li} :\circlearrowleft(S_i, \mathtt{t})\}_{i\in I}\\
&\circlearrowleft(\mu \mathtt{t}.S_1, \mathtt{t}) = \mathtt{t}\\
&\circlearrowleft(\mu \mathtt{t}'.S_1, \mathtt{t}) = \mu \mathtt{t}'.\circlearrowleft(S_1, \mathtt{t})\\
&\circlearrowleft(\mathsf{end}, \mathtt{t}) = \mathsf{end}\\
&\circlearrowleft(\mathtt{t}', \mathtt{t}) = \mathtt{t}'
\end{aligned}$$

**Definition 13 (Top)**

$$\mathtt{Top}(S) = \begin{cases} \mathtt{t} & \text{if } S = \mu \mathtt{t}.S' \\ \emptyset & \text{otherwise} \end{cases}$$

*We say that* $S$ *is free from* unguarded nested recursions *if*

- *either* $\mathtt{Top}(S) = \emptyset$ *or* $S = \mu \mathtt{t}.S'$ *and* $\mathtt{Top}(S') = \emptyset$*, and*
- *all syntactic subterms of* $S$ *are free from unguarded nested recursions.*

As noted in the commentary of Definition Definition 1, we assume without loss of generality that protocols are free from unguarded nested recursions.

**Lemma G.6** *If* $S_1$ *and* $S_2$ *have guarded nested recursions and* $T_L; T_R; A \vdash S_1 \circ S_2 \triangleright \mu \mathtt{t}.S$ *then* $\mathtt{Top}(S) = \emptyset$

**Proof sketch.** Can be proved by induction on the proof of $\mu\mathtt{t}.S$. Observe that the composition rules never concatenate recursions from the same protocol (by assumption that $S_1$ is free from unguarded nested recursions) or from different protocols (by assumption that $S_1$ is free from unguarded nested recursions and by the fact that $\mu\mathtt{t}_1$ and $\mu\mathtt{t}_2$ are not sequentially composed in [rec2]).

**Lemma G.7 (Preservation - open protocols)** *If* $T_L; T_R; A \vdash S_1 \circ S_2 \triangleright S$ *and* $(A, S) \xrightarrow{\ell} (A', S')$ *for some* $\ell, A', S'$ *then one of the following holds:*

1. $(A, S_1) \xrightarrow{\ell} (A', S_1')$ *and* $T_L; T_R\underline{@}; A' \vdash \circlearrowleft (S_1', \mathtt{Top}(S_1))@ \circ S_2 \triangleright \circlearrowleft (S', \mathtt{Top}(S))$*, for some* $@$ *substitution of* $\mathtt{t}' \in \mathtt{Top}(S_1) \setminus fn(S')$ *with* $\mathtt{t} \in T_R$ *and* $\underline{@}$ *substitution of* $\mathtt{t}$ *with* $\underline{\mathtt{t}}$.
2. $(A, S_2) \xrightarrow{\ell} (A', S_2')$ *and* $T_L\underline{@}; T_R; A' \vdash S_1 \circ \circlearrowleft (S_2', \mathtt{Top}(S_2))@ \triangleright \circlearrowleft (S', \mathtt{Top}(S))$*, for some* $@$ *substitution of* $\mathtt{t}' \in \mathtt{Top}(S_2) \setminus fn(S')$ *with* $\mathtt{t} \in T_L$ *and* $\underline{@}$ *substitution of* $\mathtt{t}$ *with* $\underline{\mathtt{t}}$.
3. $(A, S_i) \xrightarrow{+1} (A', S_i')$ *with* $i \in \{1, 2\}$ *and* $S' = \overline{S}[S/\mathtt{Top}(S)]$ *and* $S_i' = \overline{S}[S_i/\mathtt{Top}(S_i)]$ *for some* $\overline{S}$

**Proof 24** *This lemma holds for all variants of composition:* $\triangleright_s$, $\triangleright_w$, $\triangleright_c$*, and* $\triangleright_{wc}$ *(recall that notation* $\triangleright$ *is used to refer to any of the aforementioned composition judgments). The proof focusses on proving* $\triangleright_{wc}$ *which is the most general case; the other cases can be obtained by simply omitting the inductive cases for rules not used by that kind of composition (e.g., for* $\triangleright_s$ *omit the [wbra] and [cbra] case).*

*By induction on the derivation of S by case analysis on last rule used.*

**[end]** *The hypothesis does not hold since* $(A, \mathtt{end}) \not\rightarrow$ *hence done.*

**[call]** *The hypothesis does not hold since* $(A, \mathtt{t}) \not\rightarrow$ *hence done.*

**[consume] -** $S = \mathsf{consume}(n).\hat{S}$ *The top of the derivation is of the following form:*

$$\frac{T_L; T_R; A \vdash \hat{S}_1 \circ S_2 \triangleright \hat{S}}{T_L; T_R; A \cup \{n\} \vdash \mathsf{consume}(n).\hat{S}_1 \circ S_2 \triangleright \mathsf{consume}(n).\hat{S}} \tag{50}$$

*By hypothesis*

$$(A \cup \{n\}, \mathsf{consume}(n).\hat{S}) \xrightarrow{\ell} (A, S')$$

*Since the only transition rule applicable to* $(A \cup \{n\}, \mathsf{consume}(n).\hat{S})$ *is* $\langle\mathtt{consume}\rangle$ *then* $\ell = \mathsf{consume}(n)$ *and* $S' = \hat{S}$

$$(A \cup \{n\}, \mathsf{consume}(n).\hat{S}) \xrightarrow{\mathsf{consume}(n)} (A, \hat{S})$$

*Similarly, by* $\langle\mathtt{consume}\rangle$

$$(A \cup \{n\}, \mathsf{consume}(n).\hat{S}_1) \xrightarrow{\mathsf{consume}(n)} (A, \hat{S}_1)$$

*The thesis (item 1) follows immediately by the premise of (equation (50)) observing that* $\mathtt{Top}(S_1) = \emptyset$ *and* $\mathtt{Top}(S) = \emptyset$ *and* $@$ *is the empty substitution.*

**Cases [pref], [assume], [assert]** *are similar to the case for [consume].*

**Cases [wbra]** *By hypothesis*

$$\frac{\begin{array}{l} I_A \cap I_B = \emptyset \quad I_A \cup I_B = I \quad I_A \neq \emptyset \\ \forall i \in I_A \quad T_L; T_R; A \vdash S_i \circ S_2 \triangleright S'_i \\ \forall i \in I_B \quad A\{S_i\} \quad T_L; T_R; A \vdash S_i \circ S_2 \not\triangleright \end{array}}{T_L; T_R; A \vdash +\{\mathrm{l_i} : S_i\}_{i\in I} \circ S_2 \triangleright +\{\mathrm{l_i} : S'_i\}_{i\in I_A} \cup +\{\mathrm{l_i} : S_i\}_{i\in I_B}} \tag{51}$$

$S = +\{\mathrm{l_i} : S'_i\}_{i\in I_A} \cup +\{\mathrm{l_i} : S_i\}_{i\in I_B}$ *can only move by* $\langle \mathtt{Branch} \rangle$ *with* $\ell = \mathrm{l}_j$ *and either* $j \in I_A$ *or* $j \in I_B$.

**Case** $j \in I_A$. $(A, +\{\mathrm{l_i} : S'_i\}_{i\in I_A} \cup +\{\mathrm{l_i} : S_i\}_{i\in I_B}) \xrightarrow{+l_i} (A, S'_j)$. *Similarly, by* $\langle \mathtt{Branch} \rangle$ *on* $S_1$

$$(A, +\{\mathrm{l_i} : S_i\}_{i\in I}) \xrightarrow{+l_j} (A, S_j)$$

*The thesis hold (item 1) as it is the premise in (equation (51)) for* $i = j \in I_A$ *observing that* $\mathtt{Top}(+\{\mathrm{l_i} : S'_i\}_{i\in I}) = \emptyset$ *and* $\mathtt{Top}(+\{\mathrm{l_i} : S_i\}_{i\in I}) = \emptyset$ *and @ is the empty substitution.*

**Case** $j \in I_B$. $(A, +\{\mathrm{l_i} : S'_i\}_{i\in I_A} \cup +\{\mathrm{l_i} : S_i\}_{i\in I_B}) \xrightarrow{+l_i} (A, S_j)$. *Similarly, by* $\langle \mathtt{branch} \rangle$ *on* $S_1$

$$(A, +\{\mathrm{l_i} : S'_i\}_{i\in I}) \xrightarrow{+l_j} (A, S_j)$$

*Thesis holds (item 3) with* $\overline{S} = S_j$ *since* $fn(S_j) \setminus fn(S) = fn(S_j) \setminus fn(S_1) = \emptyset$.

**Cases [bra]** *As the case [wbra] assuming* $I_B = \emptyset$.

**Cases [cbra]** *By hypothesis*

$$\frac{\begin{array}{l} \forall i \in I \quad J_i \neq \emptyset \quad \bigcup_{i\in I} J_i = J \\ \qquad \forall j \in J_i \quad T_L; T_R; A \vdash S_i \circ S'_j \triangleright S_{ij} \\ \qquad \forall j \in J \setminus J_i \quad T_L; T_R; A \vdash S_i \circ S'_j \not\triangleright \end{array}}{T_L; T_R; A \vdash +\{\mathrm{l_i} : S_i\}_{i\in I} \circ +'\{\mathrm{l'_j} : S'_j\}_{j\in J} \triangleright +\{\mathrm{l_i} : +'\{\mathrm{l'_j} : S_{ij}\}_{i\in J_i}\}_{i\in I}} \tag{52}$$

$S = +\{\mathrm{l_i} : +'\{\mathrm{l'_j} : S_{ij}\}_{i\in J_i}\}_{i\in I}$ *can only move by* $\langle \mathtt{Branch} \rangle$ *with* $\ell = \mathrm{l}_i$ *as follows:*

$$(A, +\{\mathrm{l_i} : +'\{\mathrm{l'_j} : S_{ij}\}_{i\in J_i}\}_{i\in I}) \xrightarrow{+l_i} (A, +'\{\mathrm{l'_j} : S_{ij}\}_{i\in J_i})$$

*Similarly, by* $\langle \mathtt{Branch} \rangle$ *on* $S_1$

$$(A, +\{\mathrm{l_i} : S_i\}_{i\in I}) \xrightarrow{+l_i} (A, S_i)$$

*The first premise in (equation (51)) can be applied as axiom in the derivation below to obtain the thesis (item 1) and observing that* $\mathtt{Top}(+\{\mathrm{l_i} : S'_i\}_{i\in I}) = \emptyset$ *and @ is the empty substitution:*

$$\dfrac{\dfrac{\dfrac{T_L; T_R; A \vdash S_i \circ S'_j \triangleright S_{ij}}{T_R; T_L; A \vdash S'_j \circ S_i \triangleright S_{ij}}\,[sym]}{T_R; T_L; A \vdash +'\{\mathrm{l'_j} : S'_j\}_{j\in J} \circ S_i \triangleright +'\{\mathrm{l'_j} : S_{ij}\}_{i\in J_i}}\,[bra]}{T_L; T_R; A \vdash S_i \circ +'\{\mathrm{l'_j} : S'_j\}_{j\in J} \triangleright +'\{\mathrm{l'_j} : S_{ij}\}_{i\in J_i}}\,[sym]$$

**Case [sym]** *The last rule applies is of the following form:*

$$\frac{T_L;\, T_R;\, A \vdash S_2 \circ S_1 \triangleright S}{T_L;\, T_R;\, A \vdash S_1 \circ S_2 \triangleright S}$$

*By hypothesis* $(A,S) \xrightarrow{\ell} (A',S')$.

*By induction one of the following holds:*

1. *if* $(A,S_2) \xrightarrow{\ell} (A',S_2')$ *then* $T_L;\, T_R;\, A' \vdash\circlearrowleft (S_2', \mathtt{Top}(S_2))@ \circ S_1 \triangleright \circlearrowleft (S', \mathtt{Top}(S))$ *which yields the thesis when applied as a premise of [sym]. The case for*
2. *if* $\ell \in \{\oplus l, \& l\}$ *and* $(A,S_1) \xrightarrow{\ell} (A',S_1')$ *and* $S' = \overline{S}[S/fn(\overline{S})\backslash fn(S)]$ *and* $S_1' = \overline{S}[S_1/fn(\overline{S})\backslash fn(S_1)]$ *for some* $\overline{S}$ *then the thesis (item 3) holds after applying [sym].*
3. *if* $\ell \in \{\oplus l, \& l\}$ *and* $(A,S_2) \xrightarrow{\ell} (A',S_2')$ *the case is similar to case (2).*
4. *if* $(A,S_1) \xrightarrow{\ell} (A',S_1')$ *the case is similar to case (1).*

**Case [rec1] -** $S = \mu \mathtt{t}_1.\hat{S}$ *By hypothesis*

$$\frac{T_L, \mathtt{t}_1;\, T_R;\, A \vdash \hat{S}_1 \circ \mu \mathtt{t}_2.\hat{S}_2 \triangleright \hat{S}}{T_L;\, T_R;\, A \vdash \mu \mathtt{t}_1.\hat{S}_1 \circ \mu \mathtt{t}_2.\hat{S}_2 \triangleright \mu \mathtt{t}_1.\hat{S}} \tag{53}$$

*and*

$$\frac{(A,\hat{S}) \xrightarrow{\ell} (A',S')}{(A,\mu \mathtt{t}_1.\hat{S}) \xrightarrow{\ell} (A',S'[\mu \mathtt{t}_1.\hat{S}/\mathtt{t}_1])} \tag{54}$$

*By induction, considering the premise of equation* (53) *we have one of the following three cases:*

**Case $S_1$ moves and composition is preserved.** *If* $(A,\hat{S}_1) \xrightarrow{\ell} (A',S_1')$ *and*

$$T_L, \mathtt{t}_1;\, T_R;\, A' \vdash\circlearrowleft (S_1', \mathtt{Top}(\hat{S}_1))@ \circ S_2 \triangleright \circlearrowleft (S', \mathtt{Top}(\hat{S}))$$

*Since* $\mu \mathtt{t}_1.\hat{S}_1$ *is free from unguarded recursion we have* $\mathtt{Top}(\hat{S}_1) = \emptyset$ *hence* @ *is empty substitution and we get*

$$T_L, \mathtt{t}_1;\, T_R;\, A' \vdash S_1' \circ S_2 \triangleright S' \tag{55}$$

*By* $\langle \mathtt{rec} \rangle$, $(A, \mu \mathtt{t}_1.S_1) \xrightarrow{\ell} (A', S_1'[\mu \mathtt{t}_1.S_1])$. *The thesis to prove is therefore*

$$T_L;\, T_R;\, A' \vdash\circlearrowleft (S_1'[\mu \mathtt{t}_1.S_1/\mathtt{t}_1], \mathtt{Top}(\hat{S}_1))@ \circ S_2 \triangleright \circlearrowleft (S'[\mu \mathtt{t}_1.S/\mathtt{t}_1], \mathtt{Top}(\hat{S}))$$

*Observing that* $\mathtt{Top}(S_1) = \mathtt{t}_1$, *the derivation above is equivalent to*

$$T_L;\, T_R;\, A' \vdash\circlearrowleft (S_1'[\mu \mathtt{t}_1.S_1/\mathtt{t}_1], \mathtt{t}_1)@ \circ S_2 \triangleright \circlearrowleft (S'[\mu \mathtt{t}_1.S/\mathtt{t}_1], \mathtt{t}_1)$$

*that is*

$$T_L;\, T_R;\, A' \vdash S_1'@ \circ S_2 \triangleright S'$$

*Observing that* @ *is the empty substitution since in* $\mathtt{Top}(S_1) = \mathtt{t}_1$ *and* $\mathtt{t}_1 \in fn(S')$ *the above follows immediately by equation* (55).

**Case $S_2$ moves and composition is preserved.** *If* $(A, S_2) \xrightarrow{\ell} (A', S_2')$ *and*

$$T_L, \mathtt{t}_1; \emptyset; A' \vdash \hat{S}_1 \circ \circlearrowleft (S_2', \mathtt{Top}(S_2))@ \triangleright \circlearrowleft (S', \mathtt{Top}(\hat{S}))$$

*then by Lemma G.6,* $\mathtt{Top}(S) = \emptyset$ *and @ is empty. Therefore, the above is equivalent to*

$$T_L, \mathtt{t}_1; T_R; A' \vdash \hat{S}_1 \circ \circlearrowleft (S_2', \mathtt{Top}(S_2)) \triangleright S' \tag{56}$$

*The thesis*

$$T_L; T_R; A' \vdash \mu\mathtt{t}_1.\hat{S}_1 \circ \circlearrowleft (S_2', \mathtt{Top}(S_2))@ \triangleright \circlearrowleft (S'[\mu\mathtt{t}_1.\hat{S}/\mathtt{t}_1], \mathtt{Top}(S))$$

*is equivalent to equation* (56) *since:* $\mathtt{Top}(S) = \mathtt{t}_1$, $\circlearrowleft (S'[\mu\mathtt{t}_1.\hat{S}/\mathtt{t}_1], \mathtt{t}_1) = S'$ *and @ is the empty substitution (*$fn(S') = \mathtt{t}_1$ *which is not a name in* $S_2$ *as we assume bound names of* $S_1$ *and* $S_2$ *to be disjoint).*

**Case $\ell \in \{\oplus l, \& l\}$ and composition is not preserved.** *By induction either* $S_1$ *or* $S_2$ *makes a transition with label* $\ell$*. We show the case in which* $S_1$ *moves, as the case in which* $(A, \hat{S}_2) \xrightarrow{\ell} (A', S')$ *is symmetric.*

*Assume by induction* $(A, \hat{S}_1) \xrightarrow{\ell} (A', S')$*. Then: (1) since* $\mathtt{Top}(\hat{S}_1) = \emptyset$ *then* $fn(S') \setminus fn(\hat{S}_1) = \emptyset$*, and (2) by Lemma G.6* $\mathtt{Top}(\hat{S}_1) = \emptyset$ *then* $\overline{S} = S'$ *and* $fn(S') \setminus fn(\hat{S}) = \emptyset$*. So, by* $\langle\mathtt{rec}\rangle$

$$(A, \mu\mathtt{t}_1.S_1) \xrightarrow{\ell} (A', S'[\mu\mathtt{t}_1.S_1/\mathtt{t}_1] \qquad (A, \mu\mathtt{t}_1.S) \xrightarrow{\ell} (A', S'[\mu\mathtt{t}_1.S/\mathtt{t}_1])$$

*with* $fn(S') \setminus fn(\mu\mathtt{t}_1.\hat{S}) = \mathtt{t}_1$ *and* $fn(S_1') \setminus fn(\mu\mathtt{t}_1.\hat{S}_1) = \mathtt{t}_1$ *hence the thesis.*

**Case [rec2] -** $S = \mu\mathtt{t}_1.\hat{S}$ *By hypothesis*

$$\frac{T_L; T_1, \underline{\mathtt{t}}, T_2; A \vdash \hat{S}_1[\mathtt{t}/\mathtt{t}_1] \circ S_2 \triangleright S \quad \mathtt{unused}(T_2)}{T_L; T_1, \mathtt{t}, T_2; A \vdash \mu\mathtt{t}_1.\hat{S}_1 \circ S_2 \triangleright S} \tag{57}$$

*and*

$$\frac{(A, \hat{S}) \xrightarrow{\ell} (A', S')}{(A, \mu\mathtt{t}_1.\hat{S}) \xrightarrow{\ell} (A', S'[\mu\mathtt{t}_1.\hat{S}/\mathtt{t}_1])} \tag{58}$$

*By induction on the premise of equation* (57) *we have one of the following cases:*

1. *First, assume*

$$(A, \hat{S}_1) \xrightarrow{\ell} (A', S_1') \tag{59}$$

*and*

$$T_L; T_1, \underline{\mathtt{t}}, T_2; A' \vdash \circlearrowleft (S_1', \mathtt{Top}(\hat{S}_1[\mathtt{t}/\mathtt{t}_1]))@ \circ S_2 \triangleright \circlearrowleft (S', \mathtt{Top}(S)) \quad \mathtt{unused}(T_2) \tag{60}$$

*By assumption of freeness from unguarded recursions we have* $\mathtt{Top}(\underline{S_1}) = \emptyset$ *hence* $\mathtt{Top}(\underline{S_1}[\mathtt{t}/\mathtt{t}_1]) = \emptyset$*. It follows that* $@$ *is empty and equation (60) is equivalent to*

$$T_L;\, T_1, \underline{\mathtt{t}}, T_2;\, A' \vdash S_1' \circ S_2 \triangleright \circlearrowleft (S', \mathtt{Top}(S)) \tag{61}$$

*By* $\langle\mathtt{rec}\rangle$ *with premise equation (59)*

$$(A, \mu\mathtt{t}_1.\hat{S}_1[\mathtt{t}/\mathtt{t}_1]) \xrightarrow{\ell} (A', S_1'[\mu\mathtt{t}_1.\hat{S}_1[\mathtt{t}/\mathtt{t}_1]/\mathtt{t}_1])$$

*By the transition above and Lemma G.2*

$$(A, \mu\mathtt{t}_1.\hat{S}_1) \xrightarrow{\ell} (A', S_1'[\mathtt{t}_1/\mathtt{t}][\mu\mathtt{t}_1.\hat{S}_1/\mathtt{t}_1]) \tag{62}$$

*We need to prove*

$$T_L;\, T_1, \mathtt{t}, T_2;\, A' \vdash \circlearrowleft (S_1'[\mathtt{t}_1/\mathtt{t}][\mu\mathtt{t}_1.\hat{S}_1/\mathtt{t}_1], \mathtt{Top}(\mu\mathtt{t}_1.\hat{S}_1))@ \circ S_2 \triangleright \circlearrowleft (S', \mathtt{Top}(S)) \quad \mathtt{unused}(T_2) \tag{63}$$

*which, since* $\mathtt{Top}(\mu\mathtt{t}_1.\hat{S}_1) = \mathtt{t}_1$ *and applying the folding, is equivalent to*

$$T_L;\, T_1, \mathtt{t}, T_2;\, A' \vdash S_1'[\mathtt{t}_1/\mathtt{t}]@ \circ S_2 \triangleright \circlearrowleft (S', \mathtt{Top}(S)) \tag{64}$$

*In equation (64),* $@ = [\mathtt{t}/\mathtt{t}_1]$ *and* $\underline{@} = [\mathtt{t}/\underline{\mathtt{t}}]$*, hence equation (64) is equivalent to equation (61) as required.*

2. *Second, assume*

$$(A, S_2) \xrightarrow{\ell} (A', S_2') \tag{65}$$

*By induction on the premise of equation (57)*

$$T_L;\, T_1, \underline{\mathtt{t}}, T_2;\, A' \vdash \hat{S}_1[\mathtt{t}/\mathtt{t}_1] \circ \circlearrowleft (S_2', \mathtt{Top}(S_2))@ \triangleright \circlearrowleft (S', \mathtt{Top}(S)) \tag{66}$$

*for some* $@$*. Applying equation (66) as premise of [rec2] we obtain the thesis*

$$T_L;\, T_1, \underline{\mathtt{t}}, T_2;\, A' \vdash \mu\mathtt{t}_1.\hat{S}_1 \circ \circlearrowleft (S_2', \mathtt{Top}(S_2))@ \triangleright \circlearrowleft (S', \mathtt{Top}(S))$$

*as desired.*

3. *Finally, assume* $\ell \in \{\oplus l, \& l\}$*. By induction we have* $(A, \hat{S}_1) \xrightarrow{\ell} (A', S')$ *with* $\mathit{fn}(S') \setminus \mathit{fn}(\hat{S}) = \emptyset$ *(the case in which* $(A, \hat{S}_2) \xrightarrow{\ell} (A', S')$ *is symmetric). So, by* $\langle\mathtt{rec}\rangle$

$$(A, \mu\mathtt{t}_1.S_1) \xrightarrow{\ell} (A', S'[\mu\mathtt{t}_1.S_1/\mathtt{t}_1])$$

*Recall also that*

$$(A, S) \xrightarrow{\ell} (A', S')$$

*The thesis hold for* $\overline{S} = S'$ *since* $\mathit{fn}(S') \setminus \mathit{fn}(S) = \emptyset$*,* $\mathit{fn}(S_1') \setminus \mathit{fn}(\mu\mathtt{t}_1.\hat{S}_1) = \mathtt{t}_1$*.*

Recall, we denote with $\mathtt{unfold}(S,\mathtt{t})$ the one-time unfolding of $S$ with respect to $\mathtt{t}$. Namely, if there exists term $\mu\mathtt{t}.\hat{S}$ occurring syntactically in $S$, $\mathtt{unfold}(S,\mathtt{t}) = S[\hat{S}.\mu\mathtt{t}.\hat{S}/\mu\mathtt{t}.\hat{S}]$, otherwise $\mathtt{unfold}(S,\mathtt{t}) = S$.

**Lemma 6 (Preservation - closed protocols)** *Assume* $\emptyset;\emptyset;A \vdash S_1 \circ S_2 \triangleright S$*. For all* $\ell, A', S_1$ *such that* $(A,S) \xrightarrow{\ell} (A',S')$ *either*

1. $(A,S_1) \xrightarrow{\ell} (A',S_1')$ *and* $\emptyset;\emptyset;A \vdash S_1' \circ S_2' \triangleright S'$ $(S_2' = \mathtt{unfold}(S_2,\hat{\mathtt{t}})$ *for some* $\hat{\mathtt{t}}$*), or*
2. $(A,S_2) \xrightarrow{\ell} (A',S_2')$ *and* $\emptyset;\emptyset;A \vdash S_1' \circ S_2' \triangleright S'$ $(S_1' = \mathtt{unfold}(S_1',\hat{\mathtt{t}})$ *for some* $\hat{\mathtt{t}}$*), or*
3. $\ell \in \{\oplus l, \& l\}$ *and* $(A,S_i) \xrightarrow{\ell} (A', \overline{S}[S_1/fn(\overline{S})\backslash fn(S_1)])$ *with* $i \in \{1,2\}$ *and* $S' = \overline{S}[S/fn(\overline{S})\backslash fn(S)]$ *for some* $\overline{S}$

**Proof 25** *The proof focusses on proving* $\triangleright_{wc}$ *which is the most general case; the other cases can be obtained by simply omitting the inductive cases for rules not used by that kind of composition (e.g., for* $\triangleright_s$ *omit the [wbra] and [cbra] case). We proceed by induction on derivation of* $S$ *proceeding by case analysis on the last rule used.*

**Case [sym]** *The last rule applies is of the following form:*

$$\frac{\emptyset;\emptyset;A \vdash S_2 \circ S_1 \triangleright S}{\emptyset;\emptyset;A \vdash S_1 \circ S_2 \triangleright S}$$

*By induction either* $(A,S_2) \xrightarrow{\ell} (A',S_2')$ *and* $\emptyset;\emptyset;A' \vdash S_2' \circ S_1 \triangleright S'$ *which applied as a premise of [sym] yields the thesis (item 2)* $\emptyset;\emptyset;A' \vdash S_1 \circ S_2' \triangleright S'$*, or* $(A,S_1) \xrightarrow{\ell} (A',S_1')$ *and* $\emptyset;\emptyset;A' \vdash S_2 \circ S_1' \triangleright S'$ *which applied as a premise of [sym] yields the thesis (item 1)* $\emptyset;\emptyset;A' \vdash S_1' \circ S_2 \triangleright S'$*. Alternatively, case (3) applies by induction and yields the thesis as item 3 is symmetric (*$i \in \{1,2\}$*).*

**Case [consume] -** $S = \mathsf{consume}(n).\hat{S}$ *proceeds as the corresponding case in Lemma G.7.*
*The top of the derivation is of the following form:*

$$\frac{\emptyset;\emptyset;A \vdash \hat{S}_1 \circ S_2 \triangleright \hat{S}}{\emptyset;\emptyset;A \cup \{n\} \vdash \mathsf{consume}(n).\hat{S}_1 \circ S_2 \triangleright \mathsf{consume}(n).\hat{S}}$$

*By* $\langle\mathtt{consume}\rangle$

$$(A \cup \{n\}, \mathsf{consume}(n).S) \xrightarrow{\mathsf{consume}(n)} (A,S)$$

*and*

$$(A \cup \{n\}, \mathsf{consume}(n).S_1) \xrightarrow{\mathsf{consume}(n)} (A,S_1)$$

*The thesis holds as it is identical to the the premise of equation* (50).

**Cases [pref][assume][assert]** *are similar to [consume].*

**Case [wbra]** *Proceeds as the corresponding case in Lemma G.7. By hypothesis*

$$\frac{\begin{array}{l} I = I_A \cup I_B \quad I_A \cup I_B \neq \emptyset \\ \forall i \in I_A . \emptyset; \emptyset; A \vdash S_i \circ S_2 \triangleright S'_i \\ \forall i \in I_B . \emptyset; \emptyset; A \vdash S_i \circ S_2 \not\triangleright \wedge A\{S_i\} \end{array}}{\emptyset; \emptyset; A \vdash +\{\mathrm{l_i} : S_i\}_{i\in I} \circ S_2 \triangleright +\{\mathrm{l_i} : S'_i\}_{i\in I_A} \cup \{l_i : S'_i\}_{i\in I_B}}$$

*By* $\langle \mathtt{bra} \rangle$*, picking* $i \in I_A$ *which is not empty by premise of the derivation above*

$$(A, +\{\mathrm{l_i} : S_i\}_{i\in I}) \xrightarrow{+\mathrm{l}_j} (A, S_j)$$

*and*

$$(A, +\{\mathrm{l_i} : S'_i\}_{i\in I}) \xrightarrow{+\mathrm{l}_j} (A, S'_j)$$

*The thesis holds as it is identical to the the premise of the derivation above for* $j \in I_A$.

**Case [bra]** *This follows by [wbra] setting* $I_B = \emptyset$.

**Case [cbra]** *By hypothesis*

$$\frac{\begin{array}{l} \forall i \in I \quad J_i \neq \emptyset \quad \bigcup_{i\in I} J_i = J \\ \qquad \forall j \in J_i \quad \emptyset; \emptyset; A \vdash S_i \circ S'_j \triangleright S_{ij} \\ \qquad \forall j \in J \setminus J_i \quad \emptyset; \emptyset; A \vdash S_i \circ S'_j \not\triangleright \end{array}}{\emptyset; \emptyset; A \vdash +\{l_i : S_i\}_{i\in I} \circ +'\{l'_j : S'_j\}_{j\in J} \triangleright +\{l_i : +'\{l'_j : S_{ij}\}_{j\in J_i}\}_{i\in I}} \tag{67}$$

*By* $\langle \mathtt{bra} \rangle$*, picking* $i \in I$

$$(A, +\{l_i : +'\{l'_j : S'_j\}_{j\in J_i}\}_{i\in I}) \xrightarrow{+\mathrm{l}_i} (A, +'\{l'_j : S_{ij}\}_{j\in J_i})$$

*and similarly*

$$(A, +\{l_i : S_i\}_{i\in I}) \xrightarrow{+\mathrm{l}_i} (A, S_i)$$

*By applying [sym] to the second premise of equation* (67)*:*

$$\forall j \in j_i +' \{l'_j : S'_j\}_{j\in J_i} \circ S_i \triangleright +'\{l'_j : S_{ij}\}_{j\in J_i} \tag{68}$$

*By applying equation* (68) *as premise of [bra] and then [sym] we obtain the thesis (1):*

$$\emptyset; \emptyset; A \vdash S_i \circ +'\{l'_j : S'_j\}_{j\in J_i} \triangleright +'\{l'_j : S_{ij}\}_{j\in J_i}$$

**Case [rec1] -** $S = \mu\mathtt{t}_1.\hat{S}$ **and** $T_R = \emptyset$ *By hypothesis*

$$\frac{\mathtt{t}_1; \emptyset; A \vdash \hat{S}_1 \circ S_2 \triangleright S \quad A\{\mu\mathtt{t}_1.S\}}{\emptyset; \emptyset; A \vdash \mu\mathtt{t}_1.\hat{S}_1 \circ S_2 \triangleright \mu\mathtt{t}_1.\hat{S}} \tag{69}$$

*and*

$$\frac{(A, \hat{S}) \xrightarrow{\ell} (A', S')}{(A, \mu\mathtt{t}_1.\hat{S}) \xrightarrow{\ell} (A', S'[\mu\mathtt{t}_1.\hat{S}/\mathtt{t}_1])} \tag{70}$$

*By Lemma G.7 we have one of the following three cases:*

1. $(A, \hat{S}_1) \xrightarrow{\ell} (A', S_1')$ *hence by* $\langle\mathtt{rec}\rangle$

   $$(A, \mu\mathtt{t}_1.\hat{S}_1) \xrightarrow{\ell} (A', S_1'[\mu\mathtt{t}_1.\hat{S}/\mathtt{t}_1])$$

   *and*

   $$\{\mathtt{t}_1\}; \emptyset; A' \vdash \circlearrowleft (S_1', \mathtt{Top}(S_1)) \circ S_2 \triangleright \circlearrowleft (S', \mathtt{Top}(S)) \tag{71}$$

   *By assumption of nested guardedness* $\mathtt{Top}(\hat{S}_1) = \emptyset$ *and by Lemma G.6* $\mathtt{Top}(S) = \emptyset$. *Hence by equation (71) we obtain, with @ being the empty substitution:*

   $$\{\mathtt{t}_1\}; \emptyset; A' \vdash S_1' \circ S_2 \triangleright S' \tag{72}$$

   *By premise of equation (69) [rec1]* $A\{\mu\mathtt{t}_1.S\}$ *which looking at the well formedness rule [rec] can be written as*

   $$A\{\mu\mathtt{t}_1.S\}A \cup A'' \tag{73}$$

   *for some* $A''$. *By Lemma 1 equation (73) and equation (70) imply*

   $$A'\{S_1'[\mu\mathtt{t}_1.\hat{S}_1/\mathtt{t}_1]\}A''' \text{ such that } A''' \supseteq A \cup A'' \tag{74}$$

   *By Lemma G.5 since equation (69) and equation (72) and equation (74) we obtain that there exists* $\hat{\mathtt{t}}$ *such that*

   $$\emptyset; \emptyset; A' \vdash S_1'[\mu\mathtt{t}_1.S_1/\mathtt{t}_1] \circ \mathtt{unfold}(S_2, \hat{\mathtt{t}}) \triangleright S'[\mu\mathtt{t}_1.S/\mathtt{t}_1]$$

   *as desired.*

2. $(A, \hat{S}_2) \xrightarrow{\ell} (A', S_2')$ *and*

   $$\mathtt{t}_1; \emptyset; A' \vdash S_1 \circ \circlearrowleft (S_2', \mathtt{Top}(S_2)) \triangleright \circlearrowleft (S', \mathtt{Top}(\hat{S})) \tag{75}$$

   *By Lemma G.6,* $\mathtt{Top}(\hat{S}) = \emptyset$ *hence equation (75) is equivalent to*

   $$\mathtt{t}_1; \emptyset; A' \vdash S_1 \circ \circlearrowleft (S_2', \mathtt{Top}(S_2)) \triangleright S' \tag{76}$$

   *We proceed by inner induction on the syntax of* $S_2$.

- *If* $S_2 = p.\hat{S}_2$ *then* $S_2' = \hat{S}_2$*, and* $\mathtt{Top}(S_2) = 0$ *hence and* @ *is the empty substitution. Therefore, equation* (76) *is equivalent to the thesis* $\mathtt{t}_1; \emptyset; A' \vdash S_1 \circ S_2' \triangleright S'$ *as desired;*
- *If* $S_2 = a.\hat{S}_2$ *with* $a \in \{\mathsf{assert}(n), \mathsf{consume}(n), \mathsf{require}(n)\}$ *the case is similar to the prefix case above;*
- *If* $S_2 = \mathtt{end}$ *or* $S_2 = \mathtt{t}$ *then* $(A, S_2) \not\rightarrow$ *hence done.*
- *If* $S_2 = \mu\mathtt{t}_2.\hat{S}_2$ *(interesting case) then* $S_2' = \hat{S}_2'[\mu\mathtt{t}_2.S_2/\mathtt{t}_2]$ *with* $(A, \hat{S}_2) \rightarrow (A', \hat{S}_2')$ *as premise of* $\langle\mathtt{rec}\rangle$*. Since* $\mathtt{Top}(S_2) = \mathtt{t}_2$ *and* $fn(S') \not\ni \mathtt{t}_2$ *then* $@ = [\mathtt{t}_1/\mathtt{t}_2]$*. Therefore,* $\circlearrowleft (S_2', \mathtt{Top}(S_2))@ = S_2'[\mathtt{t}_1/\mathtt{t}_2]$ *and substituting this in equation* (76) *we obtain*

  $$\{\mathtt{t}_1\}; \emptyset; A' \vdash S_1 \circ S_2'[\mathtt{t}_1/\mathtt{t}_2] \triangleright S'$$

  *By applying Lemma G.4 to the above we get*

  $$\emptyset; \emptyset; A' \vdash S_1[\mu\mathtt{t}_1.\hat{S}_1/\mathtt{t}_1] \circ S_2'[\mathtt{t}_1/\mathtt{t}_2][\mu\mathtt{t}_2.\hat{S}_2/\mathtt{t}_1] \triangleright S'[\mu\mathtt{t}_1.\hat{S}/\mathtt{t}_1]$$

  *which is equivalent to*

  $$\emptyset; \emptyset; A' \vdash S_1[\mu\mathtt{t}_1.\hat{S}_1/\mathtt{t}_1] \circ S_2'[\mu\mathtt{t}_2.\hat{S}_2/\mathtt{t}_2] \triangleright S'[\mu\mathtt{t}_1.\hat{S}/\mathtt{t}_1]$$

  *as desired.*
- *By Lemma G.7 either* $S_1$ *or* $S_2$ *makes a transition with label* $\ell$ *and*

  $$(A, \mu\mathtt{t}_1.S_1) \xrightarrow{\ell} (A', S'[\mu\mathtt{t}_1.S_1/\mathtt{t}_1]) \qquad (A, \mu\mathtt{t}_1.S) \xrightarrow{\ell} (A', S'[\mu\mathtt{t}_1.S/\mathtt{t}_1])$$

  *with* $fn(S') \setminus fn(\mu\mathtt{t}_1.\hat{S}) = \mathtt{t}_1$ *and* $fn(S_1') \setminus fn(\mu\mathtt{t}_1.\hat{S}_1) = \mathtt{t}_1$ *hence the thesis.*

**Case [rec2]** $S = \mu\mathtt{t}_1.\hat{S}$ *Contradicts the hypothesis* ($T_R \neq \emptyset$) *hence done.*

# H Proofs of Fairness

**Definition 14** *Define the following context:*

$$\begin{array}{rcll} C[\cdot] & = & g.C[\cdot] & g \in \{p, \mathsf{assert}(n), \mathsf{consume}(n), \mathsf{require}(n)\} \\ & \mid & +\{\mathrm{l} : C[\cdot]\} \cup \{\mathrm{l}_i : S_i\}_{i \in I} & \\ & \mid & \mu\mathtt{t}.C[\cdot] & \\ & \mid & [\cdot] & \end{array}$$

*Write* $S = C[\cdot]$ *if* $S = C[S']$ *for some* $S'$*. Write* $C' \in C$ *(resp.* $C' \notin C$*) is there exists (resp. there exists no)* $C_1$*,* $C_2$ *such that* $C = C_1[C'[C_3[\cdot]]]$*. Define the following functions:*

$$\mathtt{clab}(g.S) = \{g\} \qquad \mathtt{clab}(+\{\mathrm{l}_i : S_i\}_{i \in I}) = \{+\mathrm{l}_i\}_{i \in I} \qquad \mathtt{clab}(\mu\mathtt{t}.S) = \mathtt{clab}(S)$$

*and*

$$\mathtt{V}(g.C[\cdot]) = g, \mathtt{V}(C[\cdot]) \qquad \mathtt{V}([\cdot]) = \epsilon \qquad \mathtt{V}(\mu\mathtt{t}.C[\cdot]) = \mathtt{V}(C[\cdot])$$
$$\mathtt{V}(+\{\mathrm{l}_j : C[\cdot]\} \cup \{\mathrm{l}_i : S_i\}_{i \in I \setminus j}) = +\mathrm{l}_j, \mathtt{V}(C[\cdot])$$

**Lemma H.1** *If* $(A,S) \xrightarrow{\ell} (A',S')$ *then* $(A,S[\hat{S}/\mathtt{t}]) \xrightarrow{\ell} (A',S'[\hat{S}/\mathtt{t}])$

**Proof 26** *(sketch) Mechanical by induction on the transition, by case analysis on the last rule used to make step* $\ell$.

**Lemma H.2** *If* $T_L;\ T_R; A \vdash S_0 \circ S_1 \triangleright S$ *and* $S_1 = \mathsf{end}$ *then* $S_0 = S$

**Proof 27** *(sketch) By induction on the proof of* $S$ *proceeding by case analysis on the last rule used, observing that the last rule used cannot be* $[\mathit{rec}1]$ *or* $[\mathit{rec}2]$ *as the only axiom that can be used if* $[\mathit{end}]$ *(i.e., not* $[\mathit{call}]$*) due to the form of* $S_1$.

**Lemma H.3** *If* $(A,S) \xrightarrow{\ell}$ *then* $\ell \in \mathtt{clab}(S)$.

**Proof 28** *(sketch) Mechanical by induction on the derivation of transition* $\xrightarrow{\ell}$ *proceeding by case analysis on the last transition rule used.*

**Lemma H.4** *If* $(A,S) \xrightarrow{\vec{r}}$ *then* $S = C[S']$ *for some* $S'$ *and* $\mathtt{V}(C) = \vec{r}$.

**Proof 29** *(sketch) First prove that* $(A,S) \xrightarrow{r}$ *implies* $S = C[S']$ *for some* $S'$ *and* $\mathtt{V}(C) = r$ *by induction on the transition proceeding by case analysis on the last transition rule used. Then by induction on the size of* $\vec{r}$ *based on the fact that contexts compositionality.*

**Lemma H.5** *If* $S = C[\,S'\,]$, $A\{S\}$, *and* $\ell \in \mathtt{clab}(S')$, *then* $(A,S) \xrightarrow{\vec{r}}\xrightarrow{\ell}$ *for some (possibly empty) vector* $\vec{r}$ *of transition labels such that* $\mathtt{V}(C) = \vec{r}$.

**Proof 30** *By induction on the syntax of* $C$.

- *Case* $C = [\,\cdot\,]$ *(and hence* $\mathtt{V}(C)$ *is the empty vector of labels). We show the case for* $S' = \mathsf{consume}(n).S''$ *and hence* $\mathtt{clab}(S) = \{\mathsf{consume}(n)\}$. *By well-assertedness of* $S$, *which in this case last applies rule [consume],* $n \in A$, *then by semantic rule* $\langle\mathtt{consume}\rangle$ *we have*

  $$(A, \mathsf{consume}(n).S') \xrightarrow{\mathsf{consume}(n)} (A \setminus \{n\}, S')$$

  *as desired. The cases for* $S' \in \{\mathsf{require}(n).S'', \mathsf{assert}(n).S'', p.S''\}$ *are similar.*
- *If* $C = \mathsf{consume}(n).C'[\,\cdot\,]$ *then we proceed with a generic* $S'$. *By well-assertedness of* $S$ *which last applies rule [consume] we have* $n \in A$ *hence by semantic rule* $\langle\mathtt{consume}\rangle$ *we have*

  $$(A, \mathsf{consume}(n).C'[S']) \xrightarrow{\mathsf{consume}(n)} (A \setminus \{n\}, C'[S'])$$

  *By Lemma 1 (well-assertedness is preserved by transition) we have*

  $$A \setminus \{n\}\{C'[\,\hat{C}[\,S'\,]\,]\}$$

  *By induction* $(A \setminus \{n\}, C'[\,S'\,]) \xrightarrow{\vec{r}}\xrightarrow{\ell}$ *with* $\ell \in \mathtt{lab}(S')$ *and* $\mathtt{V}(C') = \vec{r}$ *hence*

  $$(A, \mathsf{consume}(n).C'[S']) \xrightarrow{\mathsf{consume}(n)}\xrightarrow{\vec{r}}\xrightarrow{\ell}$$

  *with* $\ell \in \mathtt{lab}(S')$ *and* $\mathtt{V}(C) = \mathsf{consume}(n), \mathtt{V}(C') = \mathsf{consume}(n), \vec{r}$ *as desired.*

- *The other cases for* $C = g.C'[\cdot]$ *are similar to the case above.*
- *If* $C = +\{\mathtt{l} : C'[\cdot]\} \cup \{\mathtt{l}_i : S_i\}_{i\in I}$*. By semantic rule* $\langle\mathtt{branch}\rangle$ *we have*

$$(A, +\{\mathtt{l} : C'[S']\} \cup \{\mathtt{l}_i : S_i\}_{i\in I}) \xrightarrow{+\mathtt{l}} (A, C'[S'])$$

*By Lemma 1 (well-assertedness is preserved by transition) we have*

$$A\{C'[\,S'\,]\}$$

*By induction* $(A, C'[\,S'\,]) \xrightarrow{\vec{r}}\xrightarrow{\ell}$ *with* $\ell \in \mathtt{clab}(S')$ *and* $\mathtt{V}(S') = \vec{r}$ *hence*

$$(A, +\{\mathtt{l} : C'[S']\} \cup \{\mathtt{l}_i : S_i\}_{i\in I}) \xrightarrow{+\mathtt{l}}\xrightarrow{\vec{r}}\xrightarrow{\ell}$$

*with* $\ell \in \mathtt{clab}(S')$ *and* $\mathtt{V}(C) = +\mathtt{l}, \mathtt{V}(C) = +\mathtt{l}, \vec{r}$ *as desired.*
- *If* $C = \mu\mathtt{t}.C'[\cdot]$ *then by well-assertedness of* $A\{S\}$*. In this case the last rule applied is* $[rec]$*. We have by premise of* $[rec]$*,* $A\{C'[\,S'\,]\}$*. Hence, by Lemma 2 (well-asserted protocols are not stuck)*

$$(A, C'[\,S'\,]) \xrightarrow{\ell'} (A', C''[\,S'\,]) \tag{77}$$

*for some* $C''$ *and by Lemma 1* $A'\{C''[\,S'\,]\}$*. By induction*

$$(A', C'[\,S'\,]) \xrightarrow{\ell'}\xrightarrow{\vec{r}}\xrightarrow{\ell} \qquad \ell \in \mathtt{clab}(S') \quad \ell', \vec{r} = \mathtt{V}(C'[\,\cdot\,]) \tag{78}$$

*By* $\langle\mathtt{Rec}\rangle$ *with as premise the first transition of equation* (78)*:*

$$(A, \mu\mathtt{t}.C'[\,S'\,]) \xrightarrow{\ell} (A', C''[\,S'\,][\mu\mathtt{t}.C'[\,S'\,]/\mathtt{t}])$$

*By Lemma H.1 and equation* (78)

$$(A', C''[\,S'\,][\mu\mathtt{t}.C'[\,S'\,]/\mathtt{t}]) \xrightarrow{\vec{r}}\xrightarrow{\ell}$$

*hence*

$$(A, \mu\mathtt{t}.C'[\,S'\,]) \xrightarrow{\ell'}\xrightarrow{\vec{r}}\xrightarrow{\ell} \qquad \ell \in \mathtt{clab}(S')$$

$\mathtt{V}(\mu\mathtt{t}.C'[\,\cdot\,]) = \mathtt{V}(C'[\,\cdot\,])$ *and by induction* $\mathtt{V}(\mu\mathtt{t}.C'[\,\cdot\,]) = \ell', \vec{r}$ *as desired.*

**Lemma H.6** *If* $T_L;\ T_R;\ A \vdash S_0 \circ S_1 \triangleright S$ *then* $\forall \ell \in \mathtt{clab}(S_1) \exists C[\,\cdot\,], C_0[\,\cdot\,], S', S'_0$ *such that*

1. $S = C[\,S'\,]$ *and* $S_0 = C_0[\,S'_0\,]$
2. $\mathtt{V}(C[\,\cdot\,]) = \mathtt{V}(C_0[\,\cdot\,])$
3. $\ell \in \mathtt{clab}(S')$

**Proof 31** *We proceed by induction on the proof of* $S$*, proceeding by case analysis of the last rule applied.*

**Case [end]** *In this case* $\texttt{clab}(S_1) = \emptyset$ *hence done.*

**Case [call]** *In this case* $\texttt{clab}(S_1) = \emptyset$ *hence done.*

**Case [act]** *Fix* $\ell \in \texttt{clab}(S_1)$. *In this case* $S_0 = p.S'_0$ *and* $S = p.S'$ *then*

$$\frac{T_L;\ T_R; A \vdash S'_0 \circ S_1 \triangleright S'}{T_L;\ T_R; A \vdash p.S'_0 \circ S_1 \triangleright p.S'}$$

*By induction there exists* $C[\,S''\,] = S'$ *and* $C_0[\,S''_0\,] = S'_0$ *such that* $\texttt{V}(C[\cdot]) = \texttt{V}(C_0[\cdot])$ *and* $\ell \in \texttt{clab}(S'')$. *Hence there exists* $p.C$ *and* $p.C_0$ *such that* $C[\,S''_0\,] = S_0$ *and* $p.C[\,S''\,] = p.S' = S$ *and* $p.C_0[\,S''_0\,] = p.S'_0 = S_0$. *Moreover, since* $\texttt{V}(C[\cdot]) = \texttt{V}(C_0[\cdot])$ *then* $p, \texttt{V}(C[\cdot]) = p, \texttt{V}(C_0[\cdot])$ *and hence* $\texttt{V}(p.C[\cdot]) = \texttt{V}(p.C_0[\cdot])$. *Finally, still* $\ell \in \texttt{clab}(S'')$ *as desired.*

**Case [consume] [assert], [require]** *Are similar to [act].*

**Case [wbra]** *Fix* $\ell \in \texttt{clab}(S_1)$. *In this case* $S = +\{\mathtt{l_i} : S'_i\}_{i\in I_A} \cup \{\mathtt{l}_i : S_i\}_{i \in I_B}$, $S_0 = +\{\mathtt{l_i} : S_i\}_{i\in I}$ *and*

$$\frac{\begin{array}{l} I_A \cup I_B = I \quad I_A \cap I_B = \emptyset \\ \forall i \in I_A.\ T_L;\ T_R; A \vdash S_i \circ S_1 \triangleright S'_i \\ \forall i \in I_B.\ A\{S_i\} \quad T_L;\ T_R; A \vdash S_i \circ S_1 \not\triangleright \end{array}}{T_L;\ T_R; A \vdash +\{\mathtt{l_i} : S_i\}_{i\in I} \circ S_1 \triangleright +\{\mathtt{l_i} : S'_i\}_{i\in I_A} \cup \{\mathtt{l}_i : S_i\}_{i\in I_B}}$$

*By induction* $\forall i \in I_A$ *with* $I_A \neq \emptyset$ *there exist* $C_i[\,\hat{S}_i\,] = S_i$ *and* $C'_i[\,\hat{S}'_i\,] = S'_i$ *such that* $\texttt{V}(C_i[\cdot]) = \texttt{V}(C'_i[\cdot])$ *and* $\ell \in \texttt{clab}(\hat{S}'_i)$.

*Hence, there exists* $C_0[\,\hat{S}_i\,] = +\{\mathtt{l}_i : C_i[\,\hat{S}_i\,]\} \cup \{\mathtt{l}_j : S_j\}_{j\in I\setminus\{i\}}$ *and* $C[\,\hat{S}'_i\,] = +\{\mathtt{l}_i : C'_i[\,\hat{S}'_i\,]\} \cup \{\mathtt{l}_j : S'_j\}_{j \in I_A\setminus\{i\}} \cup \{\mathtt{l}_i : S_i\}_{i\in I_B}$. *Moreover,* $\texttt{V}(C_i[\cdot]) = \texttt{V}(C'_i[\cdot])$ *implies* $\mathtt{l}_i, \texttt{V}(C_i[\cdot]) = \mathtt{l}_i, \texttt{V}(C'_i[\cdot])$ *and hence* $\texttt{V}(C_0[\cdot]) = \texttt{V}(C[\cdot])$. *Finally, still* $\ell \in \texttt{clab}(\hat{S}'_i)$ *as desired.*

**Case [bra]** *As the case [wbra] assuming* $I_B = \emptyset$.

**Case [cbra]** *Fix* $\ell \in \texttt{clab}(S_1)$. *In this case* $S = +\{\mathtt{l_i} : +'\{\mathtt{l_j} : S_{ij}\}_{j\in J_i}\}_{i\in I}$, $S_0 = +\{\mathtt{l_i} : S_i\}_{i\in I}$ *and*

$$\frac{\begin{array}{l} \forall i \in I \quad J_i \neq \emptyset \quad \bigcup_{i\in I} J_i = J \\ \qquad\quad j \in J_i.\ T_L;\ T_R; A \vdash S_i \circ S_j \triangleright S_{ij} \\ \qquad\quad \forall j \in J \setminus J_i\ T_L;\ T_R; A \vdash S_i \circ S_j \not\triangleright \end{array}}{T_L;\ T_R; A \vdash +\{\mathtt{l_i} : S_i\}_{i\in I} \circ +'\{\mathtt{l_j} : S_j\}_{j\in J} \triangleright +\{\mathtt{l_i} : +'\{\mathtt{l_j} : S_{ij}\}_{j\in J_i}\}_{i\in I}}$$

*By induction* $\forall i \in I$ *there exist* $C_i[\,\hat{S}_i\,] = S_i$ *and* $C'_i[\,\hat{S}_{ij}\,] = S_{ij}$ *such that* $\texttt{V}(C_i[\cdot]) = \texttt{V}(C'_i[\cdot])$ *and* $\ell \in \texttt{clab}(\hat{S}_{ij})$.

*Hence, forall* $\mathtt{l}_j \in \texttt{clab}(S_0)$ *(which is non empty since* $I \neq \emptyset$*) there exist* $j \in J_i$, $C_1[\,\hat{S}_i\,] = +\{\mathtt{l}_i : C_i[\,\hat{S}'_i\,]\} \cup \{\mathtt{l}_j : S_j\}_{j\in I\setminus\{j\}}$ *and* $C[\,\hat{S}_{ij}\,] = +\{\mathtt{l}_i : C'_i[\,\hat{S}_{ij}\,]\} \cup \{\mathtt{l}_j : S_{ij}\}_{j\in J_i\setminus\{i\}}$, $\texttt{V}(C_0[\cdot]) = \texttt{V}(C[\cdot]) = \mathtt{l}_i, \texttt{V}(C_i[\cdot]) = \mathtt{l}_i, \texttt{V}(C'_i[\cdot])$ *and still* $\ell \in \texttt{clab}(\hat{S}_{ij})$.

**Case [rec1]** *Fix $\ell \in \mathtt{clab}(S_1)$. In this case $S_0 = \mu\mathtt{t}.S_0'$, $S = \mu\mathtt{t}.S'$ (and for simplicity we leave the recursive form of $S_1$ implicit as it is immaterial here). By composition rule [recI]:*

$$\frac{T_L, \mathtt{t}; T_R; A \vdash S_0' \circ S_1 \triangleright S'}{T_L; T_R; A \vdash \mu\mathtt{t}.S_0' \circ S_1 \triangleright \mu\mathtt{t}.S'}$$

*By induction there exist $C[\,\hat{S}'\,] = S'$ and $C_0[\,\hat{S}_0'\,] = S_0'$ such that $\mathtt{V}(C[\cdot]) = \mathtt{V}(C_0[\cdot])$ and $\ell \in \mathtt{clab}(\hat{S}')$. Hence there exists $\mu\mathtt{t}.C[\,\hat{S}'\,] = S$ and $\mu\mathtt{t}.C_0[\,\hat{S}_0'\,] = S_0$. Moreover, since by induction $\mathtt{V}(C[\cdot]) = \mathtt{V}(C_0[\cdot])$ and hence $\mathtt{V}(\mu\mathtt{t}.C[\cdot]) = \mathtt{V}(\mu\mathtt{t}.C_0[\cdot])$ (since $\mathtt{V}(\mu\mathtt{t}.C[\cdot]) = \mathtt{V}(C[\cdot])$ and $\mathtt{V}(\mu\mathtt{t}.C_0[\cdot]) = \mathtt{V}(C_0[\cdot])$ by definition of $\mathtt{V}()$). Finally, still $\ell \in \mathtt{clab}(\mu\mathtt{t}.\hat{S}')$ as desired.*

**Case [rec2]** *Fix $\ell \in \mathtt{clab}(S_1)$. In this case*

$$\frac{T_L; T_1, \underline{\mathtt{t}'}, T_2; A \vdash S_0'[\mathtt{t}'/\mathtt{t}] \circ S_1 \triangleright S \quad \mathtt{unused}(T_2)}{T_L; T_1, \mathtt{t}', T_2; A \vdash \mu\mathtt{t}.S_0' \circ S_1 \triangleright S}$$

*By induction there exists $C[\,\hat{S}'\,] = S$ and $C_0[\,\hat{S}_0'\,] = S_0'[\mathtt{t}'/\mathtt{t}]$ such that $\mathtt{V}(C[\cdot]) = \mathtt{V}(C_0[\cdot])$ and $\ell \in \mathtt{clab}(\hat{S}')$.*

*Hence there exists $C[\,\hat{S}'\,] = S$ and $\mu\mathtt{t}.C_0[\mathtt{t}/\mathtt{t}'][\,\hat{S}_0'[\mathtt{t}/\mathtt{t}']\,] = S_0$. By induction $\mathtt{V}(C_0) = \mathtt{V}(C)$ and by definition of $\mathtt{V}()$ (observing that $\mathtt{t}'$ does not affect the returned value), $\mathtt{V}(\mu\mathtt{t}.C_0[\mathtt{t}/\mathtt{t}']) = \mathtt{V}(C_0[\mathtt{t}/\mathtt{t}'])$ and $\mathtt{V}(C_0[\mathtt{t}/\mathtt{t}']) = \mathtt{V}(C_0)$. Hence $\mathtt{V}(C[\cdot]) = \mathtt{V}(\mu\mathtt{t}.C_0[\mathtt{t}/\mathtt{t}'])$ and still $\ell \in \mathtt{clab}(\mu\mathtt{t}.\hat{S}') = \mathtt{clab}(\hat{S}')$.*

**Case [sym]** *In this case*

$$\frac{T_L; T_R; A \vdash S_1 \circ S_0 \triangleright S}{T_L; T_R; A \vdash S_0 \circ S_1 \triangleright S} \tag{79}$$

*Assume that [sym] is applied only once. If [sym] it is applied multiple (but finite) times subsequently, say n times, then if n is even the thesis is immediate by hypothesis, and if n is odd then the case is equivalent to the one where the rule is applied once. Fix $\ell \in \mathtt{clab}(S_1)$. If $\mathtt{clab}(S_0) \neq \emptyset$ then by induction there exists $C[\,S'\,] = S$ and $C_1[\,S_1'\,]$ such that $\mathtt{V}(C[\cdot]) = \mathtt{V}(C_1[\cdot])$ and $\ell \in \mathtt{clab}(S')$. From $\mathtt{V}(C[\cdot]) = \mathtt{V}(C_1[\cdot])$, $C[\,S'\,] = S$ and $C_1[\,S_1'\,]$ it follows that $S$ and $S_1$ have the same first prefix hence*

$$\ell \in \mathtt{clab}(S)$$

*hence the thesis with contexts $[\cdot]$ for $S$ and $S_0$ and trivially $\mathtt{V}([\cdot]) = \mathtt{V}([\cdot])$. If $\mathtt{clab}(S_0) = \emptyset$ then either $S_0 = \mathrm{end}$ or $S_0 = \mathtt{t}$. In either case $S_1 = S$: by lemma H.2 if $S_0 = \mathrm{end}$ and by [call] if $S_0 = \mathtt{t}$. Hence with contexts $[\cdot]$ for $S$ and $S_0$ trivially $\mathtt{clab}(S_1) = \mathtt{clab}(S)$ and hence $\ell \in \mathtt{clab}(S)$ and $\mathtt{V}([\cdot]) = \mathtt{V}([\cdot])$*

Lemma H.7 is a stronger version of Lemma H.6 where quantification over contexts is universal rather than existential, and holds only for strong composition (not for weak one).

**Lemma H.7** *If* $T_L; T_R; A \vdash C_0[\, S_0 \,] \circ S_1 \triangleright_s S$ *and* $\mathtt{clab}(S_1) \neq \emptyset$*, then either:*

1. *there exist* $C_0'$, $C_0''$, $C[S'] = S$ *such that*
   - $C_0[\cdot] = C_0'[\, C_0''[\cdot]\,]$*, and*
   - $\mathtt{V}(C_0'[\cdot]) = \mathtt{V}(C[\cdot])$*, and*
   - $\mathtt{clab}(S') = \mathtt{clab}(S_1)$*, or*
2. *there exist* $C_0'[S_0']$, $C[S'] = S$ *such that*
   - $C_0[\, C_0'[\, S' \,]\,] = S_0$*, and*
   - $\mathtt{V}(C_0[\, C_0'[\cdot]\,]) = \mathtt{V}(C[\cdot])$*, and*
   - $\mathtt{clab}(S') = \mathtt{clab}(S_1)$

**Proof 32** *We proceed by induction on the syntax of* $C_0$*.*

**Case** $C_0[\cdot] = p.\hat{C}_0[\cdot]$ *By hypothesis*

$$\frac{T_L; T_R; A \vdash \hat{C}_0[\, S_0 \,] \circ S_1 \triangleright_s S}{T_L; T_R; A \vdash p.\hat{C}_0[\, S_0 \,] \circ S_1 \triangleright_s p.S}$$

*By induction either of the following holds:*

1. *there exist* $C_0'[\, C_0''[\cdot]\,] = \hat{C}_0[\cdot]$ *and* $C[S'] = S$ *such that* $\mathtt{V}(\hat{C}_0'[\cdot]) = \mathtt{V}(C[\cdot])$ *and* $\mathtt{clab}(S') = \mathtt{clab}(S_1)$*. Therefore there exist* $p.C_0'[\, C_0''[\cdot]\,] = C_0[\cdot]$ *and* $p.C[S'] = p.S$ *such that* $\mathtt{V}(C_0'[\cdot]) = p, \mathtt{V}(\hat{C}_0'[\cdot]) = p, \mathtt{V}(C[\cdot]) = \mathtt{V}(p.C[\cdot])$ *and* $\mathtt{clab}(S') = \mathtt{clab}(S_1)$*.*
2. *there exist* $C_0'[S_0']$, $C[S'] = S$ *such that* $\hat{C}_0[\, C_0'[\, S' \,]\,] = S_0$, $\mathtt{V}(\hat{C}_0[\, C_0'[\cdot]\,]) = \mathtt{V}(C[\cdot])$*, and* $\mathtt{clab}(S') = \mathtt{clab}(S_1)$*. Therefore there exist* $C_0'[S_0']$, $p.C[S'] = S$ *such that* $p.\hat{C}_0[\, C_0'[\, S' \,]\,] = p.S_0$, $\mathtt{V}(p.\hat{C}_0[\, C_0'[\cdot]\,]) = p, \mathtt{V}(\hat{C}_0[\, C_0'[\cdot]\,]) = p, \mathtt{V}(C[\cdot]) = \mathtt{V}(p.C[\cdot])$ *and* $\mathtt{clab}(S') = \mathtt{clab}(S_1)$*.*

*The cases for consume, assert, and require are similar to the prefix case above.*

**Case** $C_0[\cdot] = +\{\mathtt{l}_j : \hat{C}_0[\cdot]\} \cup \{\mathtt{l}_i : S_i\}_{i \in I \setminus \{j\}}$ *By hypothesis*

$$\frac{\forall i \in I \setminus \{j\} \quad T_L; T_R; A \vdash S_i \circ S_1 \triangleright_s S_i' \qquad \hat{C}_0[\hat{S}_j] = S_j \quad T_L; T_R; A \vdash \hat{C}_1[\hat{S}_j] \circ S_1 \triangleright_s S_j'}{T_L; T_R; A \vdash +\{\mathtt{l}_j : \hat{C}_0[\hat{S}_j]\} \cup \{\mathtt{l}_i : S_i\}_{i \in I \setminus \{j\}} \circ S_1 \triangleright_s +\{\mathtt{l}_i : S_i\}_{i \in I}}$$

*By induction either of the following holds:*

1. *there exist* $C_0'[\, C_0''[\cdot]\,] = \hat{C}_0[\cdot]$ *and* $C[S_j''] = S_j'$ *such that* $\mathtt{V}(C_0'[\cdot]) = \mathtt{V}(C[\cdot])$ *and* $\mathtt{clab}(S_j'') = \mathtt{clab}(S_1)$*. Therefore there exist* $+\{\mathtt{l}_j : C_0'[\, C_0''[\cdot]\,]\} \cup \{\mathtt{l}_i : S_i\}_{i \in I \setminus \{j\}} = C_0[\cdot]$ *and* $+\{\mathtt{l}_j : C[S_j'']\} \cup \{\mathtt{l}_i : S_i\}_{i \in I \setminus \{j\}} = +\{\mathtt{l}_i : S_i\}_{i \in I}$ *such that* $\mathtt{V}(+\{\mathtt{l}_j : C_0'[\cdot]\} \cup \{\mathtt{l}_i : S_i\}_{i \in I \setminus \{j\}}) = \mathtt{l}_j, \mathtt{V}(C_0'[\cdot]) = \mathtt{V}(+\{\mathtt{l}_j : C[\cdot]\} \cup \{\mathtt{l}_i : S_i'\}_{i \in I \setminus \{j\}})$ *and* $\mathtt{clab}(S_j') = \mathtt{clab}(S_1)$*.*

2. *there exist* $C_0'[S_j''']$, $C[S_j''] = S_j'$ *such that* $C_0[\, C_0'[S_j''']\,] = S_j$, $\mathtt{V}(\hat{C}_0[\, C_0'[\cdot]\,]) = \mathtt{V}(C[\cdot])$, *and* $\mathtt{clab}(S') = \mathtt{clab}(S_1)$. *Therefore there exist* $\hat{C}_0[\, C_0'[S_j''']\,]$, $+\{l_j : C[S_j'']\} \cup \{l_i : S_i\}_{i\in I\setminus\{j\}} = +\{l_i : S_i\}_{i\in I}$ *such that* $\mathtt{V}(C_0) = l_j, \mathtt{V}(\hat{C}_0[\, C_0'[\cdot]\,]) = l_j, \mathtt{V}(C[\cdot]) = \mathtt{V}(+\{l_j : C[\cdot]\} \cup \{l_i : S_i\}_{i\in I\setminus\{j\}})$ *and* $\mathtt{clab}(S') = \mathtt{clab}(S_1)$.

**Case** $C_0[\cdot] = \mu\mathtt{t}.\hat{C}_0[\cdot]$ *By induction observing that* $\mathtt{V}(\mu\mathtt{t}.\hat{C}_0[\cdot]) = \mathtt{V}(\hat{C}_0[\cdot])$.

**Case** $C_0[\cdot] = [\cdot]$ *Immediate by induction.*

**Theorem 2 (Fairness of compositions)** *If* $\emptyset; \emptyset; A \vdash S_0 \circ S_1 \rhd S$ *then* $S$ *is* fair *w.r.t.* $S_0$ *and* $S_1$ *on* $A$.

**Proof 33** *Immediately from Lemma H.8.*

**Lemma H.8 (Fairness)** *Let* $\emptyset; \emptyset; A \vdash S_0 \circ S_1 \rhd S$. *Then* $\forall i \in \{0,1\}$ *and any transition* $(A_i, S_i) \xrightarrow{\ell} (A_i', S_i')$ *there exists* $\vec{r}$ *such that: (1)*

- $(A, S_{|1-i|}) \xrightarrow{\vec{r}} (A'_{|1-i|}, S'_{|1-i|})$
- $(A, S) \xrightarrow{\vec{r}\ell} (A'', S')$
- $\emptyset; \emptyset; A'' \vdash S_0' \circ S_1' \rhd S'$.

**Proof 34** *Assume* $(A, S_1) \xrightarrow{\ell} (A_1', S_1')$. *By Lemma H.3* $\ell \in \mathtt{clab}(S_1)$ *so, by Lemma H.6, there are two contexts* $C_0$ *and* $C$ *such that the hypothesis can be rewritten as*

$$\emptyset; \emptyset; A \vdash C_0[\, S_0'\,] \circ S_1 \rhd C[\, S'\,]$$

*with*

$$\mathtt{V}(C_0[\cdot]) = \mathtt{V}(C[\cdot]) \tag{80}$$

*and*

$$\ell \in \mathtt{clab}(S') \tag{81}$$

*By equation* (80), *equation* (81) *and Lemma H.5*

$$\begin{array}{ll} (A, S_0) \xrightarrow{\vec{r}} (A_0', S_0') & (\textit{for some } A_0', S_0') \\ (A, S) \xrightarrow{\vec{r}\ell} (A', S') & (\textit{for some } A'', S') \end{array} \tag{82}$$

*It remains to prove that*

$$\emptyset; \emptyset; A'' \vdash S_0' \circ S_1' \rhd S' \tag{83}$$

*For every transition* $r \in \vec{r}$, *by case (1) of Lemma 6 the composition relation is preserved. More precisely, let* $\vec{r} = r_0, \ldots, r_n$:

$$\emptyset; \emptyset; A \vdash S_0 \circ S_1 \rhd S \ \wedge\ (A, S_0) \xrightarrow{r_0} (A^1, S_0^1) \ \wedge\ (A, S) \xrightarrow{r_0} (A^1, S^1) \Rightarrow \emptyset; \emptyset; A^1 \vdash S_0^1 \circ S_1 \rhd S^1$$
$$\ldots$$
$$\emptyset; \emptyset; A^n \vdash S_0^n \circ S_1 \rhd S^n \ \wedge\ (A^n, S_0^n) \xrightarrow{r_n} (A', S_0') \ \wedge\ (A^n, S^n) \xrightarrow{r_n} (A', S^{n+1}) \Rightarrow$$
$$\emptyset; \emptyset; A^{n+1} \vdash S_0' \circ S_1 \rhd S^{n+1}$$

*Note that, when using Lemma 6, case (1) of Lemma 6 can always apply (case 2 applies for the symmetric case in which $S_0$ moves first). Assume by contradiction that only case (3) applies (the case where continuations are not preserved), then $S_0$ and $S$ would move to a state in which they are both $\hat{S}$. By taking any $S_0$ (and hence $S$) that does not include any $\ell$ action we have a counter-example for equation (82) (second row) already proved above. Hence case (1) must always be applicable. Hence done.*

*Assume now that $(A, S_0) \xrightarrow{\ell} (A', S_0')$. Then by applying $[sys]$ after the last composition rule in the hypothesis we obtain*

$$\emptyset; \emptyset; A \vdash S_1 \circ S_0 \triangleright S'$$

*and the case is then identical to the one where $S_1$ moves, proved above.*

**Theorem 3 (Strong fairness of compositions with $\triangleright_s$)** *If $\emptyset; \emptyset; A \vdash S_0 \circ S_1 \triangleright_s S$ then $S$ is* strongly fair *with respect to $S_0$ and $S_1$ on $A$.*

**Proof 35** *Immediately from Lemma H.9.*

**Lemma H.9** *Let $\emptyset; \emptyset; A \vdash S_0 \circ S_1 \triangleright_s S$. Then $\forall i \in \{0,1\}$ and all transitions $(_, S_i) \xrightarrow{\ell} (_, S_i')$ and $(A, S_{|1-i|}) \xrightarrow{\vec{r}}$, there exist $\vec{r}'$, $\vec{r}''$ with $(A, S_{|1-i|}) \xrightarrow{\vec{r}'} (_, S_{|1-i|}')$ with either*

1. *$\vec{r}'\vec{r}'' = \vec{r}$ ($\vec{r}'$ is a prefix of $\vec{r}$), or*
2. *$\vec{r}' = \vec{r}\vec{r}''$ ($\vec{r}$ is an ex prefix of $\vec{r}'$)*

*such that $(A, S) \xrightarrow{\vec{r}'\ell} (A', S')$ and $\emptyset; \emptyset; A' \vdash S_0' \circ S_1' \triangleright_s S'$.*

**Proof 36** *We fix $i = 1$. By Lemma H.3 if $(A, S_1) \xrightarrow{\ell} (A', S_1')$ then $\ell \in \mathtt{clab}(S_1)$ and hence $\mathtt{clab}(S_1) \neq \emptyset$. Fix any $\vec{r}$ such that $(A, S_0) \xrightarrow{\vec{r}}$. By Lemma H.4 we can rewrite $S_0$ as $C_0[S_0'']$ with $\mathtt{V}(C_0[\cdot]) = \vec{r}$. By Lemma H.7, since $\mathtt{clab}(S_1) \neq \emptyset$, for $C_0$ either*

1. *there exist $C_0'$, $C_0''$, $C[S''] = S$ such that*
   - *$C_0[\cdot] = C_0'[C_0''[\cdot]]$, and*
   - *$\mathtt{V}(C_0'[\cdot]) = \mathtt{V}(C[\cdot])$, and*
   - *$\mathtt{clab}(S'') = \mathtt{clab}(S_1)$, or*
2. *there exist $C_0'[S_0']$, $C[S''] = S$ such that*
   - *$C_0[C_0'[S'']] = S_0$, and*
   - *$\mathtt{V}(C_0[C_0'[\cdot]]) = \mathtt{V}(C[\cdot])$, and*
   - *$\mathtt{clab}(S'') = \mathtt{clab}(S_1)$*

*In case (1) above, we can write* $S_0$ *as* $C'_0[S'''_0]$ *for some* $S'''_0$, $\vec{r'} = \mathtt{V}(C'_0)$, *and* $\vec{r''} = \mathtt{V}(C''_0)$. *By Lemma H.4*

$$(A, C'_0[S'''_0]) \xrightarrow{\vec{r'}} (_, S'_0)$$

*for some* $S'_0$. *Since* $\mathtt{clab}(S'') = \mathtt{clab}(S_1)$ *then* $\ell \in \mathtt{clab}(S'')$. *Since* $A\{S\}$ *by hypothesis (it is a composition) and* $\ell \in \mathtt{clab}(S'')$ *then by Lemma H.5*

$$(A, C[\,S''\,]) \xrightarrow{\vec{r'}\ell} (A', S')$$

*for some* $A'$ *and* $S'$.

*In case (2) above, we set* $\vec{r'} = \mathtt{V}(C_0[\,C'_0[\,\cdot\,]\,])$ *and we can write* $S_0$ *as* $C_0[\,C'_0[\,S'''_0\,]\,]$ *for some* $S'''_0$. *By Lemma H.4*

$$(A, C_0[\,C'_0[\,S'''_0\,]\,] \xrightarrow{\vec{r'}} (_, S'_0)$$

*for some* $S'_0$. *Since* $\mathtt{clab}(S'') = \mathtt{clab}(S_1)$ *then* $\ell \in \mathtt{clab}(S'')$. *Since* $A\{S\}$ *by hypothesis (it is a composition) and* $\ell \in \mathtt{clab}(S'')$ *then by Lemma H.5*

$$(A, C[\,S''\,]) \xrightarrow{\vec{r'}\ell} (A', S')$$

*for some* $A'$ *and* $S'$.

*In both case (1) and case (2) above, it remains to prove that*

$$\emptyset; \emptyset; A' \vdash S'_0 \circ S'_1 \triangleright_s S'$$

*For every transition* $r \in \vec{r}$, *by Lemma 6 (1) the composition relation is preserved; this can be shown proceeding as in Lemma H.8.*

*The case for* $i = 1$ *is symmetric (proceeds similarly, thanks to symmetric rules of composition and transition of protocols ensembles).*

## About the authors

**Laura Bocchi** l.bocchi@kent.ac.uk.

**Dominic Orchard** d.a.orchard@kent.ac.uk.

**A. Laura Voinea** laura.voinea@glasgow.ac.uk.